\documentclass{aa}
\usepackage{graphicx}
\usepackage{natbib}
\usepackage{float}
\usepackage{colortbl}
\usepackage[caption=false]{subfig}
\usepackage{wrapfig}
\usepackage{amsmath}
\usepackage{latexsym}
\usepackage{amssymb}
\usepackage{epsf}
\usepackage{booktabs}
\usepackage{placeins}    
\usepackage[breaklinks=true]{hyperref}
\bibpunct{(}{)}{;}{a}{}{,}

\definecolor{burgundy}{rgb}{0.50,0.00,0.13}

\graphicspath{{Figures/}}

\bibpunct{(}{)}{;}{a}{}{,}

\begin{document}
\newcommand{\Tc}[1]{T_{\mathrm{C}}^{\text{#1}}}
\newcommand{\Tceff}[1]{T_{\mathrm{C,eff}}^{\text{#1}}}

\title{Proto-planetary disk composition-dependent element volatility in the context of rocky planet formation}
\titlerunning{Proto-planetary disk composition-dependent element volatility} 

\author{Rob J. Spaargaren\inst{1}\inst{2} \and Oliver Herbort\inst{3} \and Haiyang S. Wang\inst{4} \and Stephen J. Mojzsis\inst{5}\inst{6}\inst{7} \and Paolo Sossi\inst{2}}

\institute{Kapteyn Astronomical Institute, Rijksuniversiteit Groningen, Landleven 12, 9747 AD Groningen, the Netherlands \email{spaargaren@astro.rug.nl} 
\and Institute of Geophysics, ETH Zurich, Sonneggstrasse 5, 8092 Zurich, Switzerland
\and Department of Astrophysics, University of Vienna, T\"urkenschanzstrasse 17, 1180 Vienna, Austria \email{oliver.herbort@univie.ac.at}
\and Center for Star and Planet Formation, Globe Institute, University of Copenhagen, {\O}ster Voldgade 5-7, 1350 Copenhagen, Denmark \email{haiyang.wang@sund.ku.dk}
\and Bavarian Geoinstitute for Experimental Geochemistry and Geophysics, University of Bayreuth, Universitätsstraße 30, 95447 Bayreuth, Germany
\and HUN-REN Research Centre for Astronomy and Earth Sciences, MTA Centre of Excellence, 15-17 Konkoly Thege Miklos ut, Budapest, 1121 Hungary
\and Department of Geological Sciences, University of Colorado, UCB 399, 2200 Colorado Avenue, Boulder, 80309 Colorado, USA
}

\date{Received ... / Accepted ...} 

\abstract{}
{The inferred compositions of the Solar System terrestrial (rocky) bodies are fractionated from that of the Sun, where elemental depletions in the bulk rocky bodies correlate with element volatility, expressed in its 50\% condensation temperature. However, because element volatility depends on disk gas composition, it is not mandated that elemental fractionation trends derived from the solar-terrestrial scenario apply to other planetary systems. Here, we expand upon previous efforts to quantify element volatility during disk condensation, and how this affects rocky planet compositional diversity.} 
{We simulate condensation sequences for a sample of 1,000 initial disk compositions based on observed stellar abundances. Based on these simulations, we present parametrisations of how element 50\% condensation temperatures depend on disk composition, and apply element fractionation trends with appropriate element volatilty to stellar abundances to simulate compositions of rocky exoplanets with the same volatile depletion pattern as the Earth, providing a robust and conservative lower limit to the compositional diversity of rocky exoplanets.}
{Here we show that Earth-like planets emerge from low-C/O disks ($\mathrm{C/O}\leq0.75$) and graphite-bearing planets from medium-to-high-C/O disks ($\mathrm{C/O}>0.75$). Furthermore, we identify an intermediate-C/O (0.84–1.04) class of planets characterized by Mg and Si depletion, leading to relatively high abundances of Fe, Ca, and Al. We show that devolatilisation patterns could be adapted potentially with disk composition-dependent condensation temperatures to make predictions of rocky planet bulk compositions within individual systems, although such patterns could be further modified by the dynamics of planetary accretion that remains under-constrained for most of exoplanetary systems. The outcomes of our analysis suggest that accounting for disk composition-dependent condensation temperatures means that we can expect an even broader range of possible rocky planet compositions than has previously been considered.}
{}


\maketitle

\section{Introduction} \label{sec:Intro}
The interior properties of a rocky planet govern the evolution of surface conditions over geological timescales - and thus their habitable potential - through extended interior-atmosphere interaction. These properties are largely determined by the planet's bulk composition, which controls interior structure and mantle mineralogy, and in turn sets key material properties \citep{Noack2014_structure,Stamenkovic2016,Unterborn2017,ONeill2020_GCE,Spaargaren2020}. Constraining the range of possible bulk compositions of rocky planets is therefore a necessary step in understanding the interior dynamics and hence the nature of secondary (degassed) vs. hybrid (i.e. mixed primary and secondary) atmospheres. 

It is not possible to directly sample exoplanet compositions, which requires us to rely on models that establish plausible constraints. In particular, the composition of a planet ultimately arises from the proto-planetary disk (PPD) from which it forms, which in turn shares its composition with the host star \citep[e.g.\ ][]{Lissauer1993}. Although some studies directly translate stellar abundances into rocky planet compositions for all but the most volatile elements \citep[e.g.\ ][]{Unterborn2017,Putirka2019,Bitsch2020}, it is well documented that the Earth \citep{Halliday2001,Palme2014,Wang2019}, Mars \citep{Yoshizaki2020, khan2022geophysical}, and Vesta \citep{Sossi2022} define depletion patterns in elements compared to the Sun that are correlated with element volatility \citep[as alluded to by e.g.\ ][]{Urey1953}, also known as a devolatilisation trend. The volatility of an element is frequently defined with respect to its half-condensation temperature, $\Tc{}$, which describes the temperature at which 50 mol\% of an element has condensed from a gas of solar composition \citep{Lodders2003,Wood2019}, that is;
\begin{equation}
    d_i(\Tc{i}) = 0.5
    \label{eq:Tc_definition}
\end{equation} 
where $\Tc{i}$ is the 50~\% condensation temperature and $d_i$ is the depletion of element $i$ from the gas, $d_i(T) = X^i(T)/X^i(T_0)$ where $X^i$ refers to the total mole fraction of element $i$ compared to the total amount of material in the disk (i.e.\ both gaseous and solid material), and $T_0$ is a reference high temperature at which all elements remain fully in the gas phase (i.e.\ no condensation has occurred). 

While the exact shape of the devolatilisation trend is known only for a few rocky bodies in the Solar System, observations of polluted white dwarfs provide indirect evidence of devolatilised planetary material \citep{Harrison2018,Harrison2021}. These observations indicate fractionation between refractory elements (those with high condensation temperatures, such as Al, Ca, Mg, and Fe) and moderately volatile elements (MVE; with condensation temperatures between 600\,K and 1300\,K, such as Na and K). Further, most proposed planet formation processes also impose a volatility-dependent element depletion, such as collisional erosion \citep{ONeill2008}, stochastic accretion of incompletely condensed material \citep{Sossi2022}, or pebble ablation in the proto-atmosphere \citep{Steinmeyer2023}. These findings suggest that rocky exoplanets may also exhibit volatility-dependent element fractionation relative to their host star, although the exact shapes of these patterns remain unconstrained.

A number of studies have applied the Earth-Sun devolatilization pattern to convert stellar abundances into plausible rocky exoplanet compositions \citep{Wang2019MNRAS,Wang2022_alphacentauri,Wang2022_planethoststars,Spaargaren2023}. Use of this trend, however, implicitly assumes that the volatilities of elements in the solar disk are independent of stellar composition. In reality, the volatility of an element depends upon the bulk disk composition. This is especially relevant under oxygen-poor conditions, where most oxygen exists as generally unreactive CO gas. Consequently, oxygen bound in CO(g) is no longer available to form silicates and oxides \citep{Gail2014,Pontoppidan2014}. This unavailability of oxygen leads to the formation of reduced condensates and changes the condensation temperatures of most elements \citep[e.g.\ ][]{Larimer1975,Larimer1979}. As a result, it is postulated that some rocky planets can form with a large amount of graphite \citep{Kuchner2005,Madhusudhan2012,Bond2010,Moriarty2014,Hakim2019}. Silicon carbide, however, is only stable under the most extreme oxygen-depleted disk compositions (fO$_2<\mathrm{IW}-6$; \citet{Takahashi2013,Hakim2018,Allen2020}). Similarly, the volatility of sulphur depends on the relative abundances of sulphur and iron \citep{Jorge2022}, which potentially has an impact on the metallic iron core composition of rocky planets. In addition to these chemical effects, the condensation reactions of most elements \citep[and thus element volatility;][]{Timmermann2023} are affected by the total gas pressure, which, all else being equal, increases $p_i$ ($p$ = partial pressure). Thus, accounting for the various factors that control element volatility can lead to shifts in the nature of the building blocks of rocky planets in ways that cannot be predicted by simply extrapolating from stellar abundances.

The composition of a rocky planet is shaped not only by the nature of its building blocks, but also by which of those materials it accretes. Within the Solar System, this has led to planets with various devolatilization trends \citep[cf.\ Earth, Mars, and Vesta;][]{Yoshizaki2020,khan2022geophysical,Sossi2022}, and to Mercury, whose large core size precludes description by a monotonic devolatilization pattern. Previous studies have explored such compositional diversity by coupling condensation sequence models with dynamical planet formation simulations \citep{Bond2010,Carter2012,Elser2012,Moriarty2014,Shakespeare2024,Hatalova2025}, predicting planets rich in Ca and Al, and coreless planets. These studies typically focus on a small sample of planetary systems, which are studied in detail. We here take the opposite approach, determining the condensation sequence for a range of stellar compositions, and inferring the compositions of hypothetical planets from the resulting $T_C^i$ by constraining them to follow a single, well-defined devolatilization trend. Building on earlier studies applying the Earth-Sun devolatilization trend to stellar abundances, we extend this framework by incorporating variations in elemental volatility across different systems, thereby systematically exploring one axis of rocky planet chemical diversity. By adopting the Earth-Sun devolatilisation trend as an empirically observed example of the outcome of planet formation, this approach is agnostic to the nature of the processes that gave rise to this trend. While limited in scope, it provides a physically motivated and observationally grounded framework for estimating a conservative lower bound on the chemical diversity of rocky exoplanets.

Element volatility, represented by $\Tc{}$, can be derived for arbitrary stellar compositions by simulating the condensation sequence of a gas matching those abundances. When studying rocky planet compositions in detail for individual systems \citep[e.g.\ ][]{Jorge2022,Wang2022_planethoststars}, a condensation sequence model can be employed for each set of stellar abundances to determine the shift in element volatilities. More general constraints on element volatility as a function of the bulk composition of the PPD have been placed by the models of \citet{Timmermann2023}, for a limited population of stellar abundances, with C/O ratios below 0.7. Here, we expand upon previous studies by analysing element volatility in disk compositions based on a broader set of stellar abundances, drawn from the GALAH \citep{Buder2021} catalogue. For each set of abundances, we systematically model the condensation sequence and derive $\Tc{}$ values of major rock- and metal-forming elements to derive generalised parametrisations of their 50\% condensation temperatures as functions of disk chemistry. We then apply these volatility trends to stellar abundances via the Earth-Sun devolatilisation pattern, to explore how nebular condensation at chemical equilibrium affects chemical diversity in rocky planets. The Mars-Sun and Vesta-Sun devolatilisation patterns could also, in principle, serve as baselines, but we adopt the Earth-Sun pattern because it is best constrained and directly relevant for anchoring the analysis to Earth-sized, potentially habitable planets. While this approach is limited to a subset of rocky exoplanets, those with the same devolatilisation pattern as the Earth, this approach provides a robust estimate on the lower limit of rocky exoplanet compositional diversity.

Our condensation sequence model and the sample of stellar abundances on which we base our parametrisations are introduced in Sect.\ \ref{sec:Methods}. We present the disk composition-dependent condensation behaviour in Sect.\ \ref{sec:Res_refrac_elems} and \ref{sec:Res_vol_elems} and apply those to stellar abundances to update the range of bulk rocky exoplanet compositions in Sect.\ \ref{sec:exo_bulk_comps}. In Sect.\ \ref{sec:Disc} we discuss our results and their implications, and finally present the main conclusions of our study in Sect.\ \ref{sec:Conc}.

\section{Methodology} \label{sec:Methods}
\subsection{Condensation sequence model}
\label{ssec:Methods_GGchem}
For simulating the condensation sequences of PPDs with various compositions, we use the open-source thermo-chemical equilibrium code \textsc{GGchem} \citep{Woitke2018}. This code calculates the composition of gas and condensate in chemical and phase equilibrium using the law of mass action with molecular equilibrium constants based on Gibbs free energy data, assuming local thermodynamic equilibrium. Gibbs free energy data are taken from the NIST-JANAF database \citep{Chase1982} and the SUPCRTBL database \citep{Johnson1992,Zimmer2016}. We simulate the composition of solids in equilibrium with the gas as a parcel of gas cools down from 2500\,K to 400\,K. The Earth-Sun devolatilization trend is not well constrained for elements with $\Tc{}$ below 400\,K \citep{Wang2019}. Gas pressure in PPDs tends to vary with stellar mass \citep{Andrews2013}, and condensation temperatures scale linearly with pressure \citep{Timmermann2023} for all elements except S \citep{Wasson1985}. Since this relationship allows for straightforward adjustments, we first analyse condensation behaviour at $10^{-4}$\,bar, representative for the solar disk at 1 AU \citep{Lewis1974,Fegley2000}, and then correct condensation temperature for pressure using results from Timmermann et al. The model contains 24 elements (H, He, Li, C, N, O, F, Na, Mg, Al, Si, P, S, Cl, K, Ca, Ti, V, Cr, Mn, Fe, Ni, Zr, W), 552 gaseous molecule species, and 240 condensate species. In this study, we focus on constraining $\Tc{}$ values for the nine most abundant elements in the Earth (O, Mg, Si, Fe, Ca, Al, Na, Ni, and S), as well as carbon. Accordingly, we include these ten elements, along with H, He, and additional elements (Cl, K, Ti, and N) that may form condensates with them.

Condensation temperatures, simulated with \textsc{GGchem} using proto-solar abundances from \citet{Lodders2003}, match their results within 20\,K for all elements except Ni (Table \ref{tab:Tc_validation}), and the results of \citet{Wood1993} within 25\,K for all elements except Ni and Na. \cite{Lodders2003} notes that Ni first appears in solid form in Fe-Ni alloys. However, \textsc{GGchem} currently does not include solid solutions such as metal alloys. As a result, Ni condenses as pure Ni metal in our simulations, which occurs at lower temperatures than the condensation of Fe-Ni alloys. Since most elements considered here do not condense in Fe alloys, \textsc{GGchem} accurately reproduces known condensation behaviour and is expected to reliably predict element condensation temperatures across a wide range of bulk PPD compositions.

\begin{table}[h]
\centering
\caption{Solar disk condensation temperatures based on \textsc{GGchem}.}
\label{tab:Tc_validation}
\begin{tabular}{cccc}\\\hline \hline  
Element & \textsc{GGchem} & L03 & W19 \\ \hline
O & $<$400 & 180 & 183 \\ 
Na & 941 & 958 & 1035 \\  
Mg & 1336 & 1336 & 1343 \\  
Al & 1652 & 1653 & 1652 \\  
Si & 1320 & 1310 & 1314 \\  
S & 661 & 664 & 672 \\  
Ca & 1512 & 1517 & 1535 \\  
Ti & 1571 & 1582 & 1565 \\ 
Fe & 1332 & 1334 & 1338 \\  
Ni & 1282 & 1353 & 1363 \\ \hline
\end{tabular} \\
\textbf{Notes.} Condensation temperatures $\Tc{}$ (K) of elements in the solar disk simulated with \textsc{GGchem} ($\pm$ 5\,K) using solar abundances from \citet{Lodders2003}, $\Tc{}$ values from \citet{Lodders2003} (L03), and $\Tc{}$ values from \citet{Wood2019} (W19).
\end{table} 

\subsection{Stellar abundances}
\label{ssec:Methods_StellarAbundances}
We base disc compositions on observed stellar abundances from the freely available GALAH catalogue \citep[DR 3.2,][]{Buder2021}. Abundances in the GALAH catalogue are reported as [X/Fe] = [X/H] - [Fe/H], alongside [Fe/H], normalised to solar abundances from \citet{Asplund2009}. To obtain logarithmic abundances normalised to H = 12 ($A_i$) for \textsc{GGchem}, we compute $A_i$ = $A_{\odot,i}$ + [X/H]$_i$ for solar abundances $A_{\odot}$, following \citet{Hinkel2022}. For molar ratios, such as C/O, we use non-logarithmic abundances $\epsilon_i = 10^{A_i}$, so $\mathrm{C/O}= \epsilon_{\mathrm{C}}/\epsilon_{\mathrm{O}}$. We filter data according to GALAH best practices\footnote{\url{https://www.galah-survey.org/dr3/using_the_data/}}, to include only entries with the general data quality flag flag\_sp=0, and element quality flags flag$\_$x$\_$fe = 0 for the elements C, O, Fe, Mg, and Si. We select for these five elements because C and O strongly affect solid composition \citep{Larimer1975}, and Fe, Mg, and Si are the most abundant non-volatile elements in the Earth. We remove non-main sequence stars, to prevent systematic errors, by discarding stars with low surface gravitational acceleration $g$ (log $g < 3.5$). After applying these criteria, our final dataset consists of 91,982 stars.

Of the 16 selected elements, only abundances of our five selection criteria elements (C, O Fe, Mg, and Si), Na, Al, K, Ca, Ti, and Ni are commonly available in our dataset. As input for \textsc{GGchem}, we set H to 12 and He to the solar value of 10.93, as it is an inert gas and will only have a minor effect on condensation chemistry, given the distribution of stellar He abundances in the catalogue. The elements N, S, and Cl are not included in the GALAH catalogue, so we estimate abundances of these elements based on available abundances of other elements. Elemental abundances often correlate due to nucleosynthesis \citep[e.g.\ ][]{Pignatari2023}, and these correlations are well established for S and N \citep{Nicholls2017}, though they are uncertain for Cl \citep{Brauner2023}. Thus, for Cl we use solar value scaled to stellar metallicity, calculated as the sum of $\epsilon_i$ all available elements except H and He, minus the sum of solar abundances of those same elements.

For N and S, we estimate abundances based on galactic chemical evolution (GCE) predictions. As N is typically produced alongside O and C, GCE models predict its co-evolution with O \citep{Kobayashi2020}. Accordingly, we adopt a parametrisation of [N/O] based on [O/H] \citep{Nicholls2017},
\begin{equation}
[N/O] = \log_{10} \left( 10^{-1.732} + 10^{A_O - 12 + 2.19} \right),
\end{equation}
which has been validated across galactic environments \citep{Vincenzo2018}. The production of S typically occurs simultaneous with the production of Si via explosive O-burning \citep{Pignatari2016}, and S and Si are ejected in similar supernova events \citep{Kobayashi2020}. Therefore, GCE models predict that S/Si should be approximately constant in time and equal to solar \citep{Brauner2023,Pignatari2023}. This is confirmed by observations of stellar abundances \citep[e.g.\ ][]{Chen2002}. Following previous studies using the GALAH catalogue \citep{Bitsch2020}, we set [S/Si] = 0 for all stars.

\begin{figure}
    \resizebox{\hsize}{!}{\includegraphics{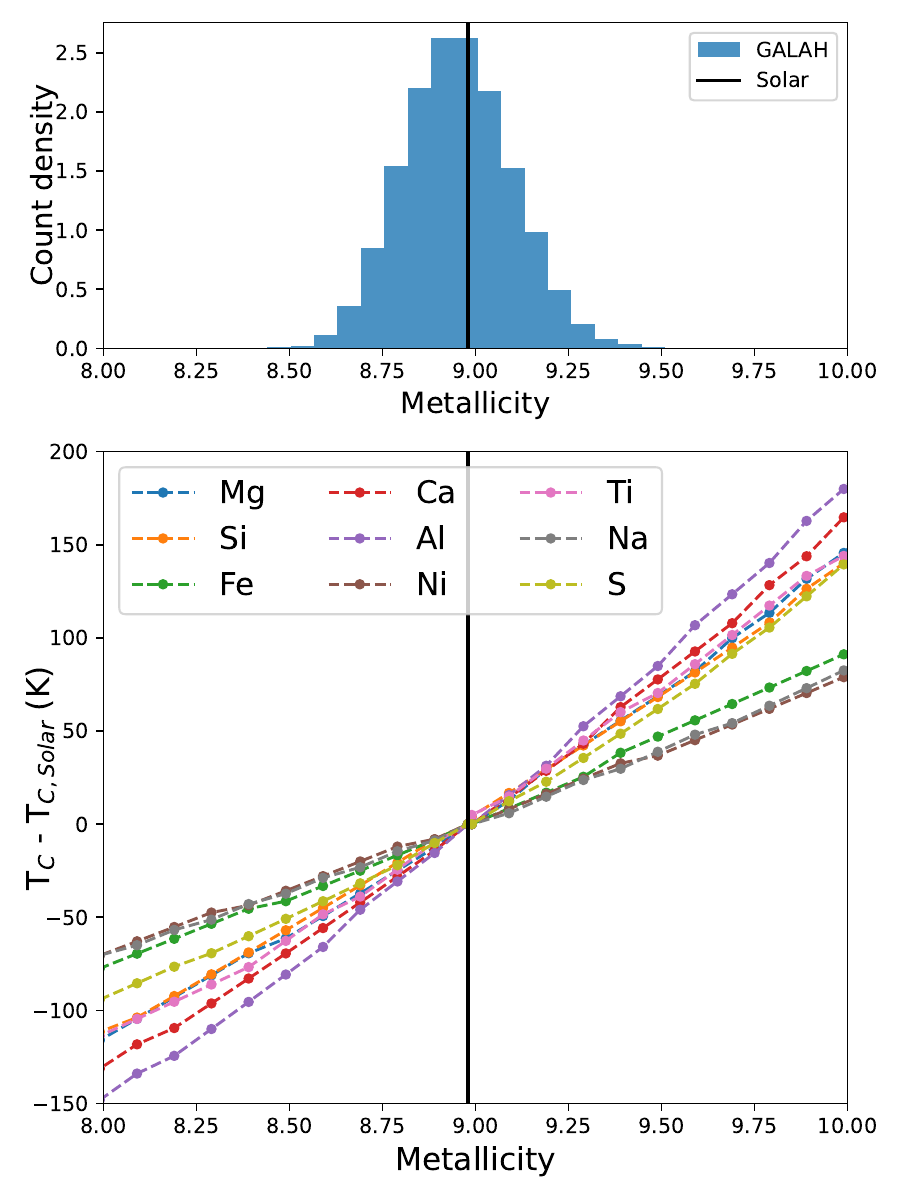}}
      \caption[Tc dependence on Metallicity]{Histogram of stellar metallicity (top) in the GALAH catalogue with solar metallicity for comparison \citep{Asplund2009}, and 50\% condensation temperature of various elements as a function of host stellar metallicity (bottom). Each column of condensation temperatures is calculated from a condensation sequence with solar abundances scaled to the corresponding metallicity. Condensation temperatures are shown relative to the simulation with unscaled solar abundances (Table \ref{tab:Tc_validation}).}
      \label{fig:Metallicity_dependence}
\end{figure}

Since all elemental abundances correlate with stellar metallicity, allowing metallicity to vary within our sample could obscure the true dependence of condensation temperatures on individual elemental abundances. As shown by \citet{Timmermann2023}, condensation temperatures tend to increase linearly with metallicity, similar to their dependence on total pressure - this result arises from the same phenomenon - an increase in the partial pressure of the major gas species. Thus, we normalise stellar abundances in our sample to solar metallicity $M_{\odot}$ by adding the difference between the metallicity of that star and the solar metallicity, $M_{\odot} - M_i$, to each star's $A_i$ values. We define metallicity as
\begin{equation}
    M_i = \log \left( \sum_i \epsilon_i \right),
    \label{eq:Metallicity}
\end{equation}
for all elements $i$ with measured abundances except H and He, and solar abundance $A_{\odot,i}$ from \citet{Asplund2009}. This lets us isolate the effect of uniformly increasing all elemental abundances (i.e.\ increasing metallicity) from the effects of varying individual element abundances. We ran a series of \textsc{GGchem} simulations using solar abundances, varying metallicity $M$ from 8.0 to 10.0 (cf.\ $M_{\odot} = 8.99)$. This was achieved by adding $M - M_{\odot}$ to each abundance $A_i$, after which we computed the 50\% condensation temperatures from the resulting condensation sequences. Within the GALAH catalogue, stellar metallicities span an order of magnitude, resulting in $\Tc{}$ variations ranging from 72\,K for Ni to 165\,K for Al (Fig.\ \ref{fig:Metallicity_dependence}; Sect.\ \ref{sec:App_TcParam}). Thus, we simulate condensation sequences for metallicity-normalised abundances, determine the dependence of $\Tc{i}$ on elemental abundances, and add the metallicity dependence of $\Tc{i}$ separately. After filtering for data availability and reliability, and normalising for metallicity, we remove 780 outliers using a Grubbs test for normal distributions. This step leaves us with a dataset of 91,202 stars.

\begin{figure}
    \resizebox{\hsize}{!}{\includegraphics{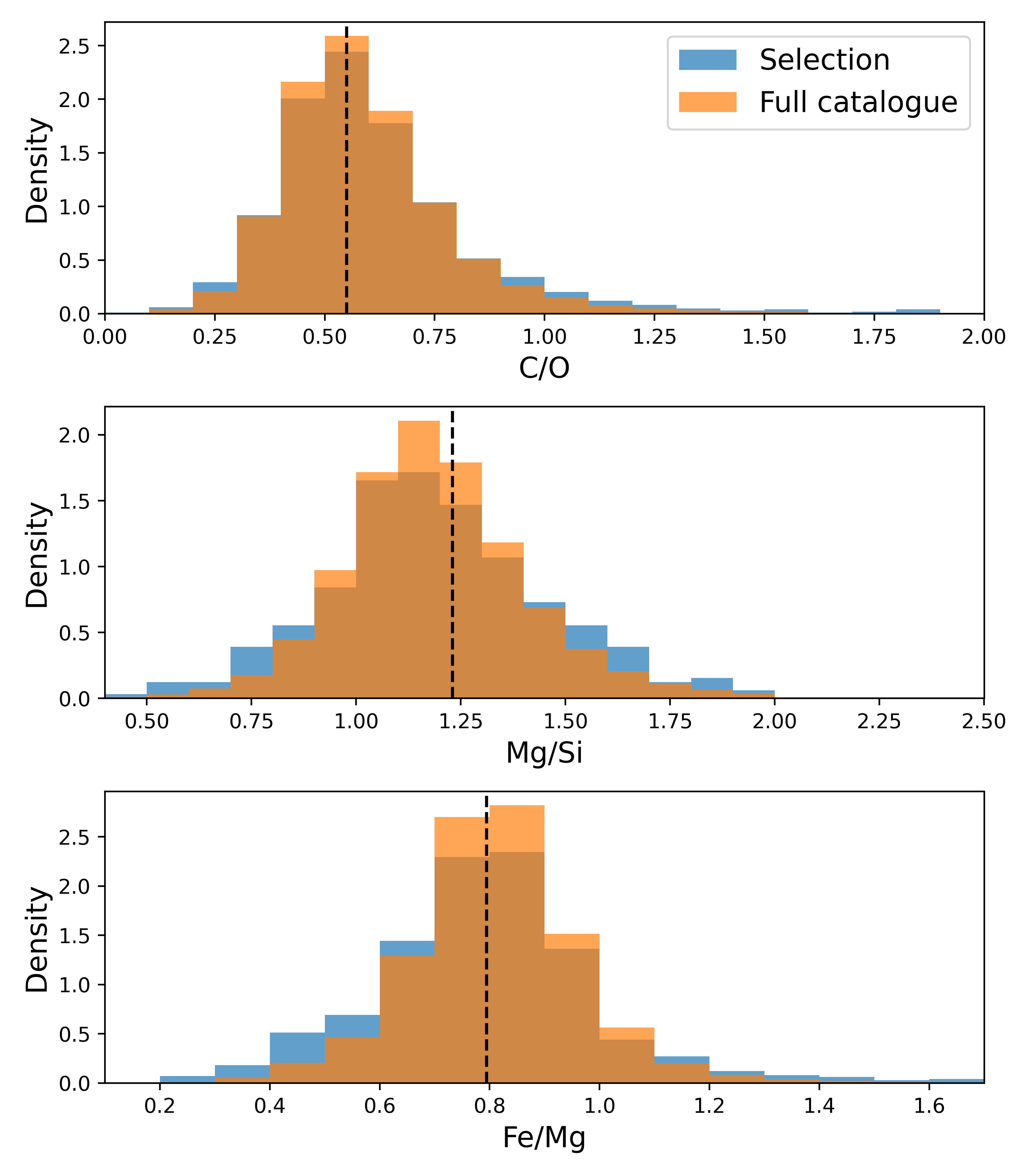}}
      \caption[Stellar abundances within catalogue and sample]{Stellar abundances within the GALAH catalogue (orange; $N=91,202$) and within the sample for which we run condensation sequence models (blue; $N=1,000$). Given our adjusted selection procedure, the selected sample places more emphasis on the tails of the distribution presented by the GALAH catalogue. Solar abundances are shown as a black dashed line \citep{Asplund2009}.}
      \label{fig:Selected_abundances}
\end{figure}
We run condensation sequence simulations for a representative subset of this dataset, which we select using a refined version of the \citet{Chaudhuri1994} algorithm. In the original algorithm, the point with the highest data density is selected. This point, alongside its $k = N/p$ neighbours, is removed, where $N=91,202$ is the dataset size and $p=1,000$ is the desired sample size. The process is repeated until the sample size is reached. This results in an over-representation of median-value points, reducing the information gained per condensation sequence simulation and under-representing the parameter space boundaries. Instead, we select the point at the lowest data density, ensuring better coverage of the outer parameter space. We use this approach to select a final dataset of 1,000 stars, which covers the same compositional space as the original catalogue, but with more emphasis on the tails of the distributions (Fig.\ \ref{fig:Selected_abundances}).

\subsection{Bulk planet compositions}
\label{ssec:Methods_PlComps}
Finally, to simulate bulk planet compositions, we include disk composition-dependent element volatility by deriving parameterisations of 50\% condensation temperatures as functions of stellar abundances from our condensation simulations. We calculate the 50\% condensation temperatures ($\Tc{i}$ for element $i$, in K) from our simulations as the highest temperature at which the gas-phase abundance of element $i$ drops below 50\% (or $\log_{10} 0.5 = -0.301$) of its initial value (Eq.\ \ref{eq:Tc_definition}). We use least squares fitting to determine the optimal parameters to explain variation in $\Tc{i}$, selecting the subset that minimises the Bayesian information criterion. We then simulate bulk rocky planet compositions by multiplying stellar compositions $\epsilon_i$ with element depletion factors $f_i$, with depletion factors based on the Earth-Sun devolatilization trend,
\begin{equation}
    \log f_i = \alpha \log \Tc{i} + \beta,
    \label{eq:devol_trend_Wang}
\end{equation} 
where $\alpha = 3.676\pm 0.142$ and $\beta = -11.556 \pm 0.436$ \citep{Wang2019}, and $\Tc{i}$ is based on our parametrisations and the stellar abundances. This states that the elemental fractionation between Star-Planet is equivalent to that between the Sun-Earth, with depletion factors based on element $\Tc{i}$, which change depending on the input composition and pressure. Using this exact relation between $\Tc{i}$ and $f_i$ results in compositions of planets with the same devolatilisation trend as Earth, providing a lower limit to rocky exoplanet compositional diversity.

\section{Condensation behaviour of refractory elements} \label{sec:Res_refrac_elems}
\subsection{Metal-forming elements}
\begin{figure*}
    \resizebox{\hsize}{!}{\includegraphics{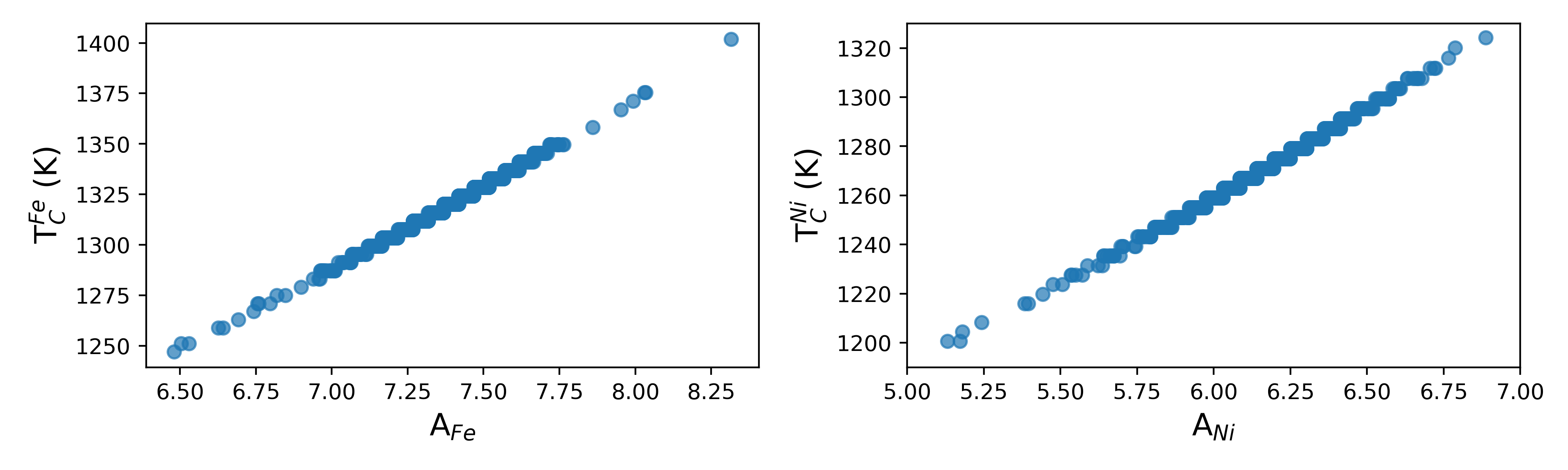}}
      \caption[Tc dependence of Fe and Ni]{Condensation temperatures ($\Tc{}$) of Fe and Ni, plotted against the logarithmic abundances of Fe and Ni (normalised to H=12; $A_{\mathrm{Fe}}$, $A_{\mathrm{Ni}}$) in those disks. Condensation temperatures are derived from condensation sequence calculations at $P=10^{-4}$\,bar.}
      \label{fig:Tc_FeNi}
\end{figure*}
For each set of stellar abundances, we simulate the condensation sequence of a disk gas with matching composition, and calculate 50\% condensation temperatures $\Tc{i}$. The primary constituents of metallic cores of rocky planets, Fe and Ni, exhibit a near-linear relationship between their condensation temperatures and abundances (Fig.\ \ref{fig:Tc_FeNi}). In our model, Fe and Ni exist as monatomic gases and condense primarily as pure metals, Fe$^0$(s) and Ni$^0$(s), with their condensation temperatures scaling with their partial pressures. Increasing Fe and Ni abundances raise the partial pressures of monatomic Fe and Ni gases, enhancing their condensation temperatures - similar to the effects of disk pressure and metallicity \citep{Timmermann2023}. Fe and Ni can form oxides in O-rich disks, but within our sample, this occurs only at lower temperatures \citep[below 600\,K, see also][]{Lewis1974} and does not affect their 50~\% condensation temperatures.

\subsection{Refractory rock-forming elements}
\begin{figure}
    \centering
    \resizebox{\hsize}{!}{\includegraphics{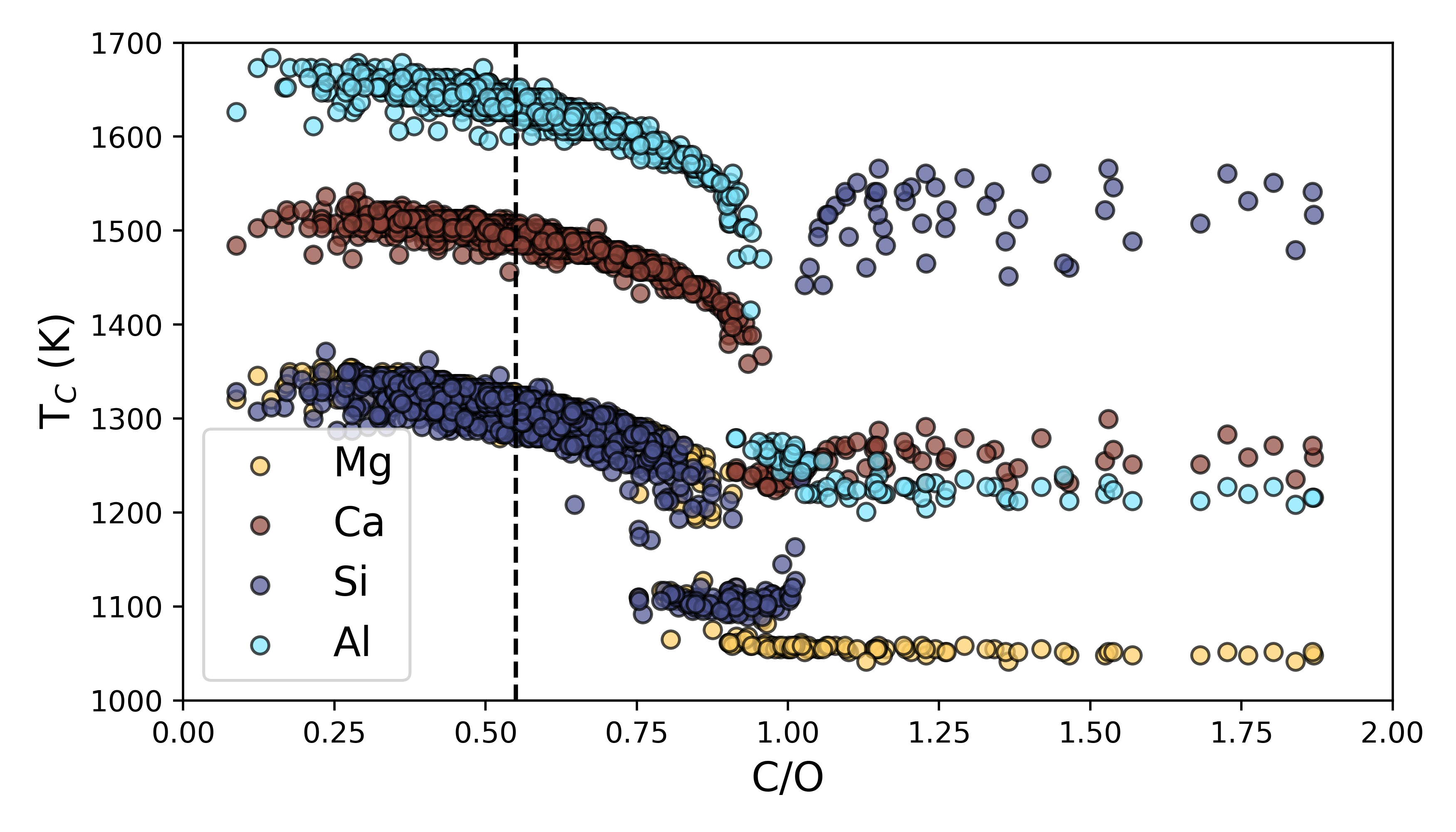}}
    \vspace{0.5cm}
    \resizebox{\hsize}{!}{\includegraphics{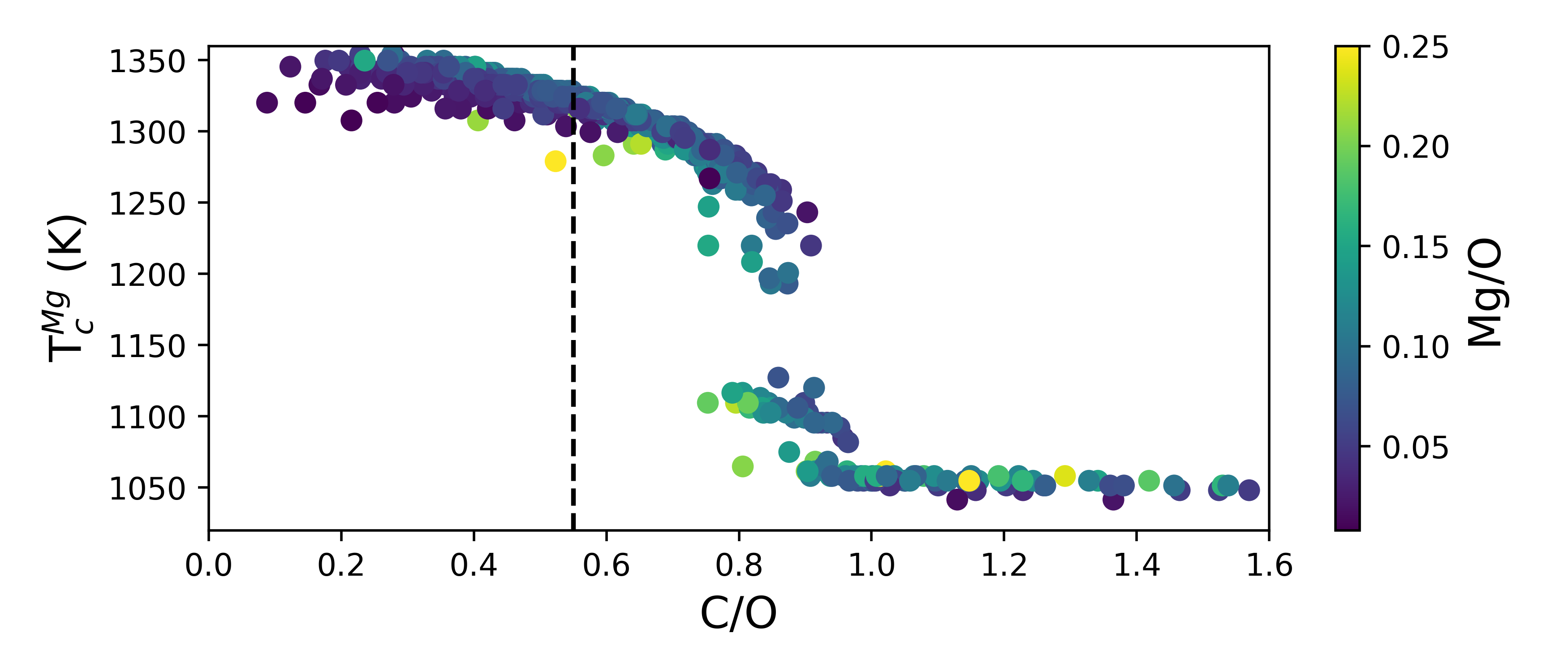}}
    \caption[Tc dependence of Mg, Si, Ca, Al]{Condensation temperatures ($\Tc{}$) of the rock-forming elements Mg, Si, Ca, and Al, plotted against the molar C/O ratio in disks, based on stellar $\epsilon_{\text{C}}/\epsilon_{\text{O}}$ values (top). Solar C/O=0.55 \citep{Asplund2009} is indicated as a dashed vertical line. All elements show a transition from high $\Tc{}$ at low C/O to lower $\Tc{}$ at higher C/O. For Mg, colouring condensation temperature by the disk Mg/O ($=\epsilon_{\text{Mg}}/\epsilon_{\text{O}}$; bottom) reveals that the C/O value at which this transition occurs depends on Mg/O. Condensation temperatures are derived from condensation sequence calculations at $P=10^{-4}$\,bar.}
    \label{fig:Tc_MgSiCaAl}
\end{figure}

The condensation temperatures of the rock-forming elements Mg, Si, Ca, and Al depend directly on disk C/O (Fig.~\ref{fig:Tc_MgSiCaAl}). At low C/O, where sufficient oxygen is available to form silicates and oxides, $\Tc{}$ shows a weak decrease with increasing C/O. As C/O increases and oxygen becomes limited, each element transitions to O-depleted condensation at a specific C/O ratio, and their condensation temperatures drop by 150-200 K across this transition. The C/O value at which an element transitions to O-depleted condensation behaviour scales with their condensation temperature in O-rich conditions. Refractory elements transition to O-depleted condensation at higher C/O ratios (Ca: $\mathrm{C/O}=0.936$, Al: $\mathrm{C/O}=0.94$) than the moderately refractory elements (Mg: $\mathrm{C/O}=0.875$, Si: $\mathrm{C/O}=0.828$). At C/O beyond these transitions, both $\Tc{Mg}$ ($\mathrm{C/O}=0.94$) and $\Tc{Al}$ ($\mathrm{C/O}=1.039$) decrease by $\sim50$\,K before stabilizing and becoming independent of C/O. Note that Mg can condense as MgS at high C/O, but this phase is only stable over a narrow temperature range (20-100\,K; Fig.\ \ref{fig:App_high_CO_condensates}), and does not condense in sufficiently high abundances to affect $\Tc{Mg}$. In contrast, $\Tc{Ca}$ rises slightly ($\sim 40$\,K) at $\mathrm{C/O}=1.045$, while $\Tc{Si}$ increases by over 400\,K within a narrow range ($\mathrm{C/O}=1.022-1.045$) as the refractory SiC becomes more stable. Beyond this, $\Tc{Si}$ stabilises and becomes independent of C/O. At high C/O, the volatility order of rock-forming elements (aside from O) differs significantly from that at near-solar C/O. Although at near-solar C/O Si is the most volatile of these four elements, it becomes the most refractory at high C/O, followed by Ca and Al with Mg becoming the most volatile.

While C/O primarily governs the broad shift from O-rich to O-depleted condensation, the specific C/O at which each element transitions also depends on its relative abundance to oxygen ($\mathrm{X/O} = \epsilon_{\mathrm{X}}/\epsilon_{\mathrm{O}}$). Since this transition occurs when there is no longer sufficient oxygen to oxidise that element, it is inherently sensitive to X/O. For example, Mg undergoes O-depleted condensation at disk C/O as low as 0.75 in Mg-rich environments (Mg/O$>0.2$), and as high as 0.92 for Mg-poor environments (Mg/O$<0.05$; Fig.\ \ref{fig:Tc_MgSiCaAl}). The transition width reflects the amount of oxygen required to fully oxidise an element. Elements with lower abundances, such as Ca and Al, require little oxygen to fully oxidize, resulting in narrow transition ranges (0.05 C/O). In contrast, Si, which has a higher abundance and forms SiO$_2$, requires more oxygen and thus transitions over a much broader C/O range (0.26). Similarly, Mg transitions across a range of C/O values of 0.17.

\begin{figure}
    \resizebox{\hsize}{!}{\includegraphics{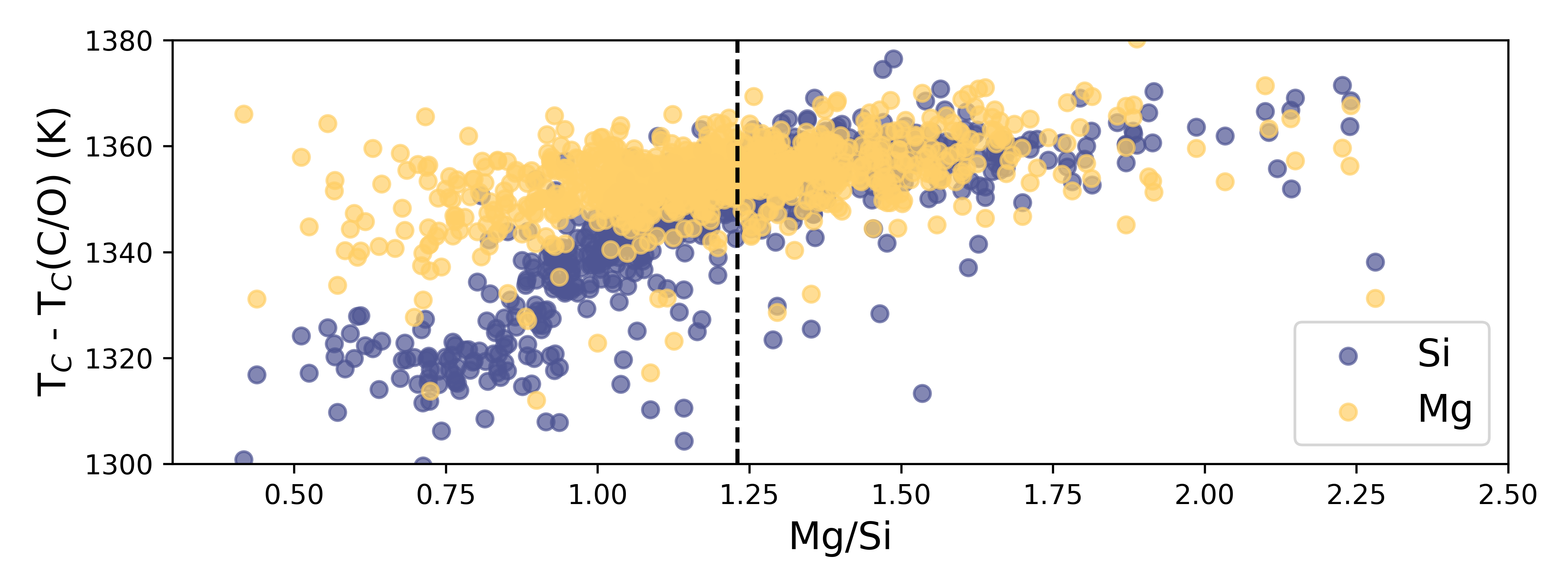}}
      \caption[Tc dependence of Mg and Si on Mg/Si]{Condensation temperatures ($\Tc{}$) of Mg and Si, after subtracting a fit to the C/O-dependence of $\Tc{Mg}$ and $\Tc{Si}$ (Fig.\ \ref{fig:Tc_MgSiCaAl}), plotted against the disk Mg/Si ratios. Here, a trend emerges between $\Tc{}$ and the molar Mg/Si ratio of the disk (bottom) for Si, but not for Mg. Solar Mg/Si=1.23 \citep{Asplund2009} is indicated as a vertical dashed line. Condensation temperatures are derived from condensation sequence calculations at $P=10^{-4}$\,bar.}
      \label{fig:Tc_MgSi_COcorrected}
\end{figure}
In addition to its dependence on C and O, $\Tc{Si}$ is also influenced by the disk molar Mg/Si-ratio, as forsterite (Mg$_2$SiO$_4$, $\Tc{}$ = 1354\,K) condenses at a higher temperature than enstatite (MgSiO$_3$, $\Tc{}$ = 1316\,K; \citealt{Lodders2003}). At Mg/Si $>$ 1.6, forsterite is both the primary Si-bearing condensate and the primary Mg-bearing condensate, leading to $\Tc{Si} \approx \Tc{Mg}$. However, at Mg/Si $<$ 1.6, more Si condenses as the less refractory enstatite, gradually shifting $\Tc{Si}$ to lower temperatures (Fig.\ \ref{fig:Tc_MgSi_COcorrected}). Forsterite still forms initially, but contains less than 50\% of available Si, and thus does not set $\Tc{Si}$ at low disk Mg/Si ratios. In contrast, $\Tc{Mg}$ is independent of Mg/Si within our sample, as forsterite condensation is only limited by Si availability at much higher Mg/Si ratios. Instead, $\Tc{Mg}$ increases with both O and Mg abundances.

Similar to Si depending on Mg/Si, the condensation of Ca is affected by the molar Ca/Al ratio. At Ca/Al $<$ 2.0, Ca condenses in highly refractory minerals such as gehlenite (Ca$_2$Al$_2$SiO$_7$) and Ti-perovskite (CaTiO$_3$). At higher Ca/Al, the less refractory larnite (Ca$_2$SiO$_4$) becomes the dominant Ca-bearing condensate, reducing $\Tc{Ca}$. While the behaviour of Ca aligns with \citet{Timmermann2023}, we find that $\Tc{Al}$ is unaffected by disk Ca abundance. Instead, $\Tc{Al}$ depends primarily on the Al abundance, as corundum (Al$_2$O$_3$), the primary Al-bearing high-T condensate, forms independently of Ca availability.

Overall, the condensation of Mg, Si, Ca, and Al is primarily governed by oxygen availability and each element’s abundance as they condense into rock-forming oxides. However, even when oxygen is abundant, condensation temperatures can vary by up to 100\,K. This variability significantly affects depletion factors (Eq.\ \ref{eq:devol_trend_Wang}), altering the bulk planetary content of these elements by as much as 30\%.

\subsection{Moderately volatile rock-forming elements}
\begin{figure}
    \resizebox{\hsize}{!}{\includegraphics{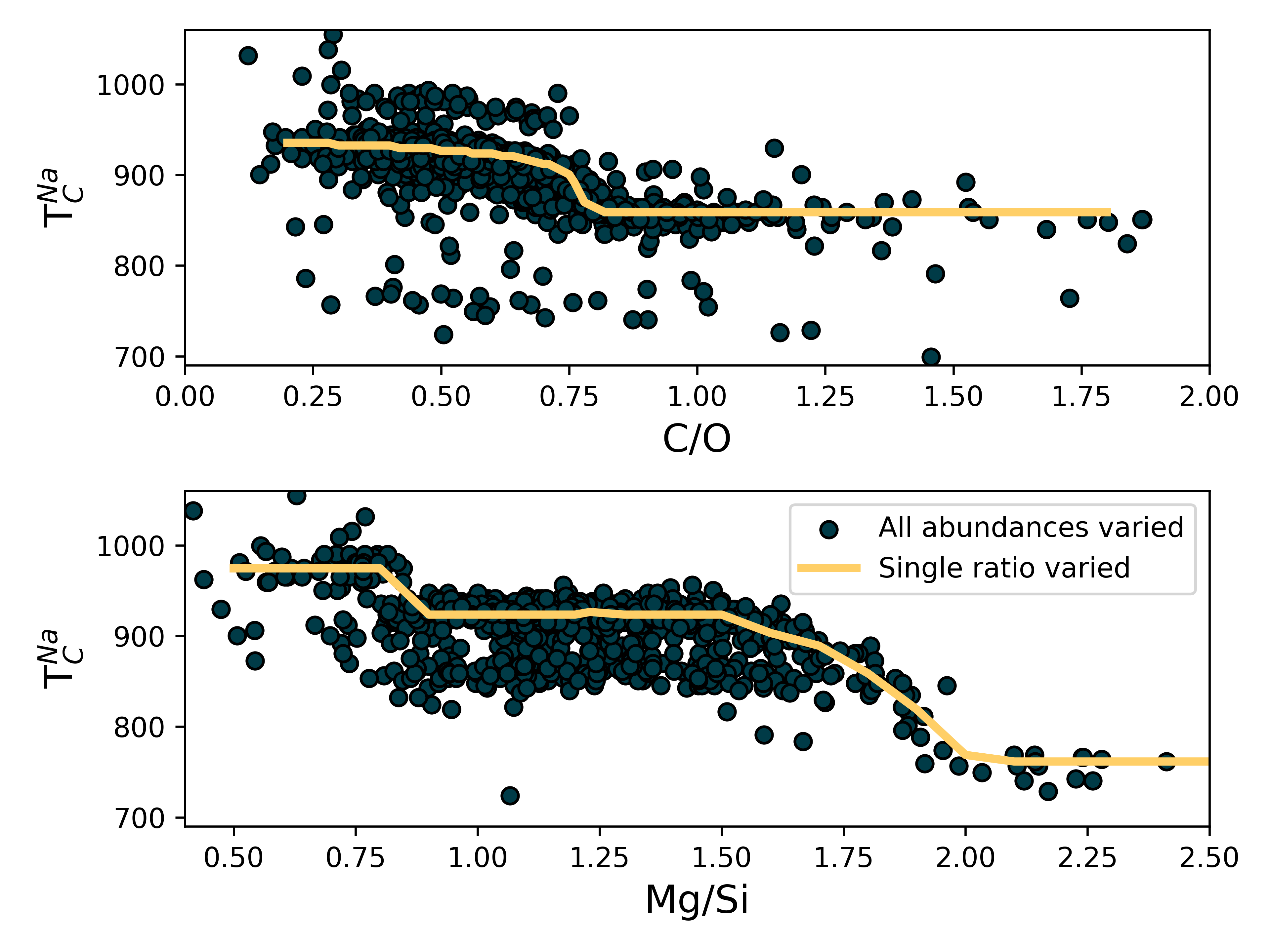}}
      \caption[Tc dependence of Na]{Condensation temperature (T$_c$) of Na, plotted against the molar disk C/O (top) and Mg/Si (bottom). Dots represent simulation results based on stellar abundance data for individual stars. Lines represent results for solar abundances with varying C (top) or varying Mg and Si while conserving bulk Mg+Si (bottom). Temperatures are derived from condensation sequence calculations at $P=10^{-4}$\,bar.}
      \label{fig:Tc_Na}
\end{figure}
Similar to the other rock-forming elements (excluding O), Na transitions from O-rich to O-depleted condensation beyond a critical C/O-ratio of 0.819, leading to a decrease in $\Tc{Na}$. However, the molar disk Mg/Si ratio influences $\Tc{Na}$ more strongly (200\,K shift) than the disk C/O ratio (100 K shift; Fig.\ \ref{fig:Tc_Na}). For most disk compositions, Na condenses primarily as albite (NaAlSi$_3$O$_8$). Albite condenses approximately 50\,K higher when quartz (SiO$_2$) is stable (Fig.\ \ref{fig:App_MgS_stability_vs_MgSi}), as excess Si promotes its formation. This occurs only at very low Mg/Si ratios (Mg/Si $\leq$ 0.83). At Mg/Si $\geq$ 1.5, albite is replaced by nepheline (NaAlSiO$_4$), which is more volatile and remains stable up to Mg/Si=1.97. Beyond this, Na instead condenses as sodium metasilicate (Na$_2$SiO$_3$), a phase with an even lower condensation temperature. This highlights how variations in disk Mg/Si ratios shape the mineralogy of planetary building blocks.

\section{Condensation behaviour of semi-volatile elements} \label{sec:Res_vol_elems}
\begin{figure*}[t]
    \includegraphics[width=0.9\textwidth]{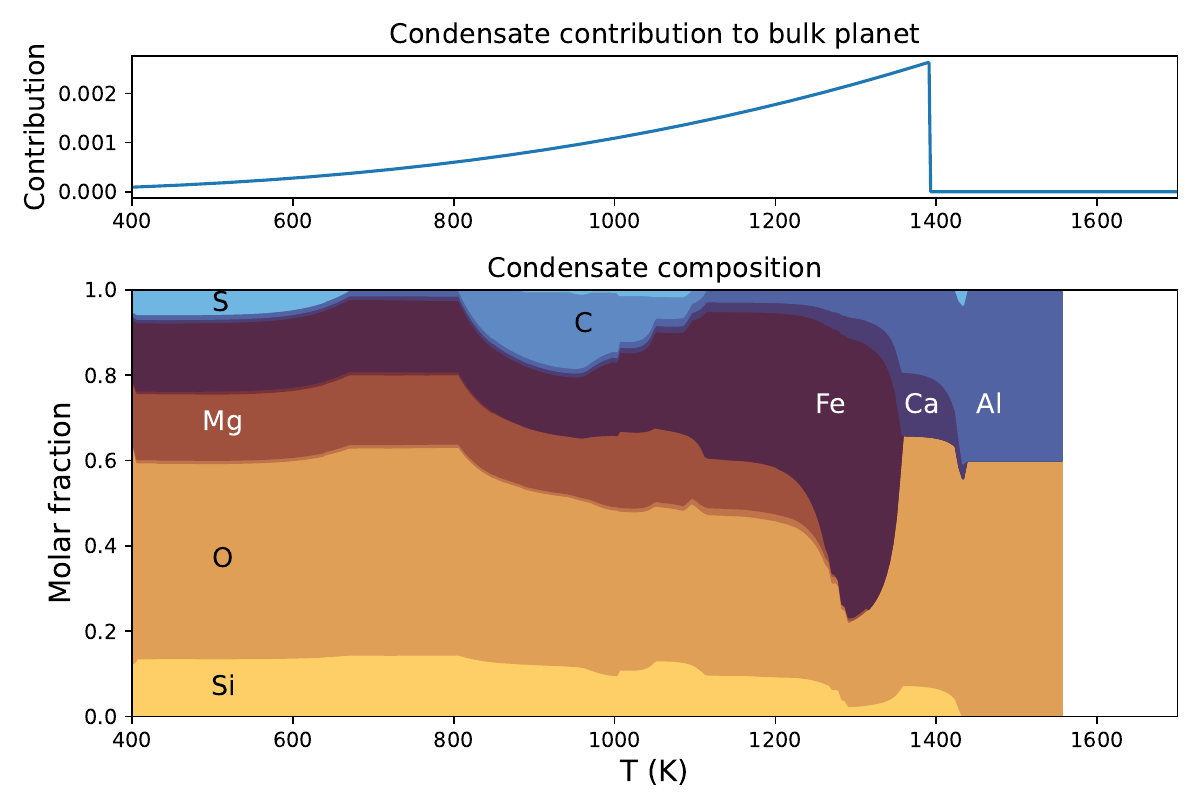}
    \medskip
    \includegraphics[width=0.9\textwidth]{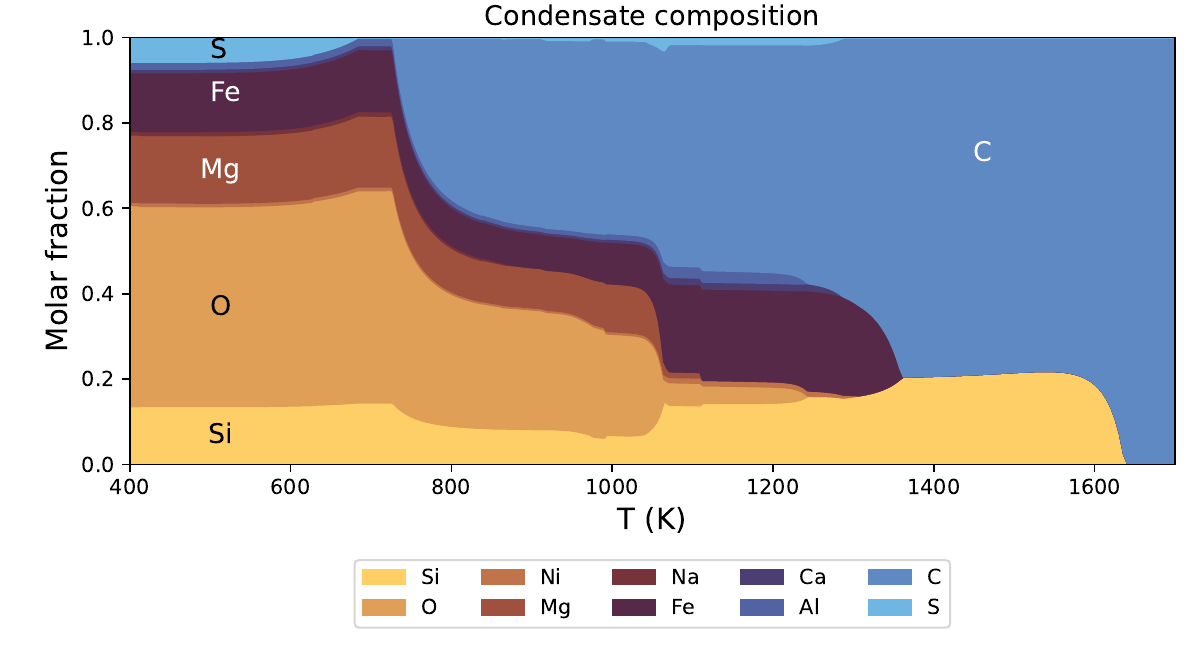}
    \caption[Solid composition vs T at C/O=0.9 and C/O=1.25]{Contribution of condensates to Earth's bulk composition as a function of temperature (top), derived by differentiating the Earth-Sun devolatilisation trend from \citet{Wang2019}. Molar solid elemental composition in chemical equilibrium with the disk gas is plotted as a function of temperature, shown here for two disks from our sample with C/O = 0.9 (middle) and C/O = 1.25 (bottom), at $P=10^{-4}$\,bar. Only elements with molar abundances greater than 0.25 mol\% are shown.}
    \label{fig:CondSeq_CO_09_125}
\end{figure*}
The elements C and O are often classified as volatile due to their low 50\% condensation temperatures. However, both elements exhibit a dual nature, condensing as both refractory and volatile condensates \citep[see discussion in][]{Wang2019}. A modest fraction of the disk oxygen budget ($\sim$20\%) condenses as silicates at temperatures as high as 1700\,K yet its 50\% condensation temperature is only reached at 180\,K \citep{Lodders2003}. Taking this figure at face value would suggest that planets are depleted in oxygen compared to their host star by more than 99\%. However, because some oxygen condenses in silicates, this does not accurately reflect planetary oxygen depletion compared to their host stars; for example, the Earth-Sun depletion factor is only $\sim$80\%. Similarly, carbon is less depleted in Earth than its 50\% condensation temperature would suggest. Moreover, we show that the condensation trends of C and S are not monotonic with decreasing temperature, as equilibria at lower temperatures can destabilise C- and S-bearing phases that condensed at higher temperatures. In disks with high molar C/O ratios ($\mathrm{C/O} > 0.8$), C initially condenses as graphite above 1100\,K (Fig.\ \ref{fig:CondSeq_CO_09_125}), but reacts with hydrogen from the disk gas to form methane at lower temperatures, returning all carbon entirely to the gas phase \citep{Larimer1975}. Similarly, at high disk C/O, sulphur initially condenses as CaS or MgS, but these condensates react with H$_2$O in the gas phase to form oxides at lower temperatures, driving sulphur back to the gas phase. Thus, the traditional 50\% $\Tc{}$ description fails to capture the condensation behaviour of semi-volatile elements like C, O, and S. Instead, we propose effective condensation temperatures ($\Tceff{}$) as a metric to characterise volatility of elements that do not exhibit monotonic condensation behaviour, enabling comparison to $\Tc{}$ values of other elements. These $\Tceff{}$ values are not mathematically equivalent to 50\% condensation temperatures, and are merely proposed as a metric for comparison of volatility. The metric is a more accurate representation of gas-solid depletion patterns than 50\% condensation temperatures (or temperature at any percentage of condensation) since it inherently captures non-monotonic behaviour based on actual condensate compositions.

We derive effective condensation temperatures by modelling how star-planet depletion factors arise from the condensation process. To determine these depletion factors, we compare the bulk composition of a hypothetical planet, which is assembled from equilibrium condensates, to that of its host star. We determine the bulk composition of this hypothetical planet by integrating the condensation sequence, which describes how solid composition evolves with temperature. This integration is weighted by a "feeding zone" function, which represents the fraction of parent gas that equilibrated with planetary building materials as a function of temperature. Since Earth's bulk composition largely reflects such an integration of the solar condensation sequence, we can derive its temperature-space feeding zone (Fig.\ \ref{fig:CondSeq_CO_09_125}) by differentiating the Earth-Sun devolatilisation trend (Eq.\ \ref{eq:devol_trend_Wang}). 

We calculate star-planet depletion factors by comparing the bulk compositions of these hypothetical planets to the composition of their host stars. Depletion factors are typically normalised to a highly refractory element like Al. However, condensation behaviour of Al is strongly influenced by disk C/O (Fig.\ \ref{fig:Tc_MgSiCaAl}). Instead, we normalise to Fe, whose condensation behaviour is more predictable (Fig.\ \ref{fig:Tc_FeNi}). The depletion factor $(f_i)$ of element $i$ is defined in terms of the Fe depletion factor $f_{\mathrm{Fe}}$,
\begin{equation*}
    f_i = \frac{X_{\mathrm{planet}}/X_{\mathrm{star}}}{Fe_{\mathrm{planet}}/Fe_{\mathrm{star}}}\cdot f_{\mathrm{Fe}},
\end{equation*}
where $f_{\mathrm{Fe}}$ is determined from Eq.\ \ref{eq:devol_trend_Wang}, using the relationship between $\Tc{Fe}$ and the stellar Fe abundance ($\Tc{Fe} = 4.776 A_{\mathrm{Fe}}^2 + 12.77 A_{\mathrm{Fe}} + 964.64$; see Sec.\ \ref{sec:App_TcParam}). Finally, we convert element depletion factors $f_i$ into effective condensation temperatures $\Tceff{}$ by applying the same devolatilisation trend used for Fe. This approach gives a useful metric to evaluate element volatility given an Earth-Sun devolatilization trend, although this $\Tceff{}$ definition is bound to the trend with which it is derived, and is not valid for bodies with devolatilization trends deviating from it. We explore the effect of varying the $\alpha$ and $\beta$ parameters in Eq.\ \ref{eq:devol_trend_Wang} on $\Tceff{}$ values in Sect.\ \ref{sec:App_Devol_heatmaps}. For the Earth-Sun devolatilization trend, we predict an effective condensation temperature of oxygen in the solar disk of 894\,K, which is within the range of 875$\pm$45\,K found by \citet{Wang2019}. In contrast, carbon does not condense in the solar disk above the lowest temperature we investigate - 400\,K - and thus the model produces $\Tceff{C}=0$\,K. Realistically, this means $\Tceff{C}<400$\,K, in line with the estimate from Wang et al.\ ($\Tceff{C} = 305 _{-135}^{+73}$\,K).

\begin{figure*}
    \resizebox{\hsize}{!}{\includegraphics{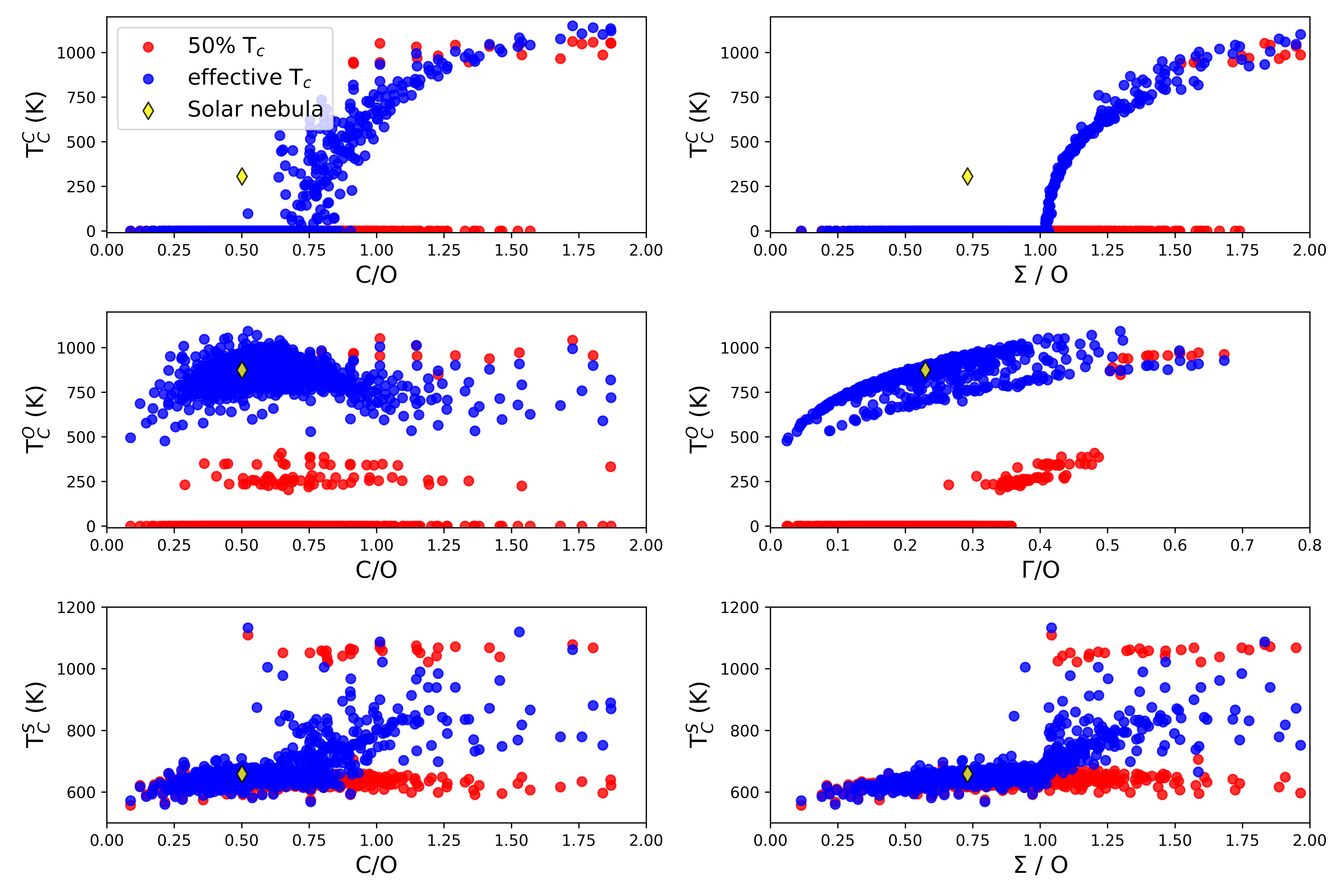}}
      \caption[Condensation temperatures of C, O, and S]{Condensation temperatures of C (top), O (middle), and S (bottom) in disks with varying compositions. Temperatures are plotted against molar C/O ratios of disks (left) and $\Sigma/$O, which is defined as the sum of all elements binding with oxygen, multiplied with their oxide stoichiometry, divided by the oxygen abundance, or $\Gamma/$O, where $\Gamma=\Sigma-\epsilon_C$ (right). The 50\% condensation temperatures are shown as red dots, while effective condensation temperatures (see text) are shown as blue dots. Condensation temperatures are derived from condensation sequence calculations at $P=10^{-4}$\,bar.}
      \label{fig:Tc_COS}
\end{figure*}

Unlike traditional 50\% condensation temperatures, which significantly underestimate the abundances of volatile elements in rocky bodies, effective condensation temperatures reveal systematic trends between the condensation of C and O and the disk C/O ratio (Fig.\ \ref{fig:Tc_COS}). Once all available oxygen is consumed in the formation of oxides and CO gas, excess carbon condenses as graphite. This begins at C/O$\approx 0.7$, with $\Tceff{C}$ increasing steadily as disk C/O rises. We quantify this competition for oxygen using the ratio of all oxygen-bonding element abundances weighted by their oxide stoichiometry, to the abundance of oxygen, as
\begin{equation}
\begin{split}
    \Sigma/O = ( \epsilon_{\mathrm{C}} + 2\epsilon_{\mathrm{Si}} + \epsilon_{\mathrm{Mg}} + \epsilon_{\mathrm{Ca}} + \\
    1.5\epsilon_{\mathrm{Al}} + 0.5\epsilon_{\mathrm{Na}})/\epsilon_{\mathrm{O}},
\end{split}
\label{eq:cations_to_oxygen_ratio}
\end{equation}
excluding cations with abundances lower than $\epsilon_{\mathrm{Na}}$. Once this ratio reaches unity, carbon can condense as graphite in a limited temperature range (Fig.\ \ref{fig:CondSeq_CO_09_125}). At C/O greater than 1.04, the stability field of graphite extends to higher temperatures (Fig.\ \ref{fig:App_C_stability_Tc}), increasingly dominating the solid-state composition (Fig.\ \ref{fig:CondSeq_CO_09_125}). In this regime, C/O becomes the primary control on $\Tceff{C}$. However, the depletion factor of C will never reach zero, as carbon reacts with H$_2$ in the disk to form CH$_4$ at lower temperatures \citep[$<$750\,K][]{Larimer1979}, preventing $\Tceff{C}$ from exceeding the devolatilisation cut-off temperature \citep[1391\,K for Earth;][]{Wang2019}. Even in C-rich disks, low-temperature condensates remain predominantly silicates and oxides. 

Because silicate and oxide condensates are prevalent in all disks at lower temperatures, regardless of C/O, $\Tceff{O}$ is always greater than 500\,K. It increases proportionally to oxide-forming element abundances, since higher concentration of these elements leads to greater oxygen condensation. Specifically, $\Tceff{O}$ is directly proportional to $\Gamma/$O, where we define $\Gamma = \Sigma - \epsilon_{\mathrm{C}}$. However, as C/O increases and carbon sequesters a greater fraction of available oxygen, $\Tceff{O}$ shifts to lower temperatures, though it remains directly proportional to $\Gamma/$O. In this high-C/O regime, $\Tceff{O}$ is lower than the temperature at which graphite destabilises.

Like carbon, sulphur condensation is influenced by oxygen availability. In oxygen-rich disks (i.e.\ $\mathrm{C/O}<0.8$), FeS is the primary S-bearing condensate, with CaS forming in smaller quantities, noting that these may exist in solid-solution in FeS were these models to be implemented. In these disks, $\Tceff{S}$ scales proportionally to $\epsilon_{\mathrm{Fe}} + \epsilon_{\mathrm{S}}$. At $\Sigma$/O$>1.005$, oxygen depletion allows MgS to condense and CaS to become more abundant. This results in an increase of $\Tceff{S}$, which continues to increase with $\Sigma$/O, and Mg and Ca abundances. However, even in these disks, $\Tceff{S}$ rarely exceeds 1000\,K. Thus, sulphur remains a moderately volatile element even under high C/O conditions. 

\section{Updated rocky exoplanet bulk compositions}\label{sec:exo_bulk_comps}
\begin{figure*}
    \includegraphics[width=0.49\hsize]{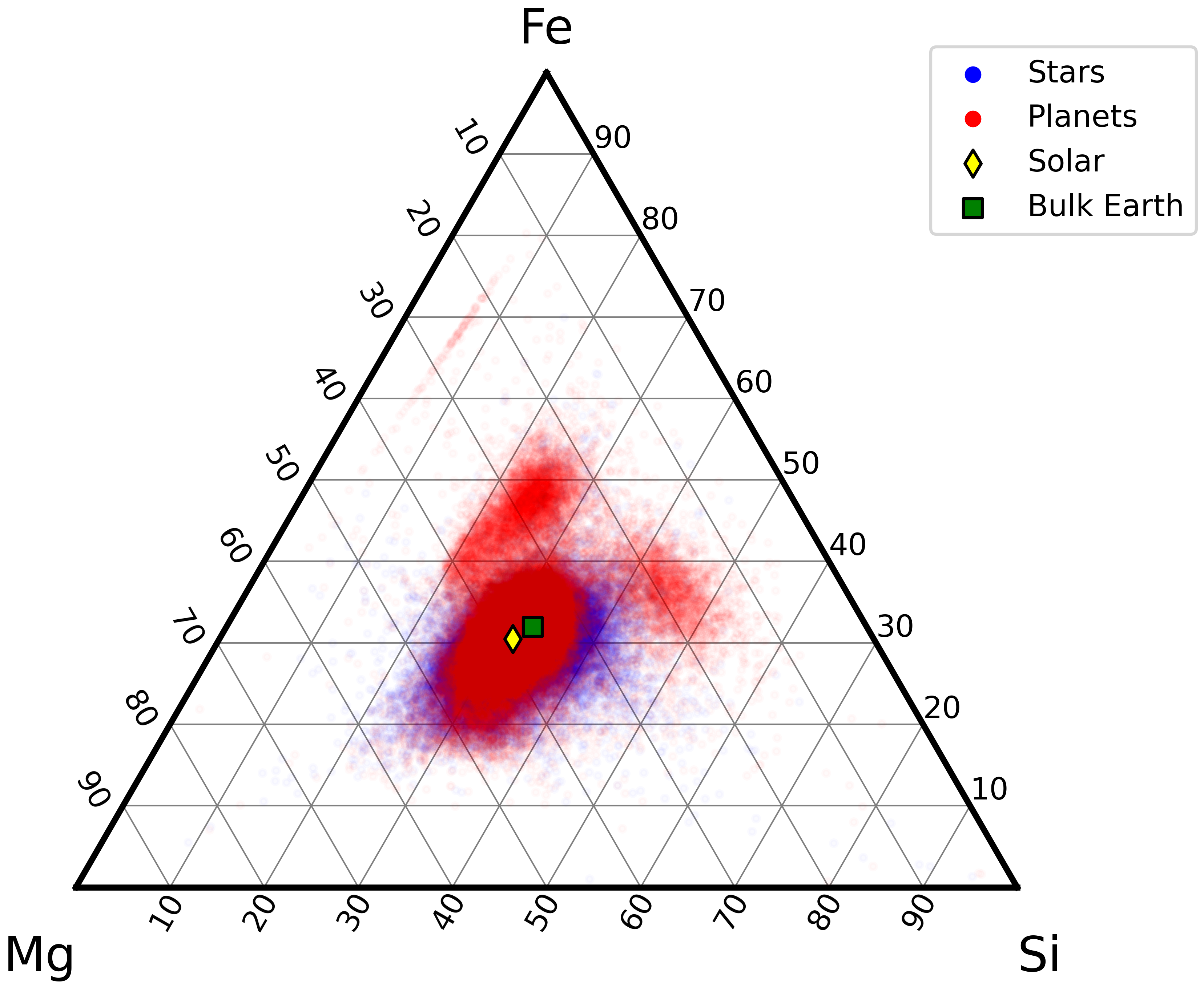}
    \includegraphics[width=0.49\hsize]{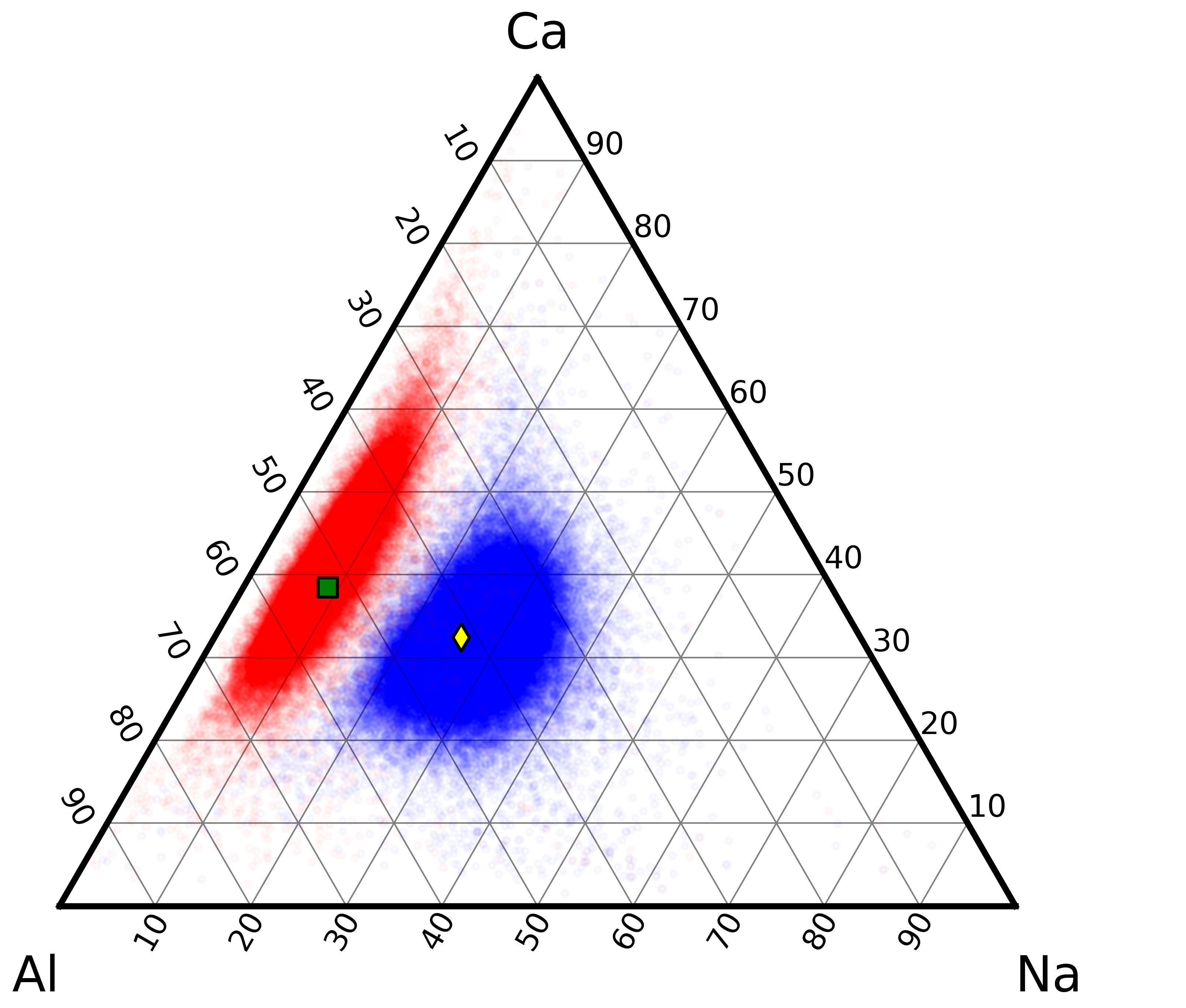}
      \caption[Stellar Planet Fe Mg Si]{Stellar (blue) and planetary (red) Fe, Mg, and Si abundances (left) and Ca, Al, and Na abundances (right), normalised to the sum of these three elements. solar abundances \citep{Asplund2009} and Earth composition \citep{Wang2018_Earthcomp} are plotted for comparison. While most planets and stars overlap, there are some clear components where Mg and/or Si are depleted.}
      \label{fig:Star_planet_triangles}
\end{figure*}
Using $\Tc{}$ parametrisations derived from our the condensation behaviour (Sect.\ \ref{sec:App_TcParam}), we refine estimates of bulk rocky exoplanet compositions (for planets with Earth-like devolatilisation trends) given stellar abundances from the GALAH catalogue (see Sect.\ \ref{ssec:Methods_PlComps}). Our composition-dependent condensation models reveal two distinct populations of planets (Fig.\ \ref{fig:Star_planet_triangles}): Planets depleted in Mg (formed at $\mathrm{C/O} > 1.04$) and planets depleted in both Mg and Si (formed at $\mathrm{C/O}= 0.84 - 1.04$; Fig.\ \ref{fig:Tc_MgSiCaAl}). Planets forming at $\mathrm{C/O} \leq 0.84$ form with Fe, Mg, and Si abundances in very similar proportions as their host star, consistent with \citet{Spaargaren2023}. Similarly, relative abundances of Ca and Al of planets are very similar to those of the host star (Fig.\ \ref{fig:Star_planet_triangles}). We also find a depletion in Na compared to the host star, in line with previous work; however, we find a population of planets where this depletion is less severe, corresponding to low-Mg/Si disks, where Na is less volatile (Fig.\ \ref{fig:Tc_Na}).

\begin{figure*}
    \centering
    \resizebox{0.8\hsize}{!}{\includegraphics{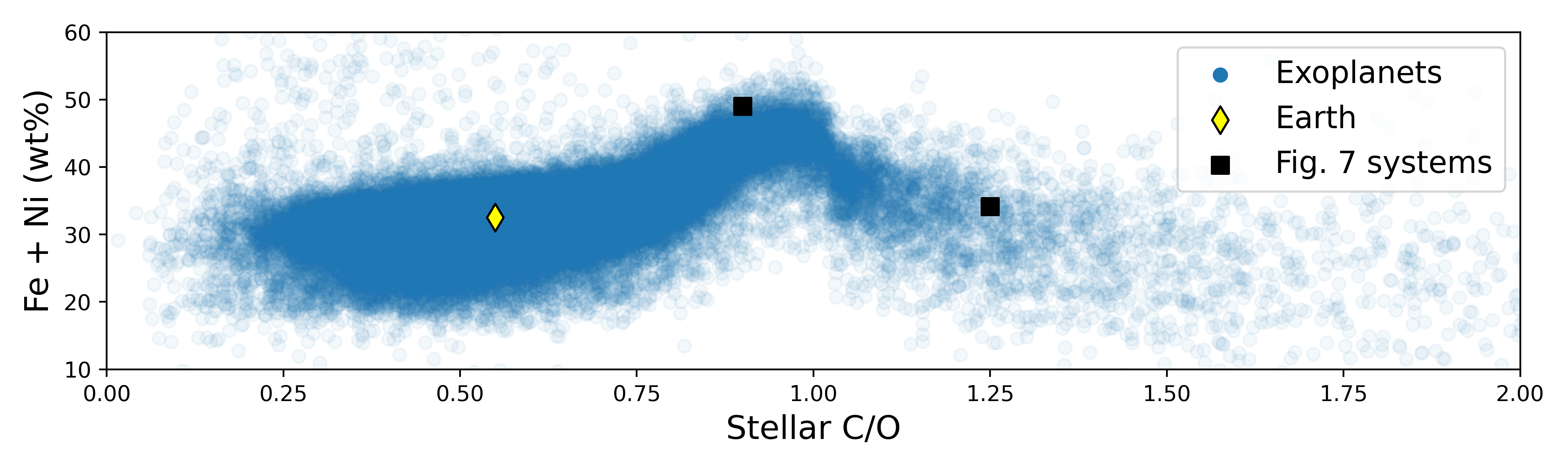}}
    \vspace{0.5cm}
    \resizebox{0.8\hsize}{!}{\includegraphics{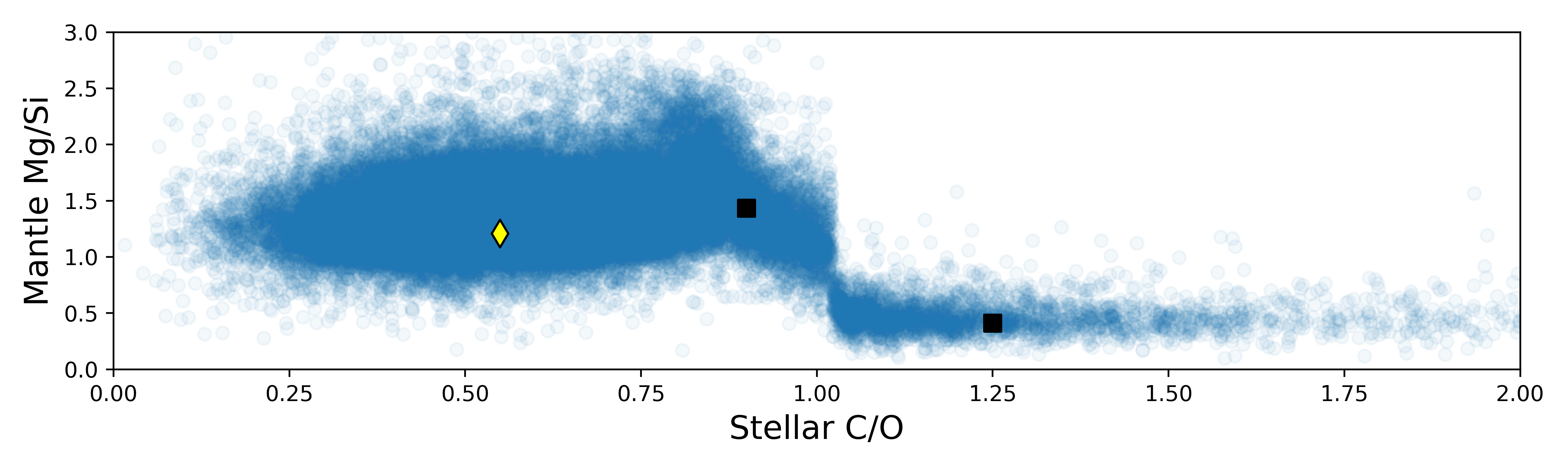}}
    \vspace{0.5cm}
    \resizebox{0.8\hsize}{!}{\includegraphics{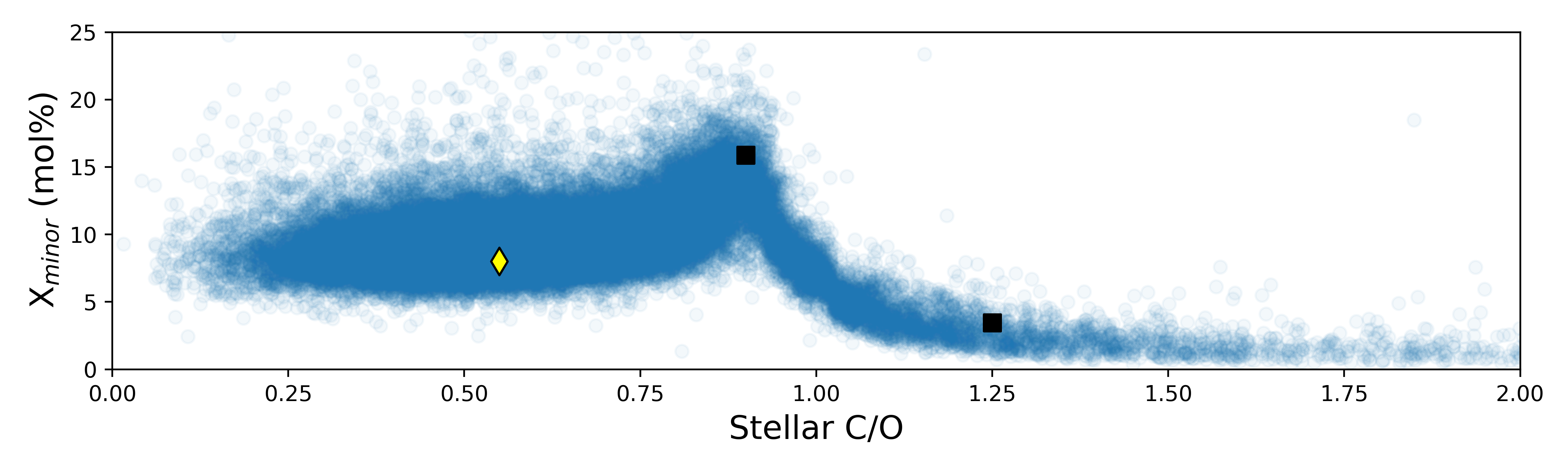}}
    \caption[C/O dependence of CMF, mantle Mg/Si, and mantle Xminor]{Predicted planet mass fractions of core-forming elements Fe and Ni, mantle molar Mg/Si ratio, and mantle minor element (all elements aside from Fe, Mg, Si, and O) molar fraction, as a function of host stellar C/O. Compositions are calculated based on stellar abundances from the GALAH catalogue by applying disk composition-dependent depletion factors to each element. Earth composition is shown (yellow diamond) from \citet{Wang2018_Earthcomp}, and solar C/O = 0.54 \citep{Asplund2009}; further, the two systems shown in Fig.\ \ref{fig:CondSeq_CO_09_125} are indicated (black squares).}
    \label{fig:Planet_comps_vs_CO}
\end{figure*}
The condensation temperatures of most rock-forming elements decrease significantly as the disk C/O increases from below 0.8 to 1.0 (Fig.\ \ref{fig:Tc_MgSiCaAl}), resulting in higher depletion factors for rock-forming elements. On the contrary, the condensation temperatures of metal-forming elements (Fe, Ni) are independent of C/O (Fig.\ \ref{fig:Tc_FeNi}). Thus, planets forming in disks with C/O between 0.8 and 1.0 have a greater fraction of metal-forming elements (Fig.\ \ref{fig:Planet_comps_vs_CO}), of 40-50 wt\% rather than 20-40 wt\%. Under the simple assumptions of pure Fe-Ni-cores, and reducing conditions so all Fe and Ni resides in the core, the Fe+Ni mass fraction corresponds to the core mass fraction (CMF); planets forming in disks with C/O between 0.8 and 1.0 are likely to have high CMFs. We also identify a population of planets with elevated bulk Mg/Si, reaching up to 2.5, in disks where Mg follows O-rich condensation while Si undergoes O-depleted condensation, decreasing $\Tc{Si}$ considerably. At C/O between 1.02 and 1.04, $\Tc{Si}$ increases drastically, and so planets forming in disks with $\mathrm{C/O}>1.04$ have Mg/Si typically below 0.5. Finally, we predict that some planets, formed in the C/O range leading to high metal-forming element abundances, also have minor element abundances ($X_{minor}$, defined as molar (Ca+Al+Na)/(Ca+Al+Na+Mg+Si)) increased from a nominal value of 5-12 mol\% to fractions up to 20 mol\%. Specifically, Ca and Al remain refractory up to $\mathrm{C/O} = 0.95$, unlike Mg and Si, and their enrichment could lead to the presence of exotic minerals such as corundum, although their enrichment is not as high as predicted for some extremely refractory planets \citep[e.g.\ ][]{Dorn2019}. Overall, planets forming at moderately high C/O (0.8 - 1.0) exhibit distinct compositions, including higher CMFs, elevated bulk Mg/Si, and increased minor element abundances compared to planets forming at solar C/O.

\section{Discussion}\label{sec:Disc}
\subsection{Condensation behaviour}
A rocky planet’s bulk composition mirrors its host star (with caveats), but a given element in the planet is depleted compared to a refractory element and its stellar abundance, proportional to its volatility. Thus, studying the condensation behaviour of extrasolar PPDs reveals how element volatility varies among planet formation environments.
For Mg, forsterite is invariably the host phase, such that $\Tc{Mg}=\Tc{Mg2SiO4}$, whereas for Si, this holds only if Mg/Si $>$ 1.5; at lower ratios, $\Tc{Si}$ shifts towards the lower condensation temperature of enstatite. We find MgO condensing at Mg/Si $\geq 2.0$, and SiO$_2$ at Mg/Si $\leq 0.87$, but these compounds condense at lower temperatures than forsterite and enstatite, and never affect $\Tc{Mg}$ or $\Tc{Si}$.

In disks with C/O$<0.8$, rock-forming elements (Mg, Si, Ca, Al, Na) condense as oxides or silicates, with their condensation temperatures scaling with oxygen availability (or oxygen fugacity), which decreases as more oxygen forms CO(g) at higher C/O. Thus, the volatility of these elements depends directly on disk C/O, where $\Tc{}$ drops by 100\,K as C/O increases from 0 to 0.8 \citep[similar to results from][]{Timmermann2023}. As C/O exceeds 0.8, oxygen depletion by CO(g) formation reaches a critical level, causing a rapid $\Tc{}$ drop (100-300\,K) for the rock-forming elements in a very narrow C/O range \citep{Larimer1975}. The exact C/O value at which this drop occurs for an element is proportional to the $\Tc{}$ of that element, as more refractory elements (Ca and Al) condense first and occupy available oxygen, limiting the formation of oxides for less refractory species \citep[Mg and Si][]{Larimer1979}. For Na, this $\Tc{}$ drop is less severe, as its lower $\Tc{}$ places it closer to the temperature at which CO reacts with disk H$_2$, releasing oxygen \citep{Larimer1975}. Unlike previous work \citep[e.g.\ ][]{Larimer1979,Lodders1993}, we find that the $\Tc{}$ drop varies with rock-forming element abundances, as these elements deplete oxygen from the gas upon condensation (e.g.\ Mg/O, Fig.\ \ref{fig:Tc_MgSiCaAl}). As a result, Mg and Si may become moderately volatile at C/O as low as 0.75, in disks with high Mg/O and Si/O. Considering elements beyond C and O is even more important when considering the volatility of C and S; these elements become significantly more refractory once oxygen abundance is insufficient to oxidise all C and rock-forming elements (i.e.\ $\Sigma$/O=1.0), which does not occur at a single C/O ratio (Fig.\ \ref{fig:Tc_COS}). Increased disk N/O is also shown to restrict oxygen availability \citep{Shakespeare2024}, though this only occurs at N/O$>0.75$, and all our disks have N/O$\leq 0.25$. While C/O serves as a useful proxy for volatility of most rock-forming elements, a full understanding requires accounting for all oxide-forming elements, whose collective abundances dictate oxygen availability and condensation pathways.

The first drop in $\Tc{}$ of rock-forming elements marks a key shift in element volatility; at higher C/O ratios, a second, lower-magnitude transition occurs, beyond which the volatility of rock-forming elements becomes largely independent of C/O (Fig.\ \ref{fig:Tc_MgSiCaAl}). This second shift occurs as the condensation temperature of graphite increases with bulk disk C/O, leading to graphite condensation buffering the gas C/O ratio to 1.0 \citep{Lodders1997}. As $\Tc{graphite}$ rises from below 1000\,K at $\mathrm{C/O} < 0.98$ to 2000\,K at $\mathrm{C/O}=2.0$ \citep{Lodders1995}, it overtakes the condensation temperatures of Ca and Al near $\mathrm{C/O}=1.04$ (Fig.\ \ref{fig:App_C_stability_Tc}). As a result, above $\mathrm{C/O}=1.04$, $\Tc{Ca}$ stabilises at 1265\,K, closely matching previous estimates (1240\,K; \citet{Larimer1979}, 1295\,K; \citet{Lodders1995}). Our model yields a lower $\Tc{Al}$ (1230\,K) than Lodders and Fegley Jr.\ (1316\,K) because the Gibbs energy datasets we use lack the condensate AlN, leading to Al$_2$O$_3$ as the primary Al condensate. However, \citet{Shakespeare2024} reports minimal AlN abundances in condensates for the range of bulk disk C/O ratios considered here, suggesting the absence of AlN does not significantly impact $\Tc{Al}$. 

Unlike Mg and Al, Si becomes increasingly refractory as C/O rises above 1.02, due to the stability of SiC, which acts as the primary Si condensate at C/O as low as 0.75 (Fig.\ \ref{fig:App_high_CO_condensates}). However, at $\mathrm{C/O}<1.02$, less than 50\% of Si condenses as SiC, as Si remains partially in the gas phase as SiO(g) and SiS(g). Only above C/O = 1.02 does increased SiC formation drive $\Tc{Si}$ upward, until it stabilises at C/O = 1.10, where $\Tc{graphite}$ overtakes $\Tc{SiC} = 1633$\,K \citep{Lodders1995}, making $\Tc{Si}$ independent of C/O. While CaS can remove some SiS(g), the universally low Ca/Si ratio limits this effect to at most 6\% of total Si, with negligible impact on $\Tc{Si}$ \citep{Lodders1995}. This highlights a limitation of using 50\% condensation temperature to define volatility, as it likely underestimates the fraction of condensed Si in the range of C/O ratios where Si follows O-depleted condensation behaviour. However, since only 50\% $\Tc{}$ correlates with observed star-planet fractionation trends, alternative measures remain more uncertain.

At high bulk disk C/O, the primary Mg-bearing condensate remains forsterite, which can only be formed once SiC becomes unstable; thus, Mg condensation is primarily controlled by Si availability. MgS can form at similar temperatures to forsterite \citep{Lodders1995}, but primarily in high-Mg/Si disks where low Si abundances limits the extent of forsterite formation, leaving excess Mg to condense as MgS (Fig.\ \ref{fig:App_MgS_stability_vs_MgSi}); thus, condensation of MgS has no effect on $\Tc{Mg}$. As a result, above $\mathrm{C/O}=0.94$, where SiC becomes significantly more stable (Fig.\ \ref{fig:App_high_CO_condensates}), $\Tc{Mg}$ stabilizes at 1060\,K, closely matching Lodders et al.\ (1080\,K). In a similar vein to MgS, previous work has found Fe$_3$C as a potential condensate, but either as a metastable compound \citep{Larimer1979}, or with condensation temperature equal to that of metallic iron \citep{Lodders1995}, and thus with no effect on iron volatility. 

Finally, we find that when disk $\Sigma$/O $> 1$, sulphur becomes more refractory, with $\Tceff{S}$ increasing by 100\,K, though it remains moderately volatile. This trend is consistent with previous studies on rocky exoplanet compositions, which show that planets forming in high-C/O disks tend to have small S budgets \citep{Bond2010,Moriarty2014}. In low-Fe/S disks ($[\mathrm{Fe/S}] < -0.67$), pyrite (FeS$_2$) can form, slightly increasing $\Tceff{S}$ \citep{Jorge2022}. However, in our sample, where $[\mathrm{Fe/S}] \geq -0.45$, higher Fe availability prevents pyrite from forming as a stable condensate. Instead, sulphur volatility is controlled by the stability of FeS for low-C/O disks and CaS and MgS for high-C/O disks. In high-C/O disks, both CaS and MgS break down at lower temperatures once more oxygen becomes available due to CO(g) destabilisation (Fig.\ \ref{fig:App_high_CO_condensates}). As a result, $\Tceff{S}$ is primarily dictated by $\Sigma$/O, as oxygen availability ultimately determines the stability of sulphur-bearing condensates.

The condensation temperatures of elements depends not only on the individual abundances, but also disk gas pressure $P$ and metallicity $M$ (Eq.\ \ref{eq:Metallicity}). Mid-plane gas pressure (this work: $P=10^{-4}$\,bar) tends to scale with distance from the host star \citep{Bitsch2015}, stellar mass \citep{Pascucci2016,Ansdell2017}, and temperature \citep{Lodders2024}, and $\Tc{}$ values increase by 40-80\,K per order of magnitude of $P$ \citep{Timmermann2023}. Similarly, $\Tc{}$ values increase proportional to $\log M$ by 70-160\,K per order of magnitude of $M$ (Fig.\ \ref{fig:Metallicity_dependence}). Rock-forming elements scale more strongly with $P$ and $M$ the more refractory they are in the solar disk. Compared to the rock-forming elements, the metal-forming elements Fe and Ni scale more strongly with $P$ and more weakly with $M$. When keeping relative element abundances (e.g.\ Fe/Si and Fe/Mg) equal to solar, an Earth-like planet can have CMF ranging from 25 wt.\% to 38 wt.\% based on $M$ and $P$ variations alone, where planets with the highest CMF exist around high-mass stars with low metallicity (taking CMF as the bulk planet Fe and Ni weight fraction). Carbon is a special case, as graphite condensation depends strongly on metallicity but not at all on gas pressure, while SiC increases moderately with both $M$ and $P$ \citep{Lodders1999,Adams2024}. Thus, planets forming around low-mass, high-metallicity stars will have greater major refractory carbon budget, if the disk $\Sigma$/O is sufficiently high. Silicates start condensing in liquid state at $P > 10^{-2}$\,bar \citep{Ebel2006}, which could dissolve various gases and would considerably alter the depletion patterns of planets; however, such pressures are only relevant for massive stars, and are therefore not relevant to rocky exoplanets.

\subsection{Planet compositions}
\begin{figure*}
    \resizebox{\hsize}{!}{\includegraphics{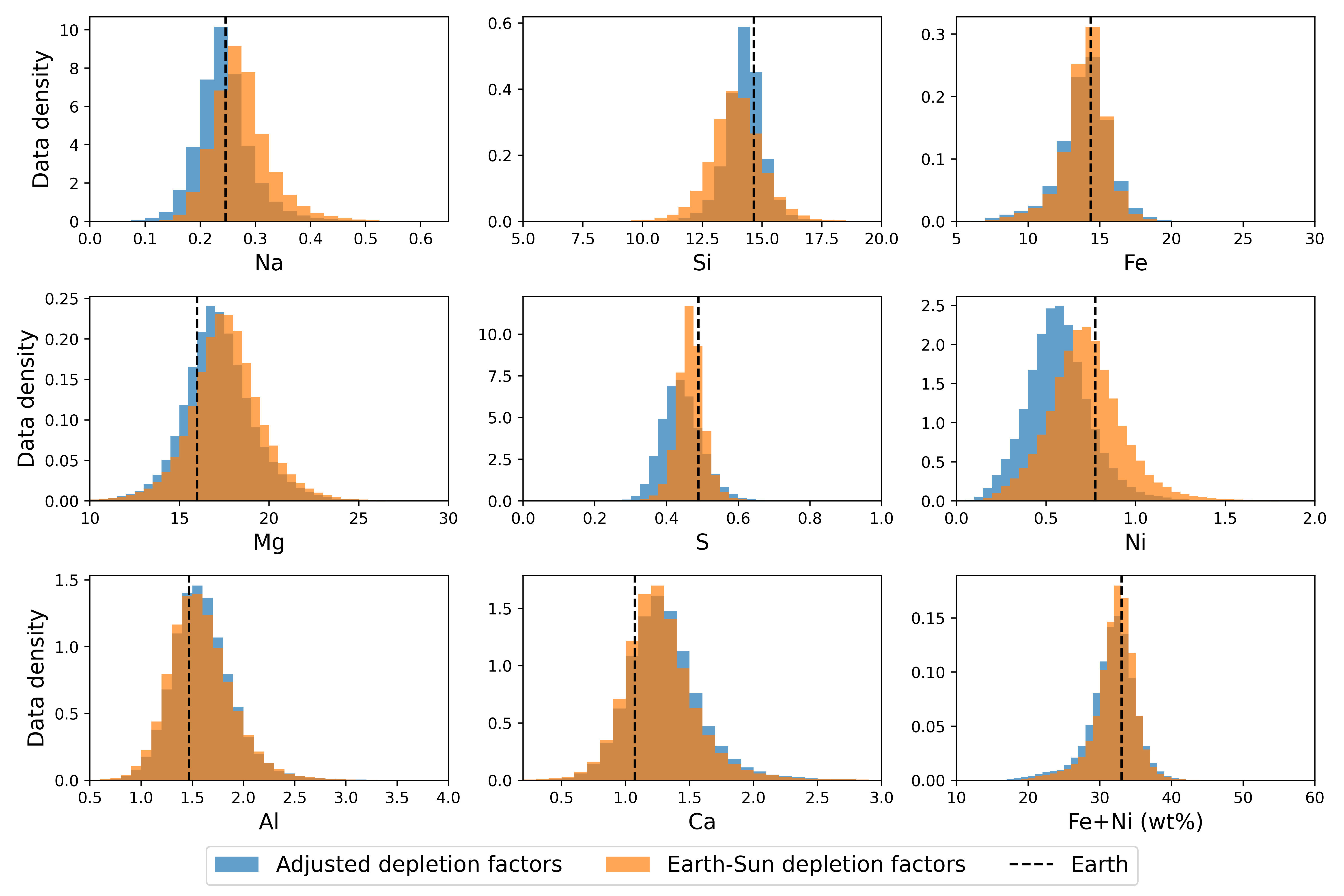}}
      \caption[Planet compositions]{Bulk planet molar compositions obtained by applying the Earth-Sun depletion factors to stellar abundances (orange), and by applying depletion factors adjusted for condensation behaviour (blue), excluding planets containing C. The Fe+Ni mass fraction of the bulk planet is indicative of the planets core mass fraction. Earth composition from \citet{Wang2018_Earthcomp} is added for comparison (dashed line).}
      \label{fig:Depletion_comparison}
\end{figure*}
Variations in disk composition alter the volatility of key elements, influencing the bulk compositions of the planets that form within them. We account for this dependence by substituting $\Tc{}$ values into Eq.\ \ref{eq:devol_trend_Wang} that depend on disk chemistry, rather than using $\Tc{}$ values from \citet{Lodders2003}. For planets that do not accrete carbon, incorporating disk composition-dependent $\Tc{}$ values primarily lowers their Na and S abundances (Fig.\ \ref{fig:Depletion_comparison}), as both elements are more volatile on average than in the solar disk. Further, we underestimate planet Ni abundances, given that \textsc{GGchem} lacks Fe-Ni alloys, which may lead to a slight underestimation of metallic core fractions (by $\sim 0.3\%$). For other elements, we closely match bulk planet compositions between scenarios that do and do not account for disk composition-dependent condensation temperatures. Focusing on lithophile elements (i.e.\ elements that preferentially stay in the silicate mantle during core-mantle differentiation), only Na, the least abundant of the lithophile elements considered here, is affected by disk chemistry-dependent volatility. As a result, the mantle mineralogy of rocky exoplanets in disks with $\mathrm{C/O} < 0.75$ remains similar to the range previously identified by \citet{Spaargaren2023}. Following this range, we find that 0.8\% of rocky planets form Mg-depleted mantle mineralogy, where pure SiO$_2$ minerals (quartz, stishovite) enter the stable mantle mineral assemblage. This is a markedly smaller fraction than previous estimates, ranging from 10 to 50\% \citep[e.g.\ ][]{Brewer2016,Stonkute2020,Clark2021}, since these studies classify planets with $\mathrm{Mg/Si}<1.0$ as Mg-depleted, while petrological simulations show that this only occurs for $\mathrm{Mg/Si}<0.75$ \citep{Spaargaren2023}. The enhanced volatility of Si in low-Mg/Si disks reduces the occurrence of Mg-depleted mantle mineralogy further, and such planet should be exceedingly rare. Conversely, our models predict that 5\% of all rocky planets have bulk $\mathrm{Mg/Si}>1.6$, which leads to planets bearing the Mg-rich, low-viscosity mineral magnesiowüstite (MgO) throughout their entire mantle (rather than just the lower mantle, as is the case for Earth). We likely underestimate the occurrence of these planets, as we do not account for the partitioning of Si into the metallic iron core here. Overall, exoplanet compositional diversity for low-C/O disks ($\mathrm{C/O} < 0.75$) is very similar to previous estimates.

\subsubsection{Core masses}
Disk composition-dependent condensation behaviour leads to the emergence of metal-rich planets (i.e.\ planets rich in Fe and Ni). In disks with bulk C/O below 0.75, where Si condenses similarly to the solar disk, planets develop core mass fractions (CMF; here calculated as the bulk planet mass fraction of Fe+Ni) of 19-41 wt.\%, aligning with stellar abundance-based predictions \citep{Adibekyan2021,Adibekyan2024,Brinkman2024}. Between C/O of 0.84 and 1.02, the lower $\Tc{Mg}$ and $\Tc{Si}$ cause a distinct population emerges with CMF reaching up to 50 wt.\% (Fig.\ \ref{fig:Planet_comps_vs_CO}). Planets with C/O between 0.9 and 1.02, in particular, rarely have cores smaller than 40 wt.\%. Such a population of high-CMF planets could help explain observations of exoplanets with greater CMF than expected from host stellar abundances \citep{Adibekyan2024}. However, uncertainties in observational data \citep{Brinkman2024} and mass fraction of ices and volatiles, which are the result of a stochastic component of planet formation and are only minutely linked to host stellar abundances, might also explain these discrepancies. The existence of a dichotomy between these high-CMF planets and low-CMF planets forming in disks with bulk C/O below 0.75 should be tested through high-accuracy mass and radius measurements, combined with accurate measurements of host stellar C, O, Fe, Mg, and Si abundances. Such observations would provide a crucial test of our predicted condensation-driven trends in planetary composition. 

Our model predicts a dichotomy in rocky exoplanet core mass fractions (CMFs), with larger cores (40–50 wt\%) forming in high C/O disks (0.85–1.0) and smaller cores (19–41 wt\%) in low C/O disks ($<0.75$). These estimates assume all Fe and Ni reside in the core, which oversimplifies real planetary compositions. Earth, for example, has oxidized Fe in its mantle and light elements in its core. Exoplanet models predict a wide range of Fe oxidation states—from fully reduced to coreless planets \citep[e.g.\ ][]{Hatalova2025}—which our model cannot capture, as oxidized iron can only condense at temperatures below 500\,K \citep[similar to][]{Larimer1979}. Nevertheless, disk C/O likely plays a significant role in controlling variations in CMF, and it offers a testable prediction: planets formed in high-C/O disks are expected, on average, to posess higher CMFs. Moreover, our model likely underestimates this effect, as it excludes Fe-Si alloy condensation, which is likely to occur in high-C/O environments \citep[e.g.\ ][]{Ferrarotti2002}, and carbon’s siderophile behaviour \citep{Kuramoto1996,Dasgupta2013_carbonmeltpartition}. Mass-radius measurements, combined with high-accuracy stellar C and O (and ideally Fe, Mg, Si) abundances, can test this prediction across a large planet sample.


\subsubsection{Refractory C and S in planets}
\begin{figure}
\centering
    \resizebox{\hsize}{!}{\includegraphics{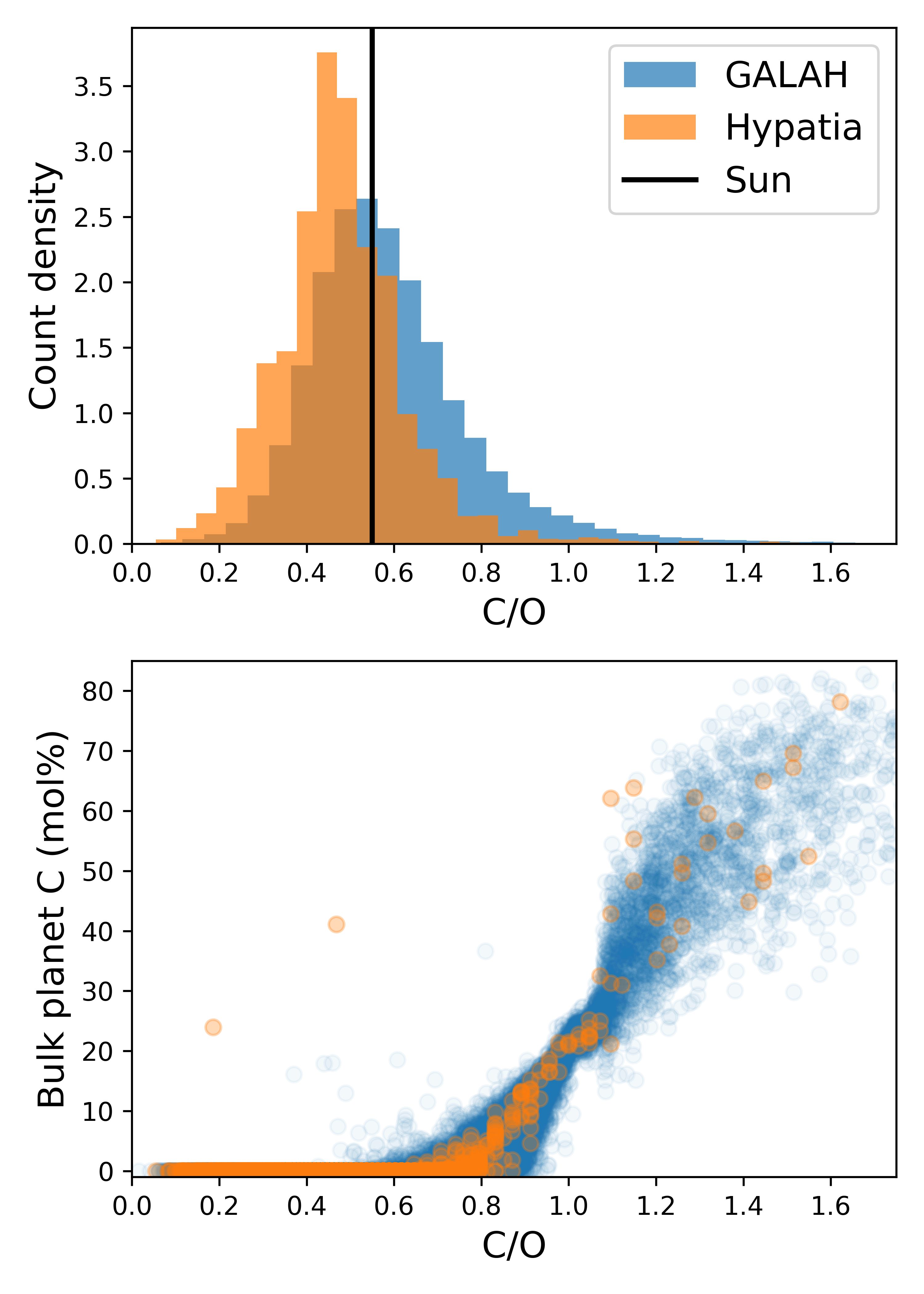}}
      \caption[GALAH and Hypatia C/O and planet C content]{Stellar C/O ratio (top) and bulk planet molar C concentration (bottom) among the Hypatia (orange) and GALAH (blue) stellar abundance catalogues.}
      \label{fig:CO_stars_C_planets}
\end{figure}
Planets have been theorized to form with a major graphite budget for C/O ratios greater than 0.8 \citep{Bond2010,Carter2012}, or as low as 0.65 when sequential condensation is considered \citep{Moriarty2014}. In comparison, we find planets forming with more than 1 mol\% C for C/O as low as 0.6, while planets can form nominally C-free for disk C/O ratios up to 0.95 (Fig.\ \ref{fig:CO_stars_C_planets}). We posit that C/O alone is insufficient as a metric to predict occurrence of carbon-bearing planets; instead, $\Sigma/$O better represents the potential of C condensation. Using $\Sigma/$O to constrain carbon condensation results in a slight increase in occurrence of carbon-bearing planets; 12.36\% of stars in the GALAH DR3.2 (after filters for data quality) could potentially host refractory carbon-bearing rocky planets (i.e.\ forms planets with more than 1 mol\% C), while 11.90\% of stars in the same sample have $\mathrm{C/O}>0.8$. For comparison, 3.71\% of stars in the Hypatia catalogue \citep{Hypatia}, which focuses on the solar neighbourhood (i.e.\ stars within 200 pc from the Sun), could potentially host refractory carbon-bearing rocky planets, while 2.84\% of stars in this catalogue have $\mathrm{C/O}>0.8$. The Hypatia catalogue contains fewer C-rich stars due to the focus on the solar neighbourhood, which is known to have a degree of chemical homogeneity \citep{Bedell2018}. These catalogues both establish that C-bearing planets, and by extension planets with elevated CMF, should exist and potentially be observable. While C/O ratios are known to be sensitive to line selection and prone to overestimation \citep[e.g.\ ][]{Teske2013,Delgado2021}, the consistently higher fraction of C-rich stars in the GALAH sample supports the likely existence of a population of carbon-rich rocky planets. By extension, these may also include planets with elevated core mass fractions.

Rocky planets bearing refractory carbon have been theorized to form graphite crusts due to the low density of graphite \citep{Madhusudhan2012,Hakim2019}, and since the bulk planet carbon budget increases with disk C/O (Fig.\ \ref{fig:CO_stars_C_planets}), it is likely that the graphite lid also grows with increasing disk C/O. While carbon is known to exhibit siderophile behaviour \citep{Kuramoto1996,Dasgupta2013_carbonmeltpartition}, graphite may not fully chemically equilibrate with the metal \citep{Keppler2019}, and the lid still forms for planets with low carbon budgets. The planet carbon budget grows quickly with disk C/O, to 20-30 mol\% at C/O=1.0, so the lid thickness likely increases over this range as well. Once the graphite stability field increases to higher temperatures at $\mathrm{C/O}>1.04$, planet carbon budgets increase to 40-80 mol\%; however, these planets always retain a oxygen-rich silicate reservoir in their interiors, since these silicates still form at temperatures below 1000\,K in high-C/O disks. Further, rocky planets primarily sampling low-temperature condensates could potentially form without accreting any refractory carbon, even in disks with $\mathrm{C/O}>1.04$, and so planet systems around high-C/O stars likely show chemical diversity, containing C-rich and C-poor planets \citep[similar to the results of ][]{Bond2010,Carter2012,Moriarty2014}.


Planets forming in high-C/O disks also accrete SiC and sulphide phases, and especially SiC is theorised to form a major mantle phase \citep[e.g.\ ][]{Bond2010}. However, stability of SiC is highly sensitive to oxidation state; it readily reacts with oxidized iron \citep{Hakim2018} and water \citep{Yoshimura1986,Allen2020} to form SiO$_2$ and graphite. Since water ice and hydrated silicates can readily form at lower temperatures, even in high-C/O environments \citep{Pekmezci2019}, it is likely that water delivery to rocky planets is still active, limiting the amount of SiC that can remain in their mantles. This is subject to the stochastic nature of water delivery, however, and these mechanisms have not been investigated in C-rich environments. The sulphides MgS and CaS are stable mantle phases for fO$_2<\mathrm{IW}-2$ \citep{Anzures2025}, although metal-silicate partitioning of S in such reducing environments is poorly known. While $\Tceff{S}$ increases with C/O, it remains a moderately volatile element with small abundances compared to Fe, Mg, and Si, and thus less than 2\% of our sample has more than 1 mol\% S in their mantles, before core-mantle partitioning has occurred. 

\subsection{Non-Earth-centric devolatilisation calculations}
\begin{figure*}
\centering
    \resizebox{\hsize}{!}{\includegraphics{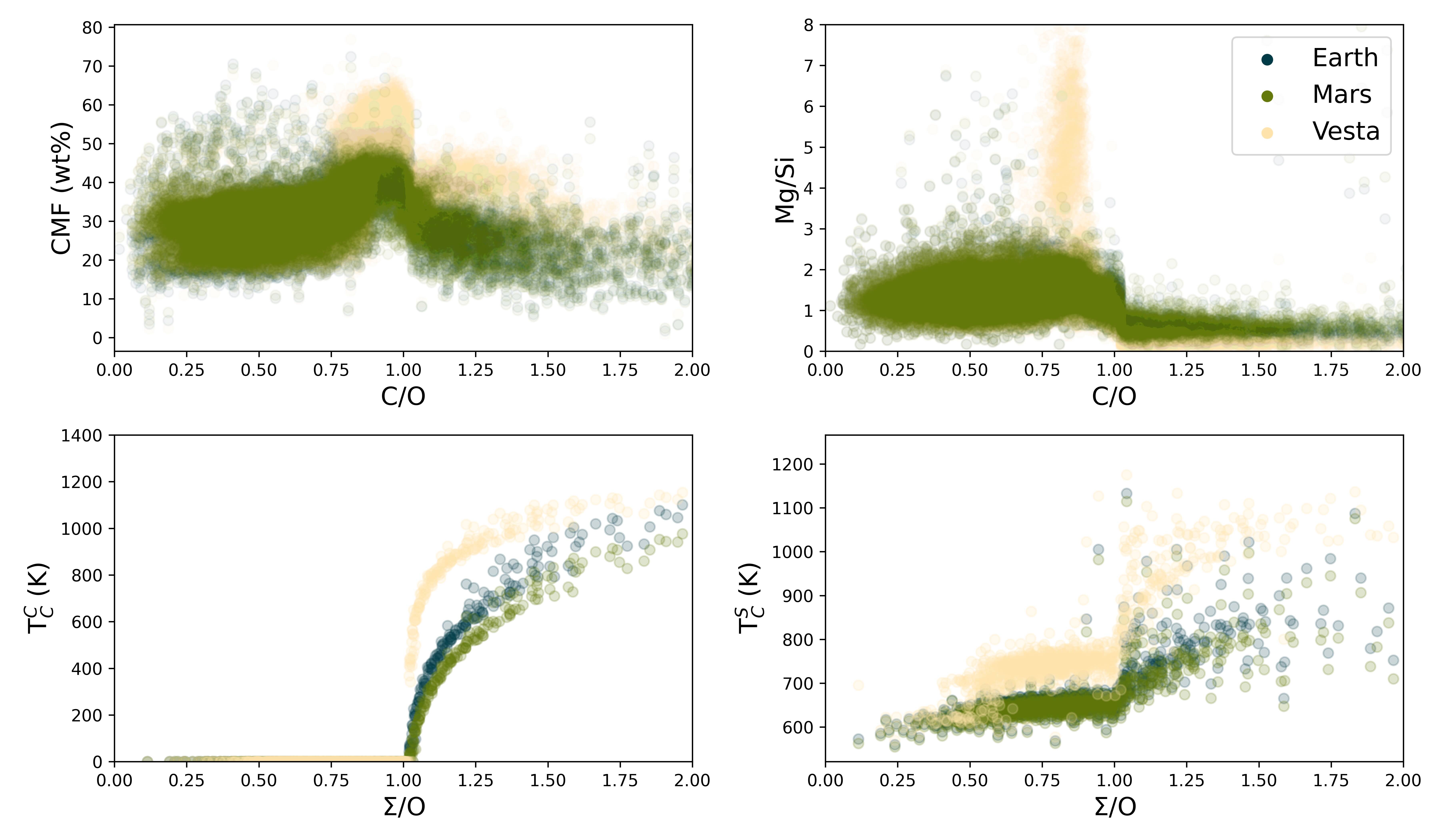}}
      \caption[Devolatilised bulk compositions of Mars and Vesta]{Core mass fractions (upper left), bulk molar Mg/Si ratio (upper right), and effective condensation temperatures of C (lower left) and S (lower right) for Earth, Mars, and Vesta. Results for Mars and Vesta are received by changing the shape of the devolatilisation curve (Eq.\ \ref{eq:devol_trend_Wang}) with estimates for Mars ($\alpha_{\text{Mars}} = 2.984 \pm 0.044$ and $\beta_{\text{Mars}} = -9.33 \pm 0.41$) and Vesta ($\alpha_{\text{Vesta}} = 10.80 \pm 0.22$ and $\beta_{\text{Vesta}} = -33.4 \pm 1.9$)}
      \label{fig:Mars_Vesta}
\end{figure*}
So far, bulk planet compositions we present are simulated using the Earth-Sun devolatilisation trend. It is not currently known how representative this trend is for exoplanets; in fact, it is reasonable to assume that exoplanet compositional diversity includes variations in devolatilisation patterns, as indicated by coupled dynamics-condensation sequence models \citep[e.g.\ ][]{Bond2010,Hatalova2025}. It is known that Mars \citep{Yoshizaki2020} and Vesta \citep{Sossi2022}, as well as most chondrites \citep[e.g.\ ][]{Vollstaedt2020}, follow different trends, and the metal-rich nature of Mercury cannot be explained by volatility-driven processes \citep{Ebel2017}. That said, our approach can be extended to bodies with different devolatilisation patterns, since element 50\% condensation temperatures are solely dependent on disk conditions (composition, pressure), and are insensitive to formation pathways. This is not true for the $\Tceff{}$ values, and to accurately determine the bulk C and S abundances of planets with different devolatilisation trends, condensate composition needs to be integrated with a different temperature-space feeding zone, following section \ref{sec:Res_vol_elems}. 

We can estimate the devolatilisation patterns of Mars and Vesta based on their bulk compositions using least squares estimation by following the procedure of \cite{Wang2019}, to explore a wider range of exoplanet compositions, loosening our strict Earth-centric devolatilization pattern assumption. We fit the devolatilisation function (Eq.\ \ref{eq:devol_trend_Wang}) to the depletion factors of lithophile elements in Mars and Vesta, to find $\alpha_{\text{Mars}} = 2.984 \pm 0.044$ and $\beta_{\text{Mars}} = -9.33 \pm 0.41$ \citep{Yoshizaki2020}, and $\alpha_{\text{Vesta}} = 10.80 \pm 0.22$ and $\beta_{\text{Vesta}} = -33.4 \pm 1.9$ \citep{Mittlefehldt2015,Sossi2022}. Mars is less volatile-depleted than Earth, and thus the effects of Mg and Si becoming more volatile at C/O between 0.8 and 1.02 is less pronounced, with CMF and Mg/Si only marginally increasing (Fig.\ \ref{fig:Mars_Vesta}). Similarly, the effective condensation temperatures of C and S are lower, since Mars naturally samples more material equilibrated with the disk gas at temperatures below the stability of graphite and in the stability of FeS. On the contrary, Vesta has a much steeper, more volatile-depleted pattern, and the volatility effects are more pronounced (c.f., a logistic-curve fit for Vesta in \cite{Sossi2022}). Exo-Vesta bodies (i.e.\ rocky bodies with the devolatilization pattern of Vesta) can have CMF of up to 70 wt.\%, Mg/Si far exceeding that of the Solar System, and the effective condensation temperatures of C and S are higher (note that at low $\Sigma$/O, $\Tceff{S}$ is controlled by MnS rather than FeS). Thus, exo-Vesta bodies forming in high-C/O disks are more C- and S-enriched than exo-Earth and exo-Mars bodies. This approach can be extended to any rocky body with an arbitrary devolatilisation trend, which can show the range of $\alpha$ and $\beta$ values (Eq.\ \ref{eq:devol_trend_Wang}) required for forming high-CMF planets or planets rich in refractory elements (Sec.\ \ref{sec:App_Devol_heatmaps}). However, $\Tceff{C}$ and $\Tceff{S}$ need to be estimated for each trend individually (Figs.\ \ref{fig:App_Tc_CS_vs_devol_low_CO}, \ref{fig:App_Tc_CS_vs_devol_high_CO}).

While our devolatilisation model (with adjusted parameters) can be extended to Mars and Vesta, applying it to exoplanets beyond those with known Solar System-like devolatilisation patterns would require planet formation models capable of predicting appropriate devolatilisation trends. Planetary compositions likely come from the stochastic accretion of smaller, variably devolatilised components \citep{Sossi2022}, with volatile loss occurring either during formation or subsequent evolutionary stages. Classical dynamical models, which simulate rocky planet formation via the accretion of incompletely condensed planetesimals \citep{Bond2010,Carter2012,Elser2012,Moriarty2014,Shakespeare2024,Hatalova2025}, are capable of predicting planets with a wide diversity of devolatilisation patterns, including refractory-rich \citep{Dorn2019}, high core-mass fraction \citep{Mah2023}, and coreless planets \citep{Hatalova2025}. In principle, these models could be used to derive depletion patterns that can then be applied to larger samples of exoplanetary systems, enabling statistical exploration of chemical diversity. However, most of these studies focus on a limited number of elements, which is usually insufficient to constrain a full devolatilisation trend. Moreover, while these studies are successful in reproducing the refractory element composition of the Earth and generally predict planetary bulk physical properties (such as mass and radius) well, many of these studies struggle with reproducing the Earth's observed depletion of Moderately Volatile Elements (MVEs), such as Na and S (see Fig.\ 2 in \citet{Bond2010a}, Fig.\ 4 in \citet{Elser2012}, and Fig.\ 5 in \citet{Moriarty2014}; cf.\ \citep{Sossi2022}). This highlights a key gap that will need to be addressed before these models can be applied to infer devolatilisation trends and assess compositional diversity across broader exoplanet samples.

Alternative scenarios, such as pebble accretion with atmospheric pebble ablation, may explain some elemental ratios \citep[e.g.\ S;][]{Steinmeyer2023}, but their general applicability remains to be tested. Other mechanisms—such as volatile loss from high disk altitudes \citep{Sengupta2022}, small molten bodies \citep{Hin2017,Norris2017,Sossi2019,Young2019}, magma oceans (\citet{Benedikt2020}, cf.\ \citet{Zhang2025}), loss during impacts \citep{ONeill2008,Pringle2014}, or accretion of chondrule-like material \citep{Vollstaedt2020}—may contribute to the observed depletion patterns, though their relative importance remains uncertain. Rather than assuming any specific formation pathway, our approach empirically constrains a lower bound on exoplanetary compositional diversity empirically, by applying the shapes of devolatilisation trends observed in Solar System bodies. This method does not aim to explain the origin of these patterns, nor does it rely on assumptions about the temperatures, timescales, or radial locations in the disk at which rocky planets form. While this method is currently limited to planets with Earth-like depletion patterns, it implicitly incorporates the full complexity of planet formation for such bodies and can be extended as models evolve to constrain devolatilisation patterns beyond the Solar System.

\subsection{Condensation assumptions}
Throughout this paper, planet compositions are assumed to follow a volatility-dependent elemental depletion trend defined under full chemical and phase equilibrium, consistent with many planet formation models \citep[e.g.\ ][]{Larimer1979,Lodders1995,Bond2010,Timmermann2023}. Most gas-solid reactions proceed rapidly above 600\,K \citep[e.g.\ Fe to FeS; ][]{Lauretta1996}, if dust grains are sufficiently small \citep[$<0.01\,\mu$m;][]{Gail1998}. However, the interiors of larger dust grains, pebbles, planetesimals, and proto-planets are shielded from direct interaction with the gas, and cease to equilibrate. Equilibrium chemistry can also cease at lower temperatures in high-C/O disks, as the hydrogenation of CO(g) and solid graphite is kinetically inhibited \citep{Lewis1980}, on timescales exceeding the dispersal of the nebular gas \citep[e.g.\ ][]{ercolanopascucci2017}. This is consistent with interstellar carbon being found mainly as CO, CO$_2$, and refractory solids \citep[e.g.\ ][]{Chiar1995,Whittet2007,Bergin2015}. 
Sequential condensation models have examined chemical evolution of condensates without continuous equilibrium \citep{Cassen1996,Petaev1998,Moriarty2014,Li2020,Shakespeare2024}, showing, for example, that graphite can condense at C/O ratios down to 0.6 \citep{Moriarty2014}. While full kinetic and shielding treatments are beyond the scope of this study, we approximate both by assuming solids remain in chemical equilibrium with the gas down to an inhibition temperature $T_0$, below which the solids are fully isolated. We follow the sequential condensation implementation in \textsc{GGchem}, as introduced by \citet{Herbort2022}, considering inhibition temperatures of 3000\,K, 1400\,K, 1000\,K, and 600\,K.

\begin{figure*}[h!]
\centering
    \resizebox{\hsize}{!}{\includegraphics{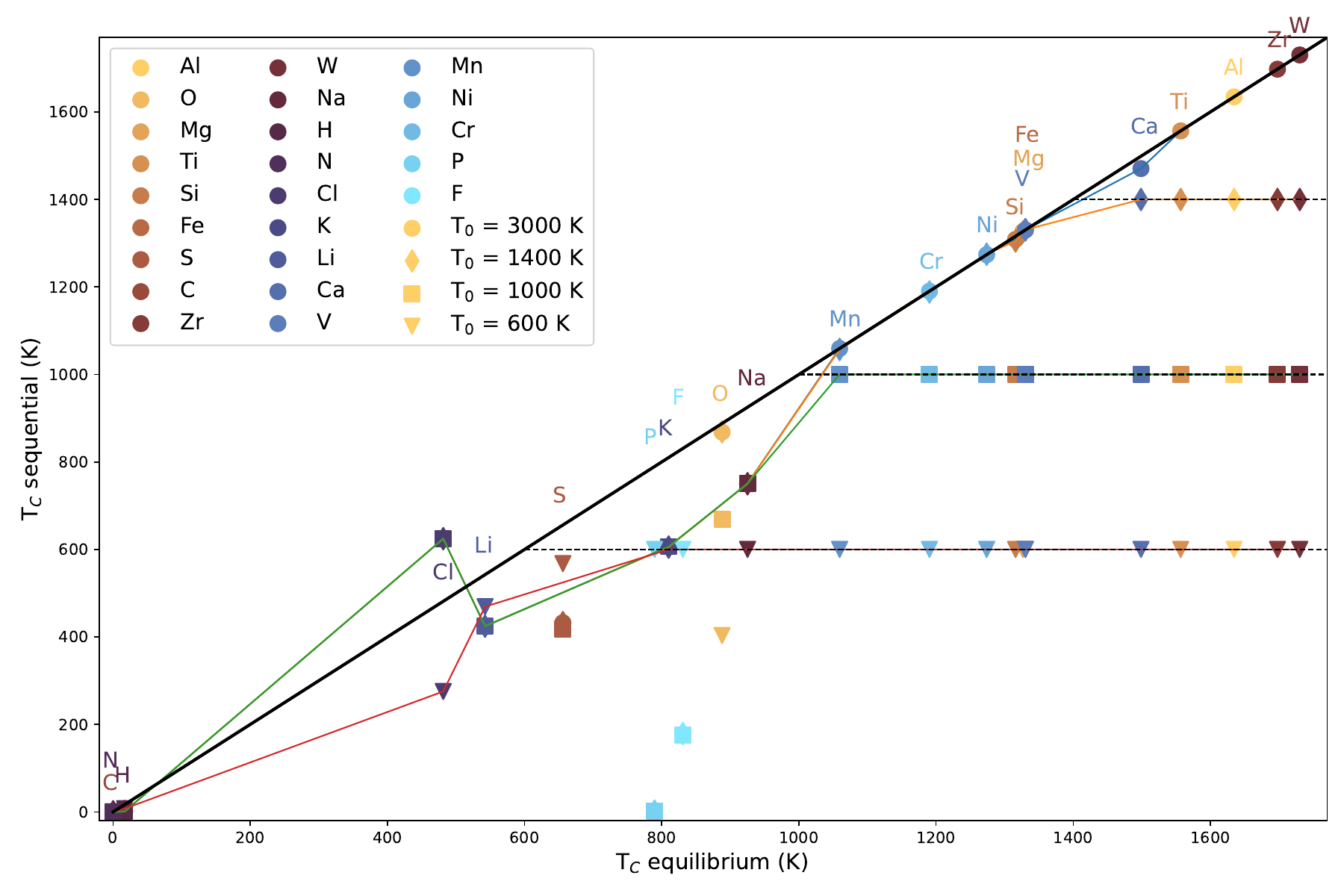}}
      \caption[Sequential condensation Tc for solar disk]{Element 50\% condensation temperatures (K) for a disk gas with solar composition, at 10$^{-4}$\,bar, under equilibrium or sequential condensation assumptions. For the sequential condensation model, the solid composition is assumed to evolve under chemical equilibrium down to $T_0$, below which solids do not interact chemically with the gas.}
      \label{fig:Tc_sequential_solar}
\end{figure*}
For a solar disk composition \citep{Asplund2009}, sequential condensation gives the same $\Tc{}$ values for elements with equilibrium $\Tc{} > 1000$\,K (Fig.\ \ref{fig:Tc_sequential_solar}). The only exception is Ca at $T_0>1800$\,K, since the initial Ca condensate under equilibrium conditions (Ca$_2$Al$_2$SiO$_7$) contains Al, which has disappeared from the gas before this condensate can form. Instead, Ca forms Ca$_2$SiO$_4$, with slightly lower $\Tc{Ca}$. Moderately volatile elements typically have lower $\Tc{}$ under sequential condensation conditions, at least if $T_0 > \Tc{}$. For example, both Na and K have sequential $\Tc{}$ several 100 K lower than their equilibrium $\Tc{}$ values, because their usual primary condensate (feldspar; (K,Na)AlSi$_3$O$_8$) requires Si and Al to form, which are both fully depleted in the disk gas. Instead, these elements condense as NaCl, KCl, and Na$_2$S. As a result, $\Tc{Cl}$ is higher under sequential condensation assumptions than in the equilibrium condensation model. The sequential condensation approach mainly affects moderately volatile elements, which was used to explain some observed depletion patterns in the Solar System \citep{Cassen2001}, although this requires a dynamical model with unrealistic parameters \citep{Ciesla2008}.

\begin{figure*}
\centering
    \resizebox{\hsize}{!}{\includegraphics{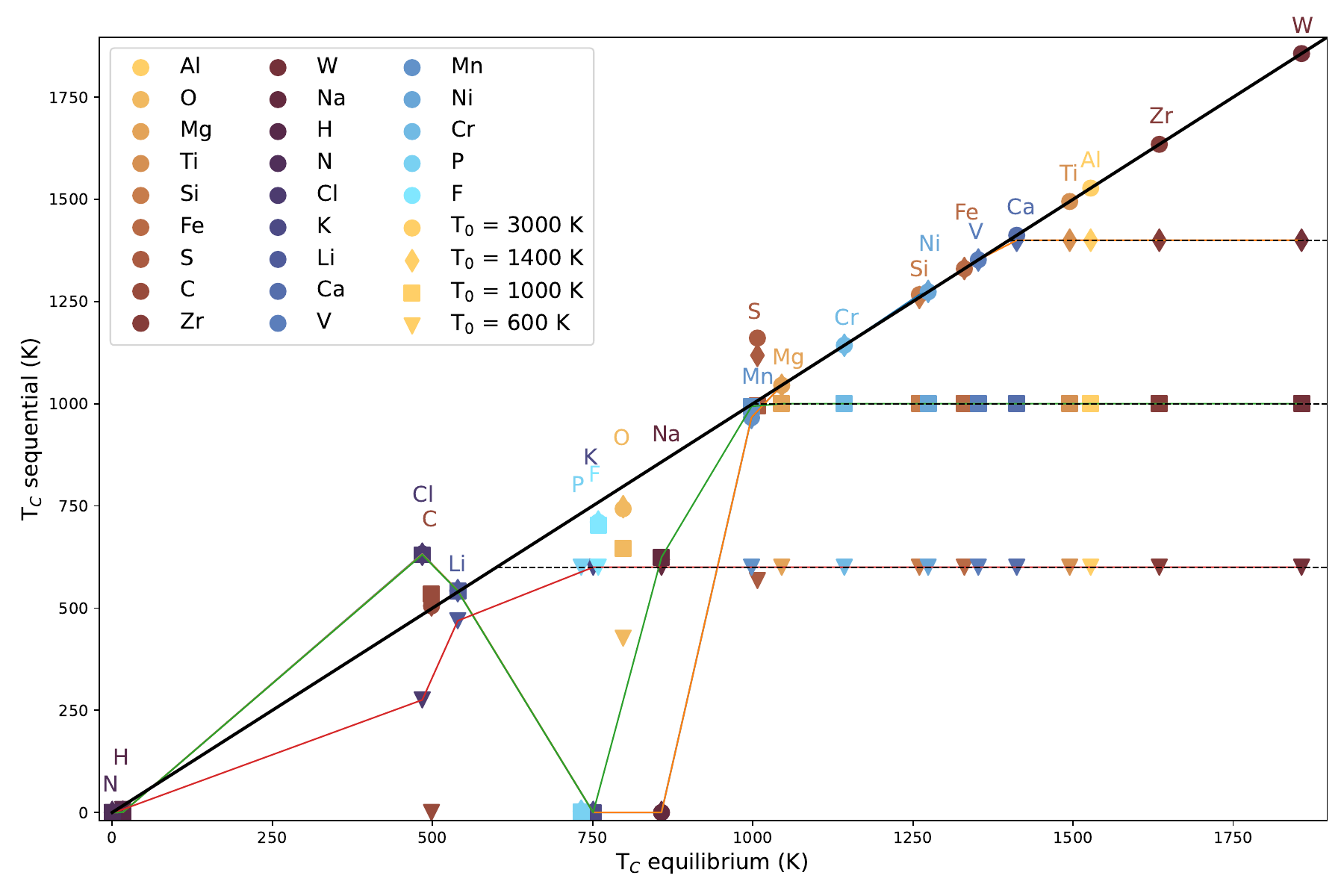}}
      \caption[Sequential condensation Tc for solar disk]{Element 50\% condensation temperatures (K) for a disk gas with $\mathrm{C/O}=0.9$ and otherwise solar composition, at 10$^{-4}$\,bar, under equilibrium or sequential condensation assumptions. For the sequential condensation model, the solid composition is assumed to evolve under chemical equilibrium down to $T_0$, below which solids do not interact chemically with the gas.}
      \label{fig:Tc_sequential_CO_090}
\end{figure*}

\begin{figure*}
\centering
    \resizebox{\hsize}{!}{\includegraphics{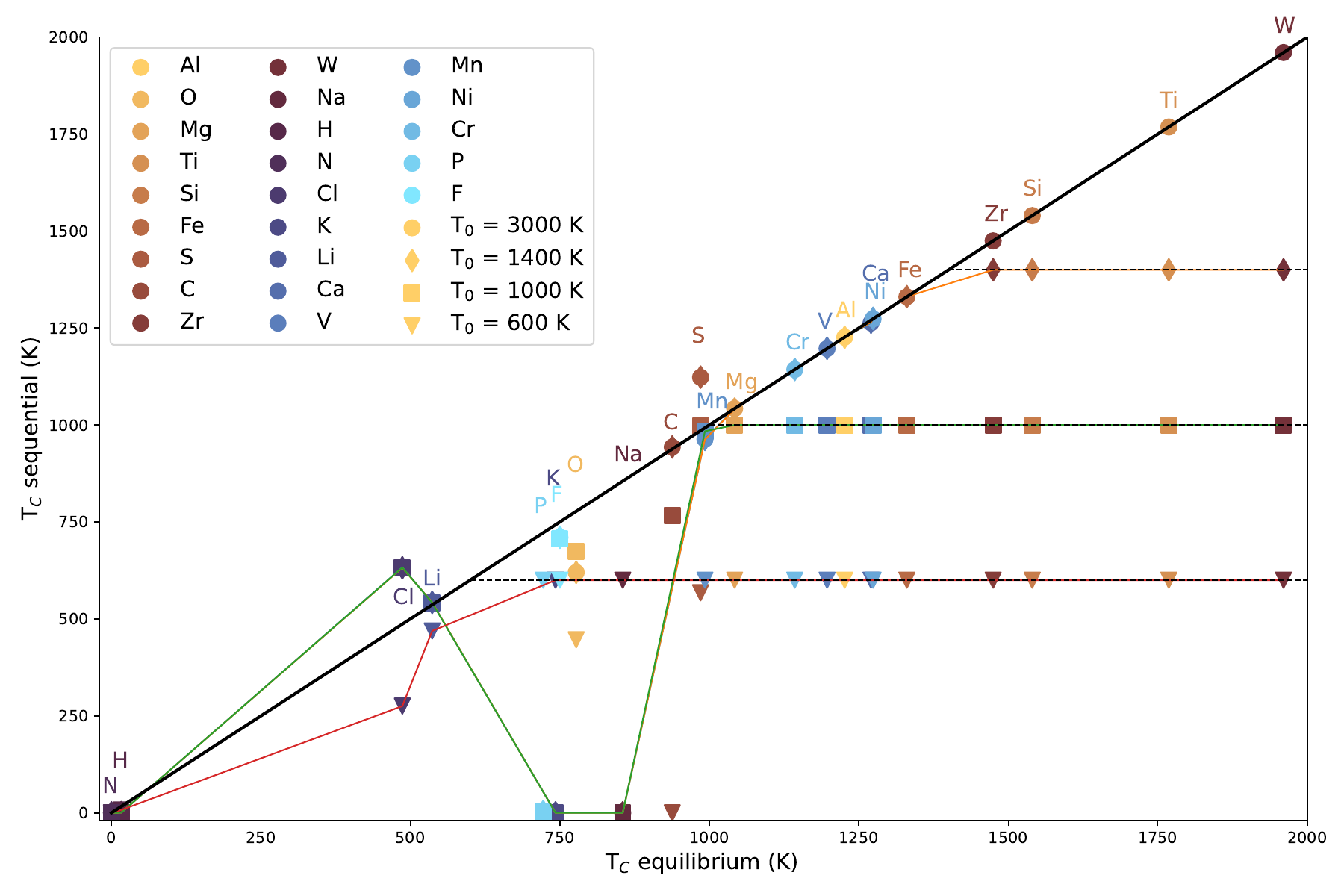}}
      \caption[Sequential condensation Tc for solar disk]{Element 50\% condensation temperatures (K) for a disk gas with $\mathrm{C/O}=1.25$ and otherwise solar composition, at 10$^{-4}$\,bar, under equilibrium or sequential condensation assumptions. For the sequential condensation model, the solid composition is assumed to evolve under chemical equilibrium down to $T_0$, below which solids do not interact chemically with the gas.}
      \label{fig:Tc_sequential_CO_125}
\end{figure*}
Under equilibrium chemistry, the volatility of elements shifts significantly as oxygen becomes depleted due to CO(g) formation. The sequential condensation model reproduces these shifts in volatility, except for Na and K (Figs.\ \ref{fig:Tc_sequential_CO_090}, \ref{fig:Tc_sequential_CO_125}); the increased condensation of S at high temperatures prevents Na$_2$S from forming, leading to all Na condensates being NaCl. This, in turn, leaves no Cl for condensing KCl, which results in K remaining gaseous. Thus, planets in high-C/O conditions may be more depleted in Na and K than we estimate assuming equilibrium condensation; this depends on how efficiently S can be freed by the conversion of MgS and CaS to oxides. Given how quickly metallic iron can be converted to FeS under contact with S-bearing gas at lower temperatures \citep{Lauretta1996}, it is likely that reactions between CaS and MgS and the disk gas are fast (full conversion of Fe to FeS takes place within 1,000 years for an Fe grain of 1 metre), and thus Na and K should be able to condense even under sequential condensation conditions, though experiments are needed to confirm this.

sulphur and carbon follow the equilibrium $\Tc{}$ values well for high $T_0$. Graphite is only stable above 700\,K (Figs.\ \ref{fig:CondSeq_CO_09_125}, \ref{fig:App_C_stability_Tc}) and thus $\Tceff{C}=0$ if $T_0 < 700$\,K. It is perhaps surprising that $\Tc{C}$ is similar to its equilibrium chemistry counterpart at higher $T_0$, since the latter incorporates the disappearance of graphite from the solid at lower temperatures. This can be explained by the stability of MgS and CaS, which would extract oxygen from CO(g) to form oxides under equilibrium chemistry conditions, and the residual C condenses as graphite; thus, less graphite can form overall under sequential chemistry conditions, but the stability of graphite to lower $T$ results in very similar $\Tceff{C}$. This increased stability of MgS and CaS also explains the higher $\Tceff{S}$ for the sequential condensation model. Thus, the sequential condensation model predicts greater stability of S and similar stability of C as the equilibrium condensation model.

Our sequential condensation models show an increased volatility of Na, K, and S in solar-like compositions compared to equilibrium chemistry. This likely holds for physically shielded solids, but for smaller dust grains the incorporation of S is not kinetically inhibited \citep{Lauretta1996}, and Na and K typically condense into solid solutions with anorthite, which forms at 1387\,K \citep{Lodders2003}, and is thus likely not kinetically inhibited. Thus, kinetic inhibition should not play a role in setting element volatility in solar-like disks.

In high-C/O environments, sulphur becomes more refractory \citep[similar to recent experimental results][]{Libourel2025} and oxygen more volatile because CO(g) is no longer destroyed by hydrogenation. While both equilibrium and sequential condensation models for high-C/O disks predict O-poor planets at high temperatures and O-rich planets at lower temperatures \citep[e.g.\ this work, ][]{Bond2010,Moriarty2014}, kinetic inhibition of CO(g) hydrogenation removes the O-rich branch, yielding only rocky planets dominated by graphite, iron, carbides, and sulphides. While equilibrium chemistry captures volatility of elements other than O, S, Na, and K well for Earth-like devolatilisation curves, it fails to do so for planets with devolatilisation patterns where element depletion only occurs for $\Tc{}$ below the temperature at which chemical equilibrium predicts CO(g) hydrogenation. Thus, for high-C/O environments, caution is required when using element volatility derived from equilibrium chemistry. For most elements and environments, however, our results indicate that kinetic inhibition does not strongly affect the equilibrium volatility-depletion correlation, and equilibrium condensation temperatures remain a reasonable approximation.

\section{Conclusions}\label{sec:Conc}
We have systematically characterised how element volatility varies with bulk proto-planetary disk composition, by running condensation sequence simulations across a large statistical sample to calculating 50\% condensation temperatures of major planet-building elements. When combined with the Earth–Sun devolatilization trend — an empirical relationship linking elemental depletion to volatility — these updated condensation temperatures enable plausible simulations of rocky exoplanet compositions for the subset of rocky planets that exhibit Earth-like devolatilization patterns. We therefore provide a robust and conservative lower limit to chemical diversity of rocky exoplanets. This approach provides a powerful means of predicting a subset of planetary compositions without requiring a complete understanding of the complex planet formation process. It allows us to assess how volatility variations shape the chemical makeup of rocky planets across diverse stellar environments. Our findings reveal key trends that govern the volatility of major planet-building elements (Fe, Mg, Si, Ca, Al, Na, Ni, C, S, and O):
\begin{itemize}
    \item Volatility of Fe and Ni only depend on their own abundances, and do not depend on other element abundances. Variations in planet core mass fraction arise primarily from the condensation behaviour of lighter elements.
    \item In oxygen-rich environments, the condensation temperatures of rock-forming elements decrease with increasing C/O. Further, each element except Si becomes less volatile as their abundance increases; The volatility of Si increases because Si-rich minerals (quartz, enstatite) condense at lower temperatures than Mg-rich minerals (forsterite).
    \item Each rock-forming element exhibits a cricital C/O transition point beyond which it follows an oxygen-depleted condensation pathway, where each element is significantly more volatile. These critical values depend on the element-to-oxygen ratio and its condensation temperature in oxygen-rich conditions.
    \item For disks with C/O$>$1.04, all element volatilities become independent of C/O. Condensation of graphite, which in these disks condenses at higher temperature than most other elements, buffers the gas to $\mathrm{C/O}=1.0$.
    \item Sulphur remains moderately volatile in all disk environments, although it becomes somewhat more refractory as oxygen is depleted. Due to the limited stability of MgS and the universally low abundance of Ca compared to Si, sulphur content of exoplanet mantles does not typically exceed 1 mol\%.
    \item Carbon condensation as graphite is controlled by the ratio of oxide-forming elements to oxygen rather than C/O alone, allowing graphite to condense at C/O as low as 0.7. Based on stellar abundances in the GALAH catalogue, 12\% of all rocky planets should form with more than 1 mol\% C in their interiors.
\end{itemize}

Notably, we identify a population of exoplanets forming in disks with C/O between 0.84 and 1.04, where the increased volatility of Mg and Si results in planets with high core mass fractions and mantles rich in refractory lithophile elements (such as Ca and Al). These planets are only mildly carbon-enriched due to the narrow temperature window for graphite condensation in this regime. Importantly, slight deviations in the shape of the devolatilisation trend, such as that of Vesta, can lead to planets with CMF consistent with super-Mercury observations. While stars with $\mathrm{C/O}>1.04$ are exceedingly rare, stars in this intermediate C/O range are likely to exist. Thus, high-CMF planets with Ca- and Al-rich mantles, which have a high mean density and should be observable with mass- and radius measurements, and may already be present among currently known rocky exoplanets.

By systematically linking condensation behaviour to disk composition, our framework enables the prediction of rocky exoplanet compositions for any specified devolatilisation trend. Because this approach does not rely on assumptions about specific planet formation pathways, it remains broadly applicable across diverse dynamical models and chemical environments. As formation models advance and begin to constrain devolatilisation patterns beyond the Solar System, our method will provide a crucial tool for translating those trends into robust predictions of planetary composition. In doing so, it expands the compositional landscape of rocky exoplanets and supports more nuanced interpretations of current and future exoplanet observations.

\begin{acknowledgements}
The authors would like to thank Tim Lichtenberg, Inga Kamp, Lena Noack, and Peter Woitke for their insightful discussions and feedback, as well as to the anonymous reviewer for their thoughtful suggestions which helped improve this article. R.J.S. has been funded by the Swiss National Science Foundation Postdoc Mobility Grant, P500PT\_217847. S.J.M. acknowledges funding from the European Research Council (ERC) under the European Union's Horizon Europe research and innovation programme (Grant agreement No 101166936 - GEOASTRONOMY).
\end{acknowledgements}

\bibliographystyle{aa}
\bibliography{Mendeley}

\begin{thebibliography}{119}
\expandafter\ifx\csname natexlab\endcsname\relax\def\natexlab#1{#1}\fi

\bibitem[{Adams \& Lodders(2024)}]{Adams2024}
Adams, G. \& Lodders, K. 2024, arXiv preprint arXiv:2411.11832

\bibitem[{Adibekyan {et~al.}(2024)Adibekyan, Deal, Dorn, Dittrich, Soares, Sousa, Santos, Bitsch, Mordasini, Barros, {et~al.}}]{Adibekyan2024}
Adibekyan, V., Deal, M., Dorn, C., {et~al.} 2024, arXiv preprint arXiv:2410.17984

\bibitem[{Adibekyan {et~al.}(2021)Adibekyan, Dorn, Sousa, Santos, Bitsch, Israelian, Mordasini, Barros, Delgado~Mena, Demangeon, {et~al.}}]{Adibekyan2021}
Adibekyan, V., Dorn, C., Sousa, S.~G., {et~al.} 2021, Science, 374, 330

\bibitem[{Allen-Sutter {et~al.}(2020)Allen-Sutter, Garhart, Leinenweber, Prakapenka, Greenberg, \& Shim}]{Allen2020}
Allen-Sutter, H., Garhart, E., Leinenweber, K., {et~al.} 2020, The Planetary Science Journal, 1, 39

\bibitem[{Andrews {et~al.}(2013)Andrews, Rosenfeld, Kraus, \& Wilner}]{Andrews2013}
Andrews, S.~M., Rosenfeld, K.~A., Kraus, A.~L., \& Wilner, D.~J. 2013, The Astrophysical Journal, 771, 129

\bibitem[{Ansdell {et~al.}(2017)Ansdell, Williams, Manara, Miotello, Facchini, van~der Marel, Testi, \& van Dishoeck}]{Ansdell2017}
Ansdell, M., Williams, J.~P., Manara, C.~F., {et~al.} 2017, The Astronomical Journal, 153, 240

\bibitem[{Anzures {et~al.}(2025)Anzures, Parman, Milliken, Namur, Cartier, McCubbin, Vander~Kaaden, Prissel, Iacovino, Lanzirotti, {et~al.}}]{Anzures2025}
Anzures, B., Parman, S., Milliken, R., {et~al.} 2025, in 56th Lunar and Planetary Science Conference (LPSC), Lunar and Planetary Institute

\bibitem[{Asplund {et~al.}(2009)Asplund, Grevesse, Sauval, \& Scott}]{Asplund2009}
Asplund, M., Grevesse, N., Sauval, A.~J., \& Scott, P. 2009, ARA{\&}A, 47, 481

\bibitem[{Bedell {et~al.}(2018)Bedell, Bean, Mel{\'{e}}ndez, Spina, Ram{\'{i}}rez, Asplund, Alves-Brito, dos Santos, Dreizler, Yong, Monroe, \& Casagrande}]{Bedell2018}
Bedell, M., Bean, J.~L., Mel{\'{e}}ndez, J., {et~al.} 2018, The Astrophysical Journal, 865, 68

\bibitem[{Benedikt {et~al.}(2020)Benedikt, Scherf, Lammer, Marcq, Odert, Leitzinger, \& Erkaev}]{Benedikt2020}
Benedikt, M.~R., Scherf, M., Lammer, H., {et~al.} 2020, Icarus, 347, 113772

\bibitem[{Bergin {et~al.}(2015)Bergin, Blake, Ciesla, Hirschmann, \& Li}]{Bergin2015}
Bergin, E.~A., Blake, G.~A., Ciesla, F., Hirschmann, M.~M., \& Li, J. 2015, Proceedings of the National Academy of Sciences, 112, 8965

\bibitem[{Bitsch \& Battistini(2020)}]{Bitsch2020}
Bitsch, B. \& Battistini, C. 2020, AnA, 633, A10

\bibitem[{Bitsch {et~al.}(2015)Bitsch, Johansen, Lambrechts, \& Morbidelli}]{Bitsch2015}
Bitsch, B., Johansen, A., Lambrechts, M., \& Morbidelli, A. 2015, Astronomy \& Astrophysics, 575, A28

\bibitem[{Bond {et~al.}(2010{\natexlab{a}})Bond, Lauretta, \& O'Brien}]{Bond2010a}
Bond, J.~C., Lauretta, D.~S., \& O'Brien, D.~P. 2010{\natexlab{a}}, Icarus, 205, 321

\bibitem[{Bond {et~al.}(2010{\natexlab{b}})Bond, O'Brien, \& Lauretta}]{Bond2010}
Bond, J.~C., O'Brien, D.~P., \& Lauretta, D.~S. 2010{\natexlab{b}}, The Astrophysical Journal, 715, 1050

\bibitem[{Brauner {et~al.}(2023)Brauner, Masseron, Garc\'ia-Hern\'andez, Pignatari, Womack, Lugaro, \& Hayes}]{Brauner2023}
Brauner, M., Masseron, T., Garc\'ia-Hern\'andez, D., {et~al.} 2023, Astronomy and Astrophysics, 673, A123

\bibitem[{Brewer \& Fischer(2016)}]{Brewer2016}
Brewer, J.~M. \& Fischer, D.~A. 2016, The Astrophysical Journal, 831, 20

\bibitem[{Brinkman {et~al.}(2024)Brinkman, Polanski, Huber, Weiss, Valencia, \& Plotnykov}]{Brinkman2024}
Brinkman, C.~L., Polanski, A.~S., Huber, D., {et~al.} 2024, The Astronomical Journal, 168, 281

\bibitem[{Buder {et~al.}(2021)Buder, Sharma, Kos, Amarsi, Nordlander, Lind, Martell, Asplund, Bland-Hawthorn, Casey, {et~al.}}]{Buder2021}
Buder, S., Sharma, S., Kos, J., {et~al.} 2021, MNRAS, 506, 150

\bibitem[{Carter-Bond {et~al.}(2012)Carter-Bond, O'Brien, \& Raymond}]{Carter2012}
Carter-Bond, J.~C., O'Brien, D.~P., \& Raymond, S.~N. 2012, The Astrophysical Journal, 760, 44

\bibitem[{Cassen(1996)}]{Cassen1996}
Cassen, P. 1996, Meteoritics \& Planetary Science, 31, 793

\bibitem[{Cassen(2001)}]{Cassen2001}
Cassen, P. 2001, Meteoritics \& Planetary Science, 36, 671

\bibitem[{Chase~Jr {et~al.}(1982)Chase~Jr, Curnutt, Downey~Jr, McDonald, Syverud, \& Valenzuela}]{Chase1982}
Chase~Jr, M., Curnutt, J., Downey~Jr, J., {et~al.} 1982, Journal of Physical and Chemical Reference Data, 11, 695

\bibitem[{Chaudhuri(1994)}]{Chaudhuri1994}
Chaudhuri, B. 1994, Pattern recognition letters, 15, 893

\bibitem[{Chen {et~al.}(2002)Chen, Nissen, Zhao, \& Asplund}]{Chen2002}
Chen, Y., Nissen, P., Zhao, G., \& Asplund, M. 2002, Astronomy \& Astrophysics, 390, 225

\bibitem[{Chiar {et~al.}(1995)Chiar, Adamson, Kerr, \& Whittet}]{Chiar1995}
Chiar, J., Adamson, A., Kerr, T., \& Whittet, D. 1995, Astrophysical Journal v. 455, p. 234, 455, 234

\bibitem[{Ciesla(2008)}]{Ciesla2008}
Ciesla, F.~J. 2008, Meteoritics \& Planetary Science, 43, 639

\bibitem[{Clark {et~al.}(2021)Clark, Clert{\'e}, Hinkel, Unterborn, Wittenmyer, Horner, Wright, Carter, Morton, Spina, {et~al.}}]{Clark2021}
Clark, J.~T., Clert{\'e}, M., Hinkel, N.~R., {et~al.} 2021, MNRAS, 504, 4968

\bibitem[{Dasgupta {et~al.}(2013)Dasgupta, Chi, Shimizu, Buono, \& Walker}]{Dasgupta2013_carbonmeltpartition}
Dasgupta, R., Chi, H., Shimizu, N., Buono, A. A.~S., \& Walker, D. 2013, Geochimica et Cosmochimica Acta, 102, 191

\bibitem[{Delgado~Mena {et~al.}(2021)Delgado~Mena, Adibekyan, Santos, Tsantaki, Gonz{\'a}lez~Hern{\'a}ndez, Sousa, \& Bertr{\'a}n~de Lis}]{Delgado2021}
Delgado~Mena, E., Adibekyan, V., Santos, N., {et~al.} 2021, Astronomy and Astrophysics, 655, A99

\bibitem[{Dorn {et~al.}(2019)Dorn, Harrison, Bonsor, \& Hands}]{Dorn2019}
Dorn, C., Harrison, J.~H., Bonsor, A., \& Hands, T.~O. 2019, MNRAS, 484, 712

\bibitem[{Ebel {et~al.}(2006)Ebel, Lauretta, \& McSween}]{Ebel2006}
Ebel, D.~S., Lauretta, D., \& McSween, H. 2006, Meteorites and the early solar system II, 1, 253

\bibitem[{Ebel \& Stewart(2017)}]{Ebel2017}
Ebel, D.~S. \& Stewart, S.~T. 2017, arXiv preprint arXiv:1712.08234

\bibitem[{Elser {et~al.}(2012)Elser, Meyer, \& Moore}]{Elser2012}
Elser, S., Meyer, M.~R., \& Moore, B. 2012, Icarus, 221, 859

\bibitem[{Ercolano \& Pascucci(2017)}]{ercolanopascucci2017}
Ercolano, B. \& Pascucci, I. 2017, Royal Society Open Science, 4, 170114

\bibitem[{Fegley(2000)}]{Fegley2000}
Fegley, B. 2000, Space Science Reviews, 92, 177

\bibitem[{Ferrarotti \& Gail(2002)}]{Ferrarotti2002}
Ferrarotti, A. \& Gail, H.-P. 2002, Astronomy \& Astrophysics, 382, 256

\bibitem[{Gail(1998)}]{Gail1998}
Gail, H.-P. 1998, Astronomy and Astrophysics, v. 332, p. 1099-1122 (1998), 332, 1099

\bibitem[{Gail \& Sedlmayr(2014)}]{Gail2014}
Gail, H.-P. \& Sedlmayr, E. 2014, Physics and chemistry of circumstellar dust shells No.~52 (Cambridge University Press)

\bibitem[{Hakim {et~al.}(2019)Hakim, Spaargaren, Grewal, Rohrbach, Berndt, Dominik, \& {Van Westrenen}}]{Hakim2019}
Hakim, K., Spaargaren, R., Grewal, D.~S., {et~al.} 2019, Astrobiology, 19, 867

\bibitem[{Hakim {et~al.}(2018)Hakim, Van~Westrenen, \& Dominik}]{Hakim2018}
Hakim, K., Van~Westrenen, W., \& Dominik, C. 2018, Astronomy \& Astrophysics, 618, L6

\bibitem[{Halliday \& Porcelli(2001)}]{Halliday2001}
Halliday, A.~N. \& Porcelli, D. 2001, Earth and Planetary Science Letters, 192, 545

\bibitem[{Harrison {et~al.}(2021)Harrison, Bonsor, Kama, Buchan, Blouin, \& Koester}]{Harrison2021}
Harrison, J.~H., Bonsor, A., Kama, M., {et~al.} 2021, MNRAS

\bibitem[{Harrison {et~al.}(2018)Harrison, Bonsor, \& Madhusudhan}]{Harrison2018}
Harrison, J.~H., Bonsor, A., \& Madhusudhan, N. 2018, Monthly Notices of the Royal Astronomical Society, 479, 3814

\bibitem[{Hatalova {et~al.}(2025)Hatalova, Brodholt, Brasser, Shan, \& Werner}]{Hatalova2025}
Hatalova, P., Brodholt, J., Brasser, R., Shan, Y., \& Werner, S. 2025, Astronomy and Astrophysics, 694

\bibitem[{{Herbort} {et~al.}(2022){Herbort}, {Woitke}, {Helling}, \& {Zerkle}}]{Herbort2022}
{Herbort}, O., {Woitke}, P., {Helling}, C., \& {Zerkle}, A.~L. 2022, \aap, 658, A180

\bibitem[{Hin {et~al.}(2017)Hin, Coath, Carter, Nimmo, Lai, {Pogge von Strandmann}, Willbold, Leinhardt, Walter, \& Elliott}]{Hin2017}
Hin, R.~C., Coath, C.~D., Carter, P.~J., {et~al.} 2017, Nature, 549, 511

\bibitem[{Hinkel {et~al.}(2014)Hinkel, Timmes, Young, Pagano, \& Turnbull}]{Hypatia}
Hinkel, N.~R., Timmes, F.~X., Young, P.~A., Pagano, M.~D., \& Turnbull, M.~C. 2014, The Astronomical Journal, 148, 54

\bibitem[{Hinkel {et~al.}(2022)Hinkel, Young, \& Wheeler~III}]{Hinkel2022}
Hinkel, N.~R., Young, P.~A., \& Wheeler~III, C.~H. 2022, arXiv preprint arXiv:2210.10800

\bibitem[{Johnson {et~al.}(1992)Johnson, Oelkers, \& Helgeson}]{Johnson1992}
Johnson, J.~W., Oelkers, E.~H., \& Helgeson, H.~C. 1992, Computers \& Geosciences, 18, 899

\bibitem[{Jorge {et~al.}(2022)Jorge, Kamp, Waters, Woitke, \& Spaargaren}]{Jorge2022}
Jorge, D., Kamp, I., Waters, L., Woitke, P., \& Spaargaren, R. 2022, AnA, 660, A85

\bibitem[{Keppler \& Golabek(2019)}]{Keppler2019}
Keppler, H. \& Golabek, G. 2019, Geochemical Perspectives Letters, 11, 12

\bibitem[{Khan {et~al.}(2022)Khan, Sossi, Liebske, Rivoldini, \& Giardini}]{khan2022geophysical}
Khan, A., Sossi, P.~A., Liebske, C., Rivoldini, A., \& Giardini, D. 2022, Earth and Planetary Science Letters, 578, 117330

\bibitem[{Kobayashi {et~al.}(2020)Kobayashi, Karakas, \& Lugaro}]{Kobayashi2020}
Kobayashi, C., Karakas, A.~I., \& Lugaro, M. 2020, The Astrophysical Journal, 900, 179

\bibitem[{Kuchner \& Seager(2005)}]{Kuchner2005}
Kuchner, M.~J. \& Seager, S. 2005, arXiv preprint astro-ph/0504214

\bibitem[{Kuramoto \& Matsui(1996)}]{Kuramoto1996}
Kuramoto, K. \& Matsui, T. 1996, Journal of Geophysical Research: Planets, 101, 14909

\bibitem[{Larimer(1975)}]{Larimer1975}
Larimer, J.~W. 1975, Geochimica et Cosmochimica Acta, 39, 389

\bibitem[{Larimer \& Bartholomay(1979)}]{Larimer1979}
Larimer, J.~W. \& Bartholomay, M. 1979, Geochimica et Cosmochimica Acta, 43, 1455

\bibitem[{Lauretta {et~al.}(1996)Lauretta, Kremser, \& Fegley~Jr}]{Lauretta1996}
Lauretta, D.~S., Kremser, D.~T., \& Fegley~Jr, B. 1996, Icarus, 122, 288

\bibitem[{Lewis(1974)}]{Lewis1974}
Lewis, J.~S. 1974, Science, 186, 440

\bibitem[{Lewis \& Prinn(1980)}]{Lewis1980}
Lewis, J.~S. \& Prinn, R.~G. 1980, Astrophysical Journal, Part 1, vol. 238, May 15, 1980, p. 357-364., 238, 357

\bibitem[{Li {et~al.}(2020)Li, Huang, Petaev, Zhu, \& Steffen}]{Li2020}
Li, M., Huang, S., Petaev, M.~I., Zhu, Z., \& Steffen, J.~H. 2020, Monthly Notices of the Royal Astronomical Society, 495, 2543

\bibitem[{Libourel {et~al.}(2025)Libourel, Mokhtari, Rohani, Bourdon, Ganino, Lagadec, Venn{\'e}gu{\`e}s, Guigoz, Cauneau, \& Fulcheri}]{Libourel2025}
Libourel, G., Mokhtari, M., Rohani, V.-J., {et~al.} 2025, Nature Astronomy, 9, 90

\bibitem[{Lissauer(1993)}]{Lissauer1993}
Lissauer, J.~J. 1993, Annual review of astronomy and astrophysics, 31, 129

\bibitem[{Lodders(2003)}]{Lodders2003}
Lodders, K. 2003, the Astrophysical Journal, 591, 1220

\bibitem[{Lodders {et~al.}(subm.)Lodders, Fegley, Mezger, \& Ebel}]{Lodders2024}
Lodders, K., Fegley, B., Mezger, K., \& Ebel, D. subm., arXiv preprint arXiv:2411.01362

\bibitem[{Lodders \& Fegley~Jr(1993)}]{Lodders1993}
Lodders, K. \& Fegley~Jr, B. 1993, Earth and Planetary Science Letters, 117, 125

\bibitem[{Lodders \& Fegley~Jr(1995)}]{Lodders1995}
Lodders, K. \& Fegley~Jr, B. 1995, Meteoritics, 30, 661

\bibitem[{Lodders \& Fegley~Jr(1997)}]{Lodders1997}
Lodders, K. \& Fegley~Jr, B. 1997, in AIP Conference Proceedings, Vol. 402, American Institute of Physics, 391--423

\bibitem[{Lodders \& Fegley~Jr(1999)}]{Lodders1999}
Lodders, K. \& Fegley~Jr, B. 1999, in Symposium-International Astronomical Union, Vol. 191, Cambridge University Press, 279--290

\bibitem[{Madhusudhan {et~al.}(2012)Madhusudhan, Lee, Mousis, Kanani, \& Mousis}]{Madhusudhan2012}
Madhusudhan, N., Lee, K. K.~M., Mousis, O., Kanani, L. K.~M., \& Mousis, O. 2012, the Astrophysical Journal Letters, 759, 5pp.

\bibitem[{Mah \& Bitsch(2023)}]{Mah2023}
Mah, J. \& Bitsch, B. 2023, Astronomy \& Astrophysics, 673, A17

\bibitem[{Mittlefehldt(2015)}]{Mittlefehldt2015}
Mittlefehldt, D.~W. 2015, Geochemistry, 75, 155

\bibitem[{Moriarty {et~al.}(2014)Moriarty, Madhusudhan, \& Fischer}]{Moriarty2014}
Moriarty, J., Madhusudhan, N., \& Fischer, D. 2014, The Astrophysical Journal, 787, 81

\bibitem[{Nicholls {et~al.}(2017)Nicholls, Sutherland, Dopita, Kewley, \& Groves}]{Nicholls2017}
Nicholls, D.~C., Sutherland, R.~S., Dopita, M.~A., Kewley, L.~J., \& Groves, B.~A. 2017, Monthly Notices of the Royal Astronomical Society, 466, 4403

\bibitem[{Noack {et~al.}(2014)Noack, Godolt, {Von Paris}, Plesa, Stracke, Breuer, \& Rauer}]{Noack2014_structure}
Noack, L., Godolt, M., {Von Paris}, P., {et~al.} 2014, Planetary and Space Science, 98, 14

\bibitem[{Norris \& Wood(2017)}]{Norris2017}
Norris, C.~A. \& Wood, B.~J. 2017, Nature, 549, 507

\bibitem[{O'Neill {et~al.}(2020)O'Neill, Lowman, \& Wasiliev}]{ONeill2020_GCE}
O'Neill, C., Lowman, J., \& Wasiliev, J. 2020, Icarus, 352, 114025

\bibitem[{O'Neill \& Palme(2008)}]{ONeill2008}
O'Neill, H. S.~C. \& Palme, H. 2008, Phil.\ Trans.\ R.\ Soc.\ A, 366, 4205

\bibitem[{Palme \& O'Neill(2014)}]{Palme2014}
Palme, H. \& O'Neill, H. 2014, {Cosmochemical Estimates of Mantle Composition}, 2nd edn., Vol.~3 (Elsevier Ltd.), 1--39

\bibitem[{Pascucci {et~al.}(2016)Pascucci, Testi, Herczeg, Long, Manara, Hendler, Mulders, Krijt, Ciesla, Henning, {et~al.}}]{Pascucci2016}
Pascucci, I., Testi, L., Herczeg, G.~J., {et~al.} 2016, The Astrophysical Journal, 831, 125

\bibitem[{Pekmezci {et~al.}(2019)Pekmezci, Johnson, Lunine, \& Mousis}]{Pekmezci2019}
Pekmezci, G., Johnson, T., Lunine, J., \& Mousis, O. 2019, The Astrophysical Journal, 887, 3

\bibitem[{Petaev \& Wood(1998)}]{Petaev1998}
Petaev, M.~I. \& Wood, J.~A. 1998, Meteoritics \& Planetary Science, 33, 1123

\bibitem[{Pignatari {et~al.}(2016)Pignatari, Herwig, Hirschi, Bennett, Rockefeller, Fryer, Timmes, Ritter, Heger, Jones, {et~al.}}]{Pignatari2016}
Pignatari, M., Herwig, F., Hirschi, R., {et~al.} 2016, The Astrophysical Journal Supplement Series, 225, 24

\bibitem[{Pignatari {et~al.}(2023)Pignatari, Trueman, Womack, Gibson, C{\^o}t{\'e}, Turrini, Sneden, Mojzsis, Stancliffe, Fong, {et~al.}}]{Pignatari2023}
Pignatari, M., Trueman, T.~C., Womack, K.~A., {et~al.} 2023, Monthly Notices of the Royal Astronomical Society, 524, 6295

\bibitem[{Pontoppidan {et~al.}(2014)Pontoppidan, Salyk, Bergin, Brittain, Marty, Mousis, \& {\"O}berg}]{Pontoppidan2014}
Pontoppidan, K.~M., Salyk, C., Bergin, E.~A., {et~al.} 2014, Protostars and Planets VI, 363

\bibitem[{Pringle {et~al.}(2014)Pringle, Moynier, Savage, Badro, \& Barrat}]{Pringle2014}
Pringle, E.~A., Moynier, F., Savage, P.~S., Badro, J., \& Barrat, J.-A. 2014, Proceedings of the National Academy of Sciences, 111, 17029

\bibitem[{Putirka \& Rarick(2019)}]{Putirka2019}
Putirka, K.~D. \& Rarick, J.~C. 2019, American Mineralogist, 104, 817

\bibitem[{Sengupta {et~al.}(2022)Sengupta, Estrada, Cuzzi, \& Humayun}]{Sengupta2022}
Sengupta, D., Estrada, P.~R., Cuzzi, J.~N., \& Humayun, M. 2022, The Astrophysical Journal, 932, 82

\bibitem[{Shakespeare {et~al.}(2024)Shakespeare, Li, Huang, Zhu, \& Steffen}]{Shakespeare2024}
Shakespeare, C.~J., Li, M., Huang, S., Zhu, Z., \& Steffen, J.~H. 2024, arXiv preprint arXiv:2408.07761

\bibitem[{Sossi {et~al.}(2019)Sossi, Klemme, O'Neill, Berndt, \& Moynier}]{Sossi2019}
Sossi, P.~A., Klemme, S., O'Neill, H. S.~C., Berndt, J., \& Moynier, F. 2019, Geochimica et Cosmochimica Acta, 260, 204

\bibitem[{Sossi {et~al.}(2022)Sossi, Stotz, Jacobson, Morbidelli, \& O’Neill}]{Sossi2022}
Sossi, P.~A., Stotz, I.~L., Jacobson, S.~A., Morbidelli, A., \& O’Neill, H. S.~C. 2022, Nature Astronomy, 1

\bibitem[{Spaargaren {et~al.}(2020)Spaargaren, Ballmer, Bower, Dorn, \& Tackley}]{Spaargaren2020}
Spaargaren, R.~J., Ballmer, M.~D., Bower, D.~J., Dorn, C., \& Tackley, P.~J. 2020, Astronomy and Astrophysics, 643, A44

\bibitem[{Spaargaren {et~al.}(2023)Spaargaren, Wang, Mojzsis, Ballmer, \& Tackley}]{Spaargaren2023}
Spaargaren, R.~J., Wang, H.~S., Mojzsis, S.~J., Ballmer, M.~D., \& Tackley, P.~J. 2023, The Astrophysical Journal, 948, 53

\bibitem[{Stamenkovi{\'{c}} \& Seager(2016)}]{Stamenkovic2016}
Stamenkovi{\'{c}}, V. \& Seager, S. 2016, The American Astronomical Society, 825, 18 pp.

\bibitem[{Steinmeyer {et~al.}(2023)Steinmeyer, Woitke, \& Johansen}]{Steinmeyer2023}
Steinmeyer, M.-L., Woitke, P., \& Johansen, A. 2023, Astronomy \& Astrophysics, 677, A181

\bibitem[{Stonkut{\.e} {et~al.}(2020)Stonkut{\.e}, Chorniy, Tautvai{\v{s}}ien{\.e}, Drazdauskas, Minkevi{\v{c}}i{\=u}t{\.e}, Mikolaitis, Kjeldsen, von Essen, Pak{\v{s}}tien{\.e}, \& Bagdonas}]{Stonkute2020}
Stonkut{\.e}, E., Chorniy, Y., Tautvai{\v{s}}ien{\.e}, G., {et~al.} 2020, The Astronomical Journal, 159, 90

\bibitem[{Takahashi {et~al.}(2013)Takahashi, Ohtani, Terasaki, Ito, Shibazaki, Ishii, Funakoshi, \& Higo}]{Takahashi2013}
Takahashi, S., Ohtani, E., Terasaki, H., {et~al.} 2013, Physics and Chemistry of Minerals, 40, 647

\bibitem[{Teske {et~al.}(2013)Teske, Cunha, Schuler, Griffith, \& Smith}]{Teske2013}
Teske, J.~K., Cunha, K., Schuler, S.~C., Griffith, C.~A., \& Smith, V.~V. 2013, The Astrophysical Journal, 778, 132

\bibitem[{Timmermann {et~al.}(2023)Timmermann, Shan, Reiners, \& Pack}]{Timmermann2023}
Timmermann, A., Shan, Y., Reiners, A., \& Pack, A. 2023, arXiv preprint arXiv:2307.00914

\bibitem[{Unterborn {et~al.}(2017)Unterborn, Hull, Stixrude, Teske, Johnson, \& Panero}]{Unterborn2017}
Unterborn, C.~T., Hull, S.~D., Stixrude, L.~P., {et~al.} 2017, arXiv preprint arXiv:1706.10282

\bibitem[{Urey(1953)}]{Urey1953}
Urey, H.~C. 1953, Proceedings of the Royal Society of London. Series A. Mathematical and Physical Sciences, 219, 281

\bibitem[{Vincenzo \& Kobayashi(2018)}]{Vincenzo2018}
Vincenzo, F. \& Kobayashi, C. 2018, Monthly Notices of the Royal Astronomical Society, 478, 155

\bibitem[{Vollstaedt {et~al.}(2020)Vollstaedt, Mezger, \& Alibert}]{Vollstaedt2020}
Vollstaedt, H., Mezger, K., \& Alibert, Y. 2020, The Astrophysical Journal, 897, 82

\bibitem[{Wang {et~al.}(2018)Wang, Lineweaver, \& Ireland}]{Wang2018_Earthcomp}
Wang, H.~S., Lineweaver, C.~H., \& Ireland, T.~R. 2018, Icarus, 299, 460

\bibitem[{Wang {et~al.}(2019{\natexlab{a}})Wang, Lineweaver, \& Ireland}]{Wang2019}
Wang, H.~S., Lineweaver, C.~H., \& Ireland, T.~R. 2019{\natexlab{a}}, Icarus, 328, 287

\bibitem[{Wang {et~al.}(2022{\natexlab{a}})Wang, Lineweaver, Quanz, Mojzsis, Ireland, Sossi, Seidler, \& Morel}]{Wang2022_alphacentauri}
Wang, H.~S., Lineweaver, C.~H., Quanz, S.~P., {et~al.} 2022{\natexlab{a}}, AJ, 134

\bibitem[{Wang {et~al.}(2019{\natexlab{b}})Wang, Liu, Ireland, Brasser, Yong, \& Lineweaver}]{Wang2019MNRAS}
Wang, H.~S., Liu, F., Ireland, T.~R., {et~al.} 2019{\natexlab{b}}, Mon. Not. R. Astron. Soc., 482, 2222

\bibitem[{Wang {et~al.}(2022{\natexlab{b}})Wang, Quanz, Yong, Liu, Seidler, Acu{\~n}a, \& Mojzsis}]{Wang2022_planethoststars}
Wang, H.~S., Quanz, S.~P., Yong, D., {et~al.} 2022{\natexlab{b}}, MNRAS, 513, 5829

\bibitem[{Wasson(1985)}]{Wasson1985}
Wasson, J. 1985, {Meteorites: Their record of early solar-system history} (New York: W.H. Freeman and Co.), 274pp.

\bibitem[{Whittet {et~al.}(2007)Whittet, Shenoy, Bergin, Chiar, Gerakines, Gibb, Melnick, \& Neufeld}]{Whittet2007}
Whittet, D., Shenoy, S., Bergin, E., {et~al.} 2007, The Astrophysical Journal, 655, 332

\bibitem[{Woitke {et~al.}(2018)Woitke, Helling, Hunter, Millard, Turner, Worters, Blecic, \& Stock}]{Woitke2018}
Woitke, P., Helling, C., Hunter, G., {et~al.} 2018, Astronomy \& Astrophysics, 614, A1

\bibitem[{Wood(1993)}]{Wood1993}
Wood, B.~J. 1993, Earth and Planetary Science Letters, 117, 593

\bibitem[{Wood {et~al.}(2019)Wood, Smythe, \& Harrison}]{Wood2019}
Wood, B.~J., Smythe, D.~J., \& Harrison, T. 2019, Am.$\backslash$ Mineral., 104, 844

\bibitem[{Yoshimura {et~al.}(1986)Yoshimura, Kase, \& S{\=o}miya}]{Yoshimura1986}
Yoshimura, M., Kase, J.-i., \& S{\=o}miya, S. 1986, Journal of Materials Research, 1, 100

\bibitem[{Yoshizaki \& McDonough(2020)}]{Yoshizaki2020}
Yoshizaki, T. \& McDonough, W.~F. 2020, Geochim. Cosmochim. Acta, 273, 137

\bibitem[{Young {et~al.}(2019)Young, Shahar, Nimmo, Schlichting, Schauble, Tang, \& Labidi}]{Young2019}
Young, E., Shahar, A., Nimmo, F., {et~al.} 2019, Icarus, 323, 1

\bibitem[{Zhang \& Driscoll(2025)}]{Zhang2025}
Zhang, Z. \& Driscoll, P.~E. 2025, Journal of Geophysical Research: Planets, 130, e2024JE008671

\bibitem[{Zimmer {et~al.}(2016)Zimmer, Zhang, Lu, Chen, Zhang, Dalkilic, \& Zhu}]{Zimmer2016}
Zimmer, K., Zhang, Y., Lu, P., {et~al.} 2016, Computers \& geosciences, 90, 97

\end{thebibliography}

\FloatBarrier

\begin{appendix}
\onecolumn

\section{Condensation temperature parametrisations}\label{sec:App_TcParam}
Based on our sample of 1000 \textsc{GGchem} simulations, we derive parametrisations of how element condensation temperature varies with the bulk PPD composition, which can be applied to all stellar abundances without simulating the condensation sequence for each star. We attempt to find mathematical fits to the condensation temperatures of our sample with the lowest Bayesian Information Criterion (BIC), where BIC$= n\cdot \ln \left( \frac{y - \bar{y}}{n} \right) + \ln (n) \cdot p $ for data $y$, fit to the data $\bar{y}$, sample size $n$, and number of parameters required to produce the fit $p$. This process is performed for the 50\% condensation temperatures of Na, Mg, Al, Si, Ca, Fe, and Ni, and for the effective condensation temperatures of C, O, and S (see Sect.\ \ref{sec:Res_vol_elems}).

\subsection{Metallicity fits}
As mentioned in Sect.\ \ref{sec:Methods}, we simulate condensation sequences for stellar abundances with metallicity normalised to solar (metallicity $M$ as defined in Eq.\ \ref{eq:Metallicity}), and add a metallicity function to the condensation temperature fits. These metallicity functions $T_M^X$ are based on a suite of \textsc{GGchem} simulations with solar abundances scaled to a range of metallicities. We fit polynomial functions of orders 1 through 10 to the element condensation temperatures from these simulations, and select the fit with the lowest BIC:
\begin{equation}
\begin{split}
&T_M^{\text{Na}}(M) = 6.08 M^2 - 33.16 M + 732.57 \\
&T_M^{\text{Mg}}(M) = 12.65 M^2 - 96.68 M + 1173.24 \\
&T_M^{\text{Al}}(M) = 14.89 M^2 - 102.05 M + 1350.80 \\
&T_M^{\text{Si}}(M) = 10.42 M^2 - 60.90 M + 1024.73 \\
&T_M^{\text{Ca}}(M) = 13.49 M^2 - 95.13 M + 1263.85 \\
&T_M^{\text{Fe}}(M) = 5.23 M^2 - 9.02 M + 988.58 \\
&T_M^{\text{Ni}}(M) = 3.94 M^2 + 2.92 + 931.38 \\
&T_M^{\text{C}}(M) = 24.97 M^3 - 687.27 M^2 + 6353.20 M - 18764.59 \\
&T_M^{\text{O}}(M) = 4.998 M^5 - 2.118 \cdot 10^2 M^4 + 3.572 \cdot 10^3 M^3 - 2.998 \cdot 10^4 M^2 + 1.253 \cdot 10^5 M - 2.077 \cdot 10^5 \\
&T_M^{\text{S}}(M) = 2.59 M^5 - 1.085 \cdot 10^2 M^4 + 1.81 \cdot 10^3 M^3 - 1.51 \cdot 10^4 M^2 + 6.26 \cdot 10^4 M - 1.03 \cdot 10^5 \\
\end{split}
 \label{eq:Tc_metallicity_dependence}
\end{equation}
where $T$ is in K, and $M$ is log metallicity, as defined in Sect.\ \ref{sec:Methods}. The volatiles, for which we consider effective condensation temperatures, have more complex dependence on metallicity than the non-volatile elements where we consider the 50\% condensation temperatures. These metallicity-dependences are mostly derived for solar abundances scaled uniformly, except for $T_M^{\text{C}}(M)$, which is based on a metallicity series where we first adjusted the C and O abundances such that C/O = 1.30, while keeping the sum of the C and O abundances constant.

\subsection{Fe and Ni}\label{ssec:App_Tc_FeNi_fit}
Aside from metallicity, Fe and Ni only depend on their own abundances (Fig.\ \ref{fig:Tc_FeNi}). Therefore, simple polynomial fits to $A_{\text{Fe}}$ and $A_{\text{Ni}}$, which are the log molar abundances of Fe and Ni normalised to H=12, describe the condensation temperatures of Fe ($\Tc{Fe}$) and Ni ($\Tc{Ni}$) well. Their condensation temperatures in K are given by
\begin{equation}
\Tc{Fe} = 4.776 A_{\text{Fe}}^2 + 12.77 A_{\text{Fe}} + 964.64 + T_M^{\text{Fe}}(M) - T_M^{\text{Fe}}(M_{\odot})
\label{eq:Tc_Fe}
\end{equation} 
and
\begin{equation}
\Tc{Ni} = 4.396 A_{\text{Ni}}^2 + 18.97 A_{\text{Ni}} + 986.76 + T_M^{\text{Ni}}(M) - T_M^{\text{Ni}}(M_{\odot}),
\label{eq:Tc_Ni}
\end{equation}
where the metallicity dependence is given by Eq.\ \ref{eq:Tc_metallicity_dependence}, and $M_{\odot} = 8.99$.

\subsection{C/O-dependence}
The elements Mg, Si, Ca, Al, and Na all have a very similar dependence on C/O in the O-rich regime (Figs.\ \ref{fig:Tc_MgSiCaAl}, \ref{fig:Tc_Na}). The function that fits these parts the best (i.e.\ with the lowest BIC) for all elements is the asymptotic function $f_a(x)$,
\begin{equation}
    f_{asymptote}(x) = a \frac{x'^n}{b + x'^n},
\end{equation}
with a coordinate transformation $x' = \alpha x + \beta$, where typically $\alpha < 0$. This function is used to fit $\Tc{}$ values for C/O smaller than the critical C/O values for each element. The area for C/O greater than the critical values is usually characterised by a short polynomial section and a flat section.

The critical C/O value between the O-rich and the O-depleted regime also depends on X/O for element X. Some fits are better by simply adjusting this critical C/O value, while for some elements the fit becomes better by replacing C/O by $\mathrm{C/O}' = \mathrm{C/O} + x_1 \mathrm{X/O} + x_2$. These fits are again guided by finding the lowest BIC.

\subsection{Ca}
The C/O-dependence of Ca is best characterised by
\begin{equation*}
    T_{C/O}^{\text{Ca}}(\text{C/O}) = \begin{cases}
    f_{asymptote}(\text{C/O}) & \text{C/O} < \text{C/O}_{crit} \\
    3.052 \cdot 10^3\ \text{C/O}^2 - 5.81 \cdot 10^3\ \text{C/O} + 4.001 \cdot 10^3,  &\text{C/O}_{upper} > \text{C/O} \geq \text{C/O}_{crit} \\
    1265.77 & \text{C/O} \geq \text{C/O}_{upper}
    \end{cases}
\end{equation*}
for C/O$_{crit} = 0.95 - 7.245 (\text{Ca/O} - \text{Ca/O}_{\odot})$, C/O$_{upper} = 1.045$, and parameters for the asymptote function $a = 1919.6$, $b = 1.356$, $n=0.1396$, $\alpha = -1.76 \cdot 10^5$, and $\beta = 1.726 \cdot 10^5$.

Further, Ca depends on the abundance of itself, which differs in the O-rich and O-depleted settings,
\begin{equation*}
    T_{\text{Ca}}^{\text{Ca}}(Ca) = \begin{cases}
    39.52 (A_{\text{Ca}} - A_{\text{Ca},\odot}) & \text{C/O} < \text{C/O}_{crit} \\
    53.91 (A_{\text{Ca}} - A_{\text{Ca},\odot}) & \text{C/O} \geq \text{C/O}_{crit}
    \end{cases},
\end{equation*}
and on Ca/Al through a polynomial,
\begin{equation*}
    T_{\text{Ca}/\text{Al}}^{\text{Ca}}(\text{Ca/Al}) = 2.83 \text{Ca/Al}^4 - 17.39 \text{Ca/Al}^3 + 27.78 \text{Ca/Al}^2 - 12.05 \text{Ca/Al} + 1497.29,
\end{equation*}
where Ca/Al is the absolute stellar Ca/Al molar ratio, which corresponds to Ca/Al$=\epsilon_{\mathrm{Ca}} - \epsilon_{\mathrm{Al}}$. Combining all of these equations, the full Tc parametrisation of Ca is
\begin{equation}
    \Tc{Ca} = T_{C/O}^{\text{Ca}}(\text{C/O}) + T_{\text{Ca}}^{\text{Ca}}(Ca) + T_{\text{Ca}/\text{Al}}^{\text{Ca}}(\text{Ca/Al}) - T_{\text{Ca}/\text{Al}}^{\text{Ca}}(\text{Ca/Al}_{\odot}) + T_M^{\text{Ca}}(M) - T_M^{\text{Ca}}(M_{\odot})
     \label{eq:Tc_Ca}
\end{equation}

\subsection{Al}
The C/O-dependence of Al is better characterised by using C/O' = C/O $+ 4.47 \cdot \text{Al/O} - 0.022$, 
\begin{equation*}
    T_{C/O}^{Al}(\text{C/O'}) = \begin{cases}
    f_{asymptote}(\text{C/O'}) & \text{C/O'} < \text{C/O}_{crit} \\
    - 2.6649 \cdot 10^5\ \text{C/O'}^3 + 7.89\cdot 10^5\ \text{C/O'}^2 - 7.79 \cdot 10^5\ \text{C/O} + 2.575 \cdot 10^5,  &\text{C/O}_{upper} > \text{C/O'} \geq \text{C/O}_{crit} \\
    1231.46 & \text{C/O'} \geq \text{C/O}_{upper}
    \end{cases},
\end{equation*}
for C/O$_{crit} = 0.95$, C/O$_{upper} = 1.039$, and parameters for the asymptote function $a = 9.563 \cdot 10^4$, $b = 75.65$, $n = 0.03$, $\alpha = -2.23 \cdot 10^4$, and $\beta = 2.15 \cdot 10^4$.

Further, Al depends on the stellar abundances of Al and O, making the full description
\begin{equation}
    \Tc{Al} = T_{C/O}^{Al}(\text{C/O'}) + 44.67 (A_{O} - A_{O,\odot}) + 55.34 (A_{Al} - A_{Al,\odot}) + T_M^{Al}(M) - T_M^{Al}(M_{\odot})
\end{equation}

\subsection{Mg}
The C/O-dependence of Mg follows that of Al, using C/O' = C/O $+ 1.184$ Mg/O $- 0.0856$,
\begin{equation*}
    T_{C/O}^{Mg}(\text{C/O'}) = \begin{cases}
    f_{asymptote}(\text{C/O'}) & \text{C/O'} < \text{C/O}_{crit} \\
    - 5.153 \cdot 10^5\ \text{C/O'}^3 + 1.418\cdot 10^6\ \text{C/O'}^2 - 1.300 \cdot 10^6\ \text{C/O} + 3.983 \cdot 10^5,  &\text{C/O}_{upper} > \text{C/O'} \geq \text{C/O}_{crit} \\
    1061.44 - 40.65*(\text{Mg/O} - \text{Mg/O}_{\odot}) & \text{C/O'} \geq \text{C/O}_{upper}
    \end{cases},
\end{equation*}
for C/O$_{crit} = 0.88$ and C/O$_{upper} = 0.94$, with asymptote function parameters $a = 1387.91$, $b=5.697$, $n=0.588$, $\alpha = -8193.75$, and $\beta = 7787.28$.

Further, Mg depends on the Mg and O abundances, making the full description
\begin{equation}
    \Tc{Mg} = T_{C/O}^{Mg}(\text{C/O'}) + 36.33 (A_{O} - A_{O,\odot}) + 31.86 (A_{Mg} - A_{Mg,\odot}) + T_M^{Mg}(M) - T_M^{Mg}(M_{\odot})
\end{equation}

\subsection{Si}
The dependence of Si on C/O contains an additional segment, as it increases strongly in a narrow C/O-range before forming a plateau,
\begin{equation*}
    T_{C/O}^{Si}(\text{C/O}) = 
    \begin{cases}
        f_{asymptote}(\text{C/O}) + f_{Mg/Si}^{Si}(\text{Mg}/\text{Si}) + \\
        \quad f_{cations}^{Si}(\Sigma /\text{O}) 
        & \text{C/O} < \text{C/O}_{crit} \\[12pt]
        
        \begin{aligned}
        &1.142\cdot 10^4\, \text{C/O}^3 - 2.97 \cdot 10^4\, \text{C/O}^2 \\
        &+ 2.57 \cdot 10^4\, \text{C/O} - 2.765 \cdot 10^3
        \end{aligned}
        & \text{C/O}_{middle} > \text{C/O} - 405.23 A_O \geq \text{C/O}_{crit} \\[12pt]
        
        \begin{aligned}
        &-1.070 \cdot 10^6\, \text{C/O}^3 + 3.252 \cdot 10^6\, \text{C/O}^2 \\
        &- 3.287 \cdot 10^6\, \text{C/O} + 1.106 \cdot 10^6
        \end{aligned}
        & \text{C/O}_{upper} > \text{C/O} \geq \text{C/O}_{middle} \\[12pt]
        
        \begin{aligned}
        &-3.650 \cdot 10^5\, \text{Si/O}^4 + 2.261 \cdot 10^5\, \text{Si/O}^3 \\
        &- 4.990 \cdot 10^4\, \text{Si/O}^2 + 4.927 \cdot 10^3\, \text{Si/O} \\
        &+ 1.351 \cdot 10^3
        \end{aligned}
        & \text{C/O} \geq \text{C/O}_{upper}
    \end{cases}
\end{equation*}
for $\text{C/O}_{crit} = 0.85 - 1.8*(\text{Si/O} - \text{Si/O}_{\odot})$, $\text{C/O}_{middle} = 1.022$, $\text{C/O}_{upper}=1.0445$, and the parameters for the asymptote function are $a = 750.48$, $b = 5.7914 \cdot 10^{-2}$, $n = 3.333$, $\alpha = -1$, and $\beta = 1.547$. Further, the low-C/O section depends on Mg/Si and the molar ratio of all oxide-forming cations, multiplied by their oxide stoichiometry, divided by oxygen ($\Sigma/\text{O}$), where the cations are multiplied with their oxide stoichiometry and then summed up. These dependencies also have critical transitions, where 
\begin{equation*}
    f_{Mg/Si}^{Si}(\text{Mg}/\text{Si}) = \begin{cases}
    117.69 & \text{Mg}/\text{Si} < \text{Mg}/\text{Si}_{lower} \\
    -67.18 \text{Mg}/\text{Si}^2 + 206.82\text{Mg}/\text{Si} & \text{Mg}/\text{Si}_{upper} > \text{Mg}/\text{Si} \geq \text{Mg}/\text{Si}_{lower} \\
    161.89 & \text{Mg}/\text{Si} > \text{Mg}/\text{Si}_{upper} \\
    \end{cases},
\end{equation*}
for $\text{Mg}/\text{Si}_{lower} = 0.73$ and $\text{Mg}/\text{Si}_{upper} = 1.64$, and
\begin{equation*}
    f_{cations}^{Si}(\Sigma /\text{O}) = \begin{cases}
    8.523 \Sigma/\text{O} + 448.63 & \Sigma/\text{O} < \Sigma/\text{O}_{crit} \\
    -890.11 \Sigma/\text{O} + 1401.59 & \Sigma/\text{O} > \Sigma/\text{O}_{crit} \\
    \end{cases},
\end{equation*}
for $\Sigma/\text{O}_{crit} = 1.0695$.

Finally, the full dependence of Si condensation on disk chemistry is
\begin{equation}
    \Tc{Si} = T_{C/O}^{Si}(\text{C/O},\text{Mg}/\text{Si},\Sigma/\text{O}) + T_M^{Si}(M) - T_M^{Si}(M_{\odot})
\end{equation}

\subsection{Na}
The condensation behaviour of Na depends critically on both C/O and Mg/Si,  
\begin{equation*}
    T_{C/O}^{Na}(\text{C/O}) = \begin{cases}
    f_{asymptote}(\text{C/O}) & \text{C/O} < \text{C/O}_{crit} \\
    -55.844 \text{C/O} + 915.76 &\text{C/O}_{upper} > \text{C/O} \geq \text{C/O}_{crit} \\
    862.77 & \text{C/O} \geq \text{C/O}_{upper},
    \end{cases}
\end{equation*}
where $\text{C/O}_{crit} = 0.77 - 28.82 (\text{Na}/\text{O} - \text{Na}/\text{O}_{\odot})$, and with asymptote function parameters $a = 947.54$, $b=56.34$, $n=1463.62$, $\alpha=-2.52 \cdot 10^{-3}$, $\beta = 1.007$. Further,
\begin{equation*}
    T_{Mg/Si}^{Na}(\text{Mg/Si}) = \begin{cases}
    47.14 & \text{Mg/Si} < \text{Mg/Si}_{lower} \\
    -182.86 (\text{Mg/Si} - \text{Mg/Si}_{middle}) &\text{Mg/Si}_{upper} > \text{Mg/Si} \geq \text{Mg/Si}_{middle} \\
    -148.80 & \text{Mg/Si} \geq \text{Mg/Si}_{upper},
    \end{cases}
\end{equation*}
with $\text{Mg/Si}_{lower} = 0.836$, $\text{Mg/Si}_{middle}=1.499$, and $\text{Mg/Si}_{upper}=1.975$. Note that there is no Mg/Si-dependence between $\text{Mg/Si}_{lower}$ and $\text{Mg/Si}_{middle}$. Further, Na depends on Na/Cl,
\begin{equation*}
    T_{Na/Cl}^{Na}(\text{Na/Cl}) = 5.971 \cdot 10^{-2} \text{Na/Cl}^3 - 1.879 \text{Na/Cl}^2 + 18.96 \text{Na/Cl} + 8.77.76.
\end{equation*}
Finally, the full dependence of Na condensation on disk chemistry is
\begin{equation}
    \Tc{Na} = T_{C/O}^{Na}(\text{C/O}) + T_{Mg/Si}^{Na}(\text{Mg/Si}) + T_{Na/Cl}^{Na}(\text{Na/Cl}) + T_M^{Na}(M) - T_M^{Na}(M_{\odot})
\end{equation}

\subsection{C}
The condensation behaviour of C can be split into three sections: The section where no C condenses in our models ($\Tc{C}=0$\,K), the region where C starts condensing and depends mainly on $\Sigma/O$, and the region where C condensation depends primarily on C/O. It follows a similar shape for both sections where C condenses, primarily following the asymptote function, with
\begin{equation*}
    T_{O}^{C}(\text{C/O},\Sigma/\text{O}) = \begin{cases}
    0 & \Sigma\text{/O} < \Sigma\text{/O}_{crit} \\
    f_{asymptote}(\Sigma\text{/O}) + 2413.80\,\text{Si/C}^2 - 1024.77\,\text{Si/C} &\Sigma\text{/O} \geq \Sigma\text{/O}_{crit}, \text{C/O} < \text{C/O}_{crit} \\
    f_{asymptote}(\text{C/O}) + 743.10\,\text{Si/C} & \text{C/O} \geq \text{C/O}_{crit}
    \end{cases},
\end{equation*}
where $\Sigma\text{/O}_{crit}=1.0181$, $\text{C/O}_{crit}=1.079$, and with asymptote function parameters for the $\Sigma\text{/O})$ section of $a=1575.77$, $b=0.491$, $n=0.613$, $\alpha=1$, and $\beta-1.015$, and for the C/O section the asymptote function has parameters $a=1343.53$, $b=0.227$, $n=0.808$, $\alpha=1$, and $\beta=-0.899$. Adding the metallicity dependence, the full condensation behaviour parametrisation of C becomes
\begin{equation}
    \Tc{C} = T_{O}^{C}(\text{C/O},\Sigma/\text{O}) + T_M^{C}(M) - T_M^{C}(M_{\odot})
\end{equation}

\subsection{O}
The effective condensation temperature of O is strongly dependent on the presence of other condensates; adding rock-forming elements beyond the ones we consider in this work would alter $\Tceff{O}$. We recommend caution in applying this parametrisation, but will nonetheless give its general form, which is informative on its primary condensation behaviour. While carbon is the primary limiter on O condensation, as it locks oxygen away as CO gas, the primary control on $\Tceff{O}$ is the abundances of the rock-forming elements that O can form oxides with, compared to the abundance of oxygen that is available. Thus, using prior definition of $\Sigma/$O, let us define $\Gamma = \left( \Sigma - \epsilon_{\text{C}} \right)$; $\Tceff{O}$ depends primarily on $\Gamma/$O. Its dependence is different for high and low C/O, with a transition to lower temperatures starting around C/O=0.7 and continuing to C/O=1.0, to a trend with a very similar slope. We can define two trends of $\Tceff{O}$ with $\Gamma/$O, $f_{lower}\left( \Gamma /\text{O} \right)$ and $f_{upper}\left( \Gamma /\text{O} \right)$, as well as a transition function $t(\text{C/O})$, such that
\begin{equation*}
    T_{\Gamma}^O = f_{lower}\left( \Gamma /\text{O} \right) \cdot (1 - t(\text{C/O})) + f_{upper}\left( \Gamma /\text{O} \right) \cdot t(\text{C/O}),
\end{equation*}
where
\begin{equation*}
    f_{lower}\left( \Gamma /\text{O} \right) = 69.39\,\Gamma /\text{O}^2 + 605.66\,\Gamma /\text{O} + 1251.52,
\end{equation*}
\begin{equation*}
    f_{upper}\left( \Gamma /\text{O} \right) = 154.10\,\Gamma /\text{O}^2 + 588.53\,\Gamma /\text{O} + 1012.52,
\end{equation*}
and
\begin{equation*}
    t(\text{C/O}) = \frac{1}{1 + \exp \left( -13.539 \cdot(\text{C/O} - 0.847) \right)}.
\end{equation*}
Thus, the condensation behaviour of oxygen can be described by
\begin{equation}
    \Tc{O} = f_{lower}\left( \Gamma /\text{O} \right) \cdot (1 - t(\text{C/O})) + f_{upper}\left( \Gamma /\text{O} \right) \cdot t(\text{C/O}) + T_M^{O}(M) - T_M^{O}(M_{\odot}).
\end{equation}

\subsection{S}
The effective condensation temperature of S primarily depends on $\Sigma$/O, with a transition from O-rich condensation, where $\Tceff{S}$ typically stays below 700\,K, to O-depleted condensation, where $\Tceff{S}$ can reach up to 1100\,K, at $\Sigma$/O$=1.005$. If we define $\phi$ as $\log \left( \epsilon_{\mathrm{Fe}} + \epsilon_{\mathrm{S}} \right)$, it can be characterised as
\begin{equation*}
    T_{\Sigma/O}^S = \begin{cases}
          -35.40 \phi^3 + 811.91 \phi^2 - 6085.17 \phi + 15540.68 & \Sigma/\text{O} \leq 1.005 \\
        T_{\Sigma/O,upper}^S\left( \Sigma/\text{O} \right) + T_{\text{Mg/S}}^S(\text{Mg/S}) + 10480.54 \text{Ca/O} & \Sigma/\text{O} > 1.005
    \end{cases},
\end{equation*}
where $T_{\text{Mg/S}}^S(\text{Mg/S}) = 0.367 \text{Mg/S}^5 - 9.076 \text{Mg/S}^4 + 79.86 \text{Mg/S}^3 - 302.50 \text{Mg/S}^2 + 514.21 \text{Mg/S}$, and
\begin{equation*}
    T_{\Sigma/O,upper}^S\left( \Sigma/\text{O} \right) = \begin{cases}
        -4.16 \cdot 10^5 \Sigma/\text{O}^4 + 1.897 \cdot 10^6 \Sigma/\text{O}^3 - 3.24 \cdot 10^6 \Sigma/\text{O}^2 + 2.45 \cdot 10^6 \Sigma/\text{O} - 6.96 \cdot 10^5 & \Sigma/\text{O} \leq 1.268 \\
        406.06 & \Sigma/\text{O} > 1.268
    \end{cases}.
\end{equation*}
Thus, overall the condensation behaviour of S can be described by
\begin{equation}
    \Tc{O} = T_{\Sigma/O}^S \left( \Sigma/\text{O},\text{Mg/S},\text{Ca/O}, \Gamma/\text{O} \right) + T_M^{S}(M) - T_M^{S}(M_{\odot}).
\end{equation}

\section{High-C/O condensate stability}\label{sec:App_highCO}
\begin{figure*}[h!]
    \resizebox{\hsize}{!}{\includegraphics{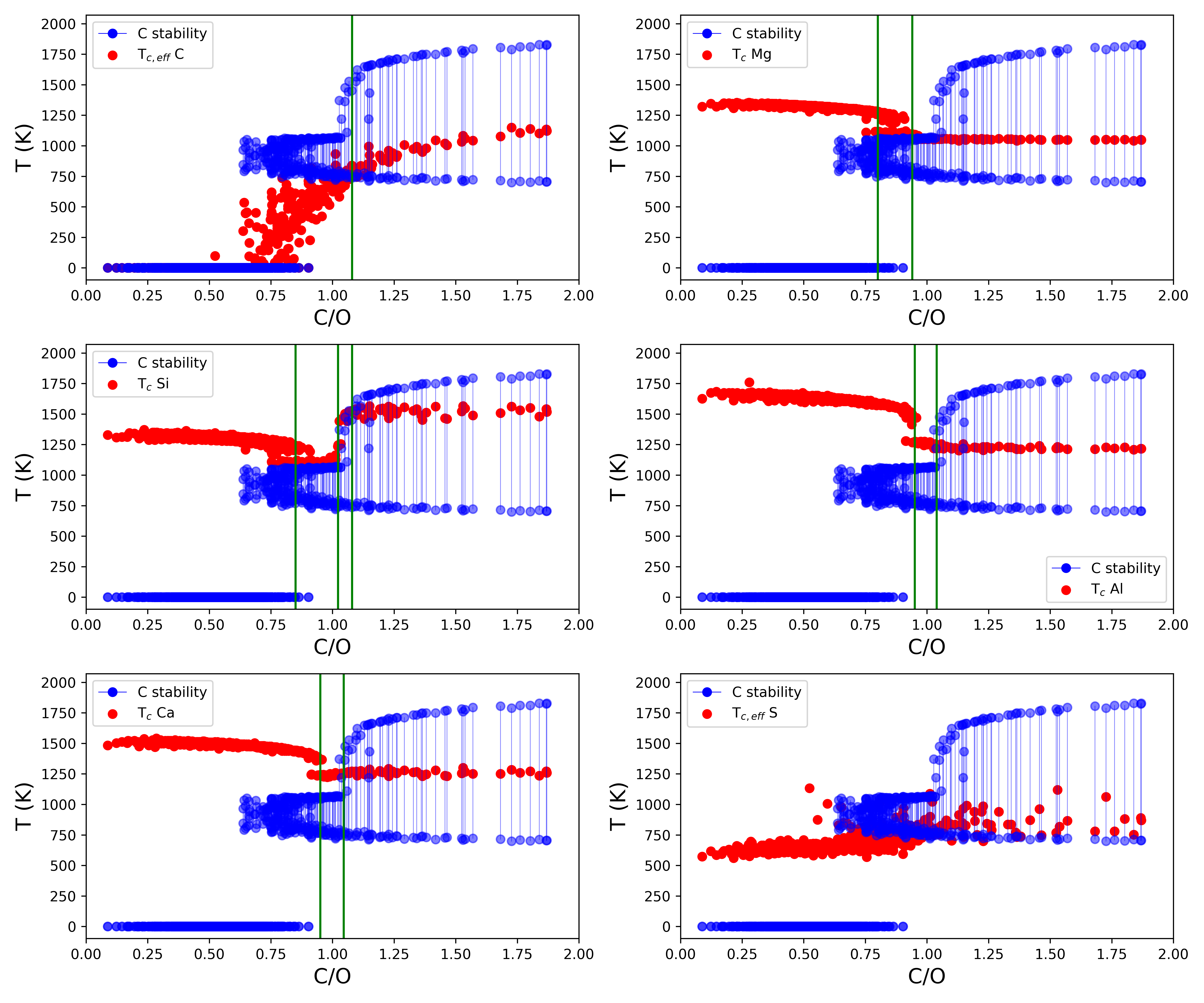}}
      \caption[Graphite stability and condensation temperatures as a function of bulk disk C/O]{Graphite condensation stability range as a function of bulk disk C/O ratio (blue). Condensation temperatures of carbon and rock-forming elements (red) are affected by increased graphite stability at C/O=1.04. Vertical green lines indicate transitions in $\Tc{}$ behaviour, as detailed in Sect.\ \ref{sec:App_TcParam}.}
      \label{fig:App_C_stability_Tc}
\end{figure*}

In most disks, including the solar disk, rock-forming elements condense as oxides and silicates, forming common minerals. However, in high-C/O disks, these elements tend to condense as sulphides and carbides instead, due to the unavailability of oxygen. This region of compositional parameter space is also characterised by the condensation of graphite, which starts occurring as soon as the summed abundances of rock-forming elements plus carbon exceeds the abundance of oxygen (i.e.\ $\Sigma/$O; Fig.\ \ref{fig:Tc_COS}). While at first the condensation temperatures of rock-forming elements depend on C/O, the gaseous C/O is buffered to 1.0 as soon as graphite becomes the primary high-temperature condensate \citep{Lodders1997}, and thus the condensation temperatures of these elements are no longer dependent on C/O. The condensation temperature of graphite starts increasing considerably at C/O=1.04, coinciding with the C/O at which $\Tc{Ca}$ and $\Tc{Al}$ become independent of C/O (Fig.\ \ref{fig:App_C_stability_Tc}). The condensation temperature of Si increases sharply once graphite becomes significantly more stable as a condensate, indicating the increased availability of C for condensation, though when the condensation temperature of graphite exceeds that of SiC, $\Tc{Si}$ also becomes independent of bulk disk C/O.

\begin{figure*}[h!]
    \resizebox{\hsize}{!}{\includegraphics{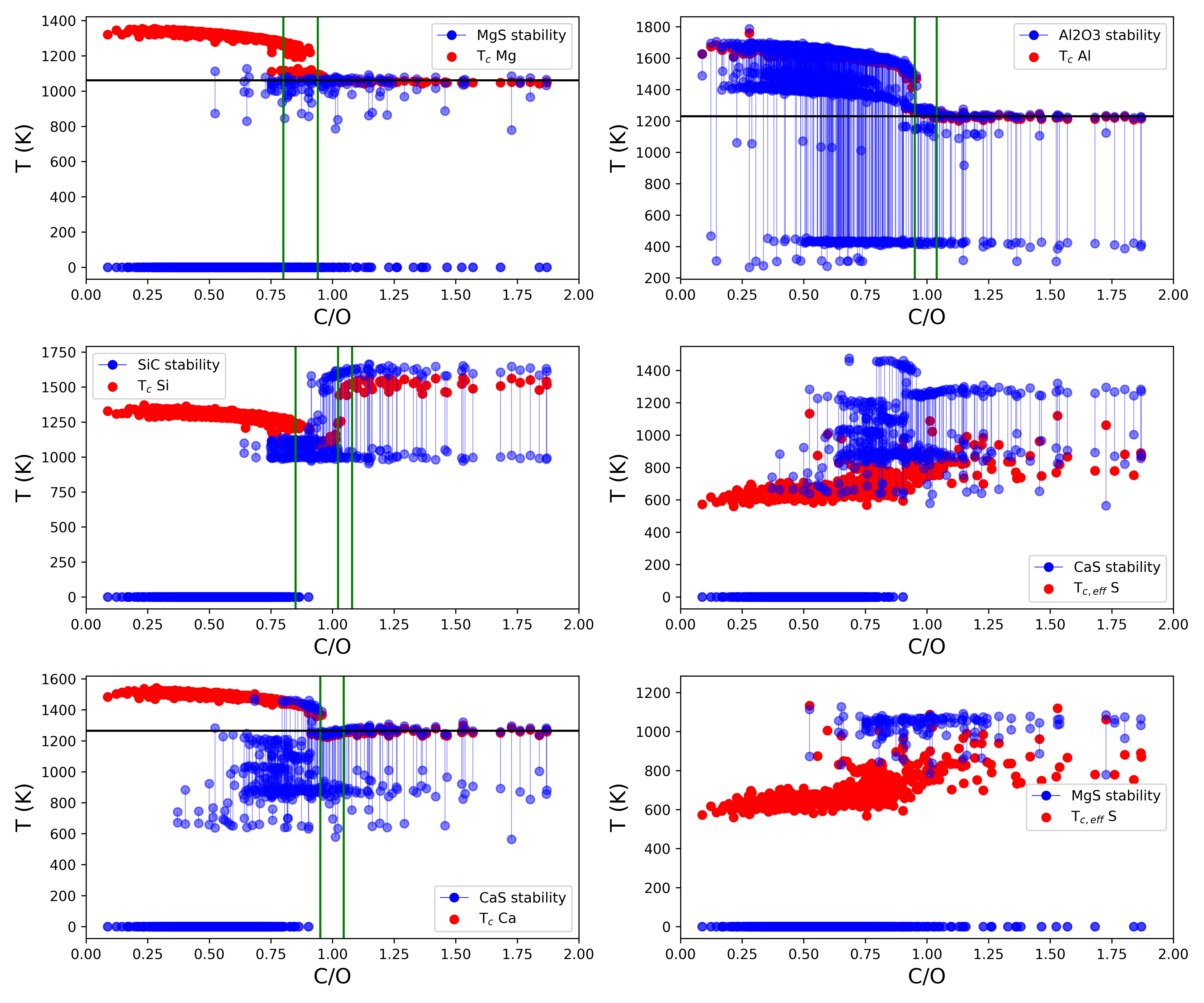}}
      \caption[High C/O condensate stability and Tc of relevant elements]{Temperature and bulk disk C/O range at which the abundance of solid MgS, SiC, CaS, and Al$_2$O$_3$ exceeds $10^{-7}$n$_{\text{H}}$, where n$_{\text{H}}$ is the molar H abundance of the gas. These solids form the primary condensates in high-C/O disks of the rock-forming elements Ca, Al, Mg, and Si, and of S. Condensation temperatures for these elements (red) are highly correlated to most of these condensate stability fields. Vertical green lines indicate transitions in $\Tc{}$ behaviour, as detailed in Sect.\ \ref{sec:App_TcParam}, and elements with constant $\Tc{}$ at high C/O display the value of $\Tc{}$ as a horizontal black line.}
      \label{fig:App_high_CO_condensates}
\end{figure*}

\begin{figure*}[h!]
    \includegraphics[width=0.49\hsize]{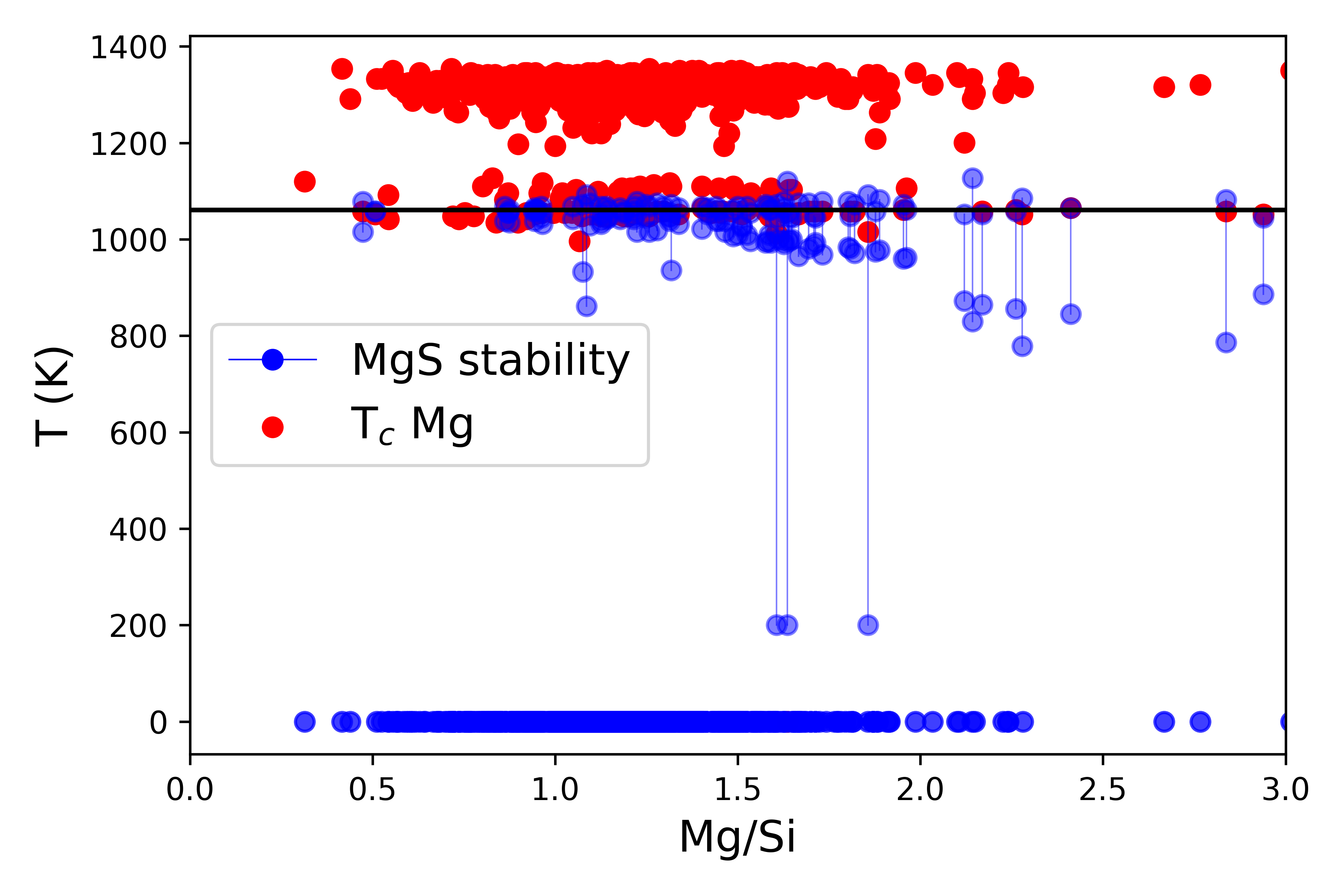}
    \includegraphics[width=0.49\hsize]{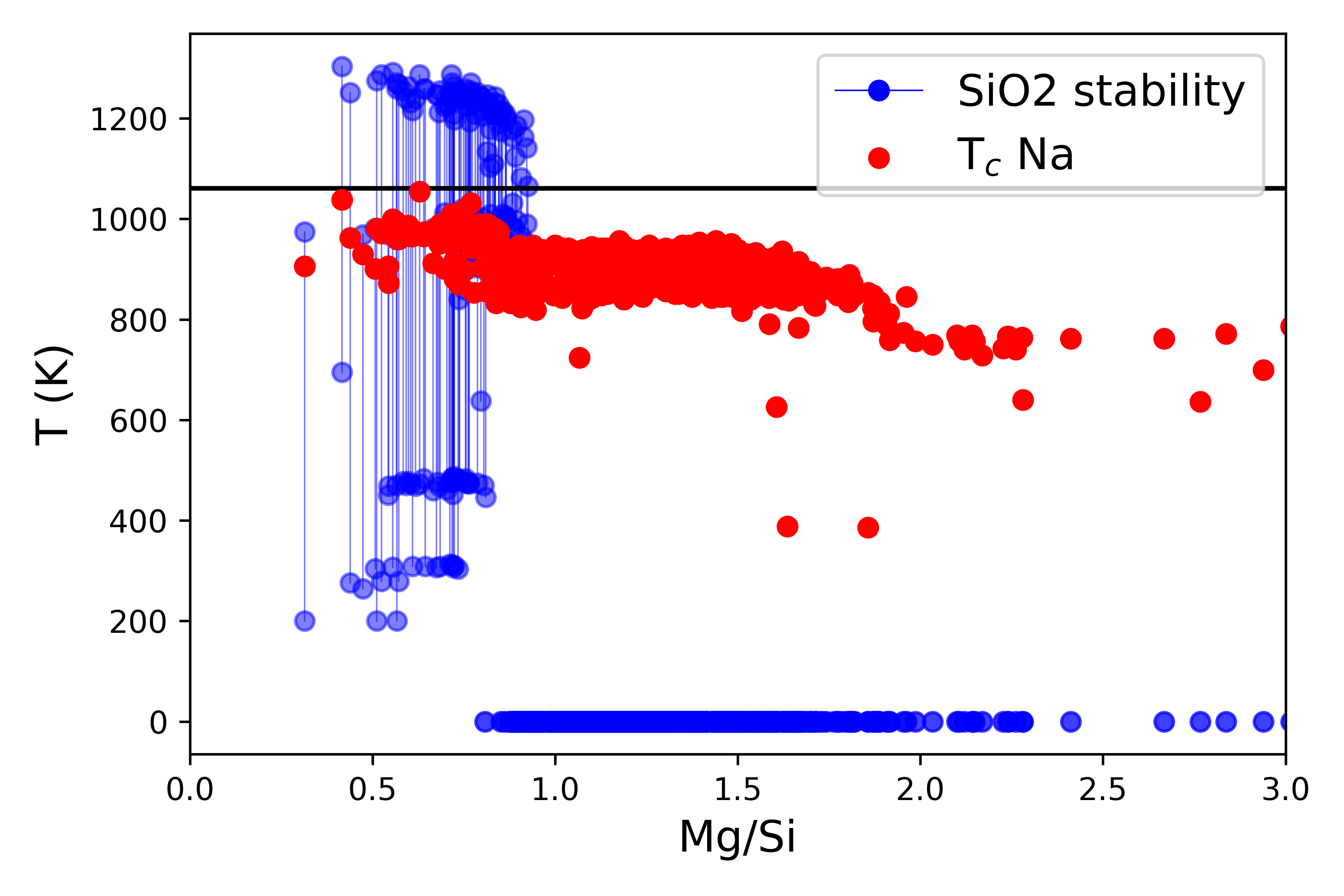}
      \caption[Stability of MgS and SiO2 as a function of bulk disk Mg/Si and temperature]{Temperature and bulk disk Mg/Si range at which the abundances of solid MgS (left) and SiO$_2$ (right) exceed $10^{-7}$n$_{\text{H}}$, where n$_{\text{H}}$ is the molar H abundance of the gas. Condensation temperature of Mg (left; red) at high C/O, indicated by a black line, is very similar to that of MgS. Condensation temperature of Na (right; red) is elevated when SiO$_2$ is stable over a temperature range stretching below $\Tc{Na}$.}
      \label{fig:App_MgS_stability_vs_MgSi}
\end{figure*}

At high C/O, both Ca and Mg tend to form sulphides rather than oxides. Specifically CaS has a wide stability field, and becomes the primary Ca-bearing condensate at C/O as low as 0.8 (Fig.\ \ref{fig:App_high_CO_condensates}). From this point, $\Tc{Ca}$ is identical to the condensation temperature of CaS. The stability of MgS is much more limited however, and Mg primarily condenses as silicates even at high C/O. The width of the MgS stability field does correlate with disk Mg/Si (Fig.\ \ref{fig:App_MgS_stability_vs_MgSi}), indicating that MgS formation is primarily linked to the availability of Si to form forsterite. Note that while we find a relation between $\Tceff{S}$ and MgS condensation, we also assume constant S/Si due to the absence of S abundances in the GALAH catalogue. It is thus not possible to discern between the effects of Mg/S and Mg/Si. The effective condensation temperature of S increases when CaS and MgS start condensing, though $\Tceff{S}$ is not clearly correlated to $\Tc{MgS}$ and $\Tc{CaS}$. For Al, the primary condensate is always Al$_2$O$_3$, and their condensation temperatures directly correlate. Finally, $\Tc{Si}$ is correlated to $\Tc{SiC}$, but condensation of SiC does not immediately take up more than half of available Si. Therefore, $\Tc{Si}$ is always lower than $\Tc{SiC}$ at high C/O. At low C/O, most elements have condensation temperatures correlating primarily with their own abundances, and the disk C/O ratio. One major exception is Na, which correlates stronger with Mg/Si than with C/O. As Mg/Si increases, the stable Na phase shifts from albite (NaAlSi$_3$O$_8$) to nepheline (NaAlSiO$_4$) to sodium metasilicate (Na$_2$SiO$_3$), which also leads to a decreasing $\Tc{Na}$. Interestingly, the stability field of albite further depends on Mg/Si, and increases by 50\,K once SiO$_2$ becomes a stable condensate at low Mg/Si (Fig.\ \ref{fig:App_MgS_stability_vs_MgSi}).

\section{Effective condensation temperatures}
\begin{figure*}[h!]
    \resizebox{\hsize}{!}{\includegraphics{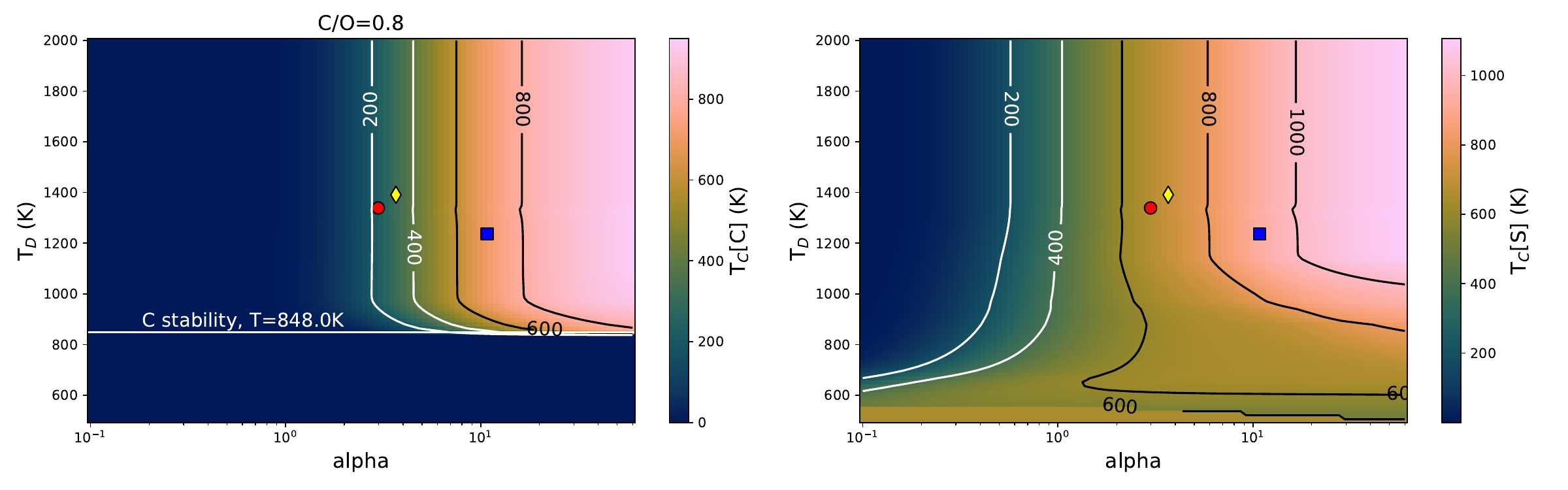}}
    \vspace{0.5cm}
    \resizebox{\hsize}{!}{\includegraphics{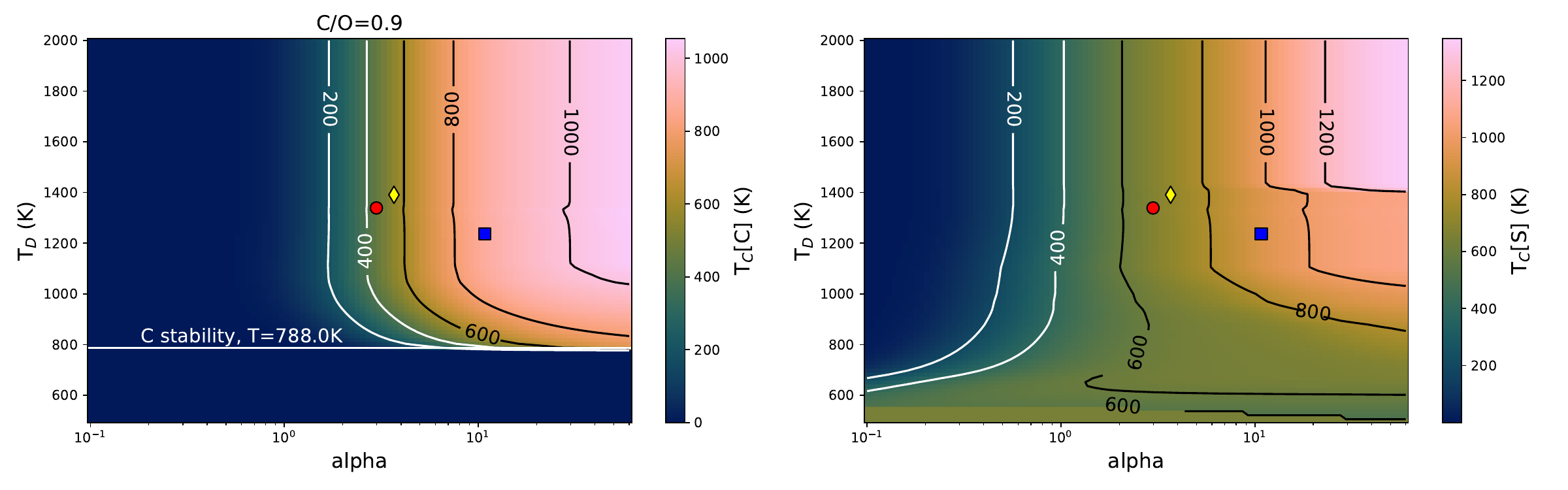}}
    \caption[Tc,eff of C and S as a function of devolatilisation trend, for C/O = 0.8 and 0.9]{Effective condensation temperatures (K) of C (left) and S (right) as a function of devolatilisation trend, in the shape of Eq.\ \ref{eq:devol_trend_Wang}. The bulk disk composition is solar \citep{Lodders2003}, adjusting only C and O abundances such that bulk disk C/O ratios are 0.8 (top) and 0.9 (bottom), while conserving $10^{A_C}+10^{A_O}$. We visualise $T_D=10^{-\beta/\alpha}$ rather than parameter $\beta$ on the y-axis. In addition, we show the lowest $T_D$ in which C may still condense. The devolatilisation trends of Earth (yellow diamond), Mars (red circle), and Vesta (blue square) are indicated.}
    \label{fig:App_Tc_CS_vs_devol_low_CO}
\end{figure*}
\begin{figure*}[h!]
    \resizebox{\hsize}{!}{\includegraphics{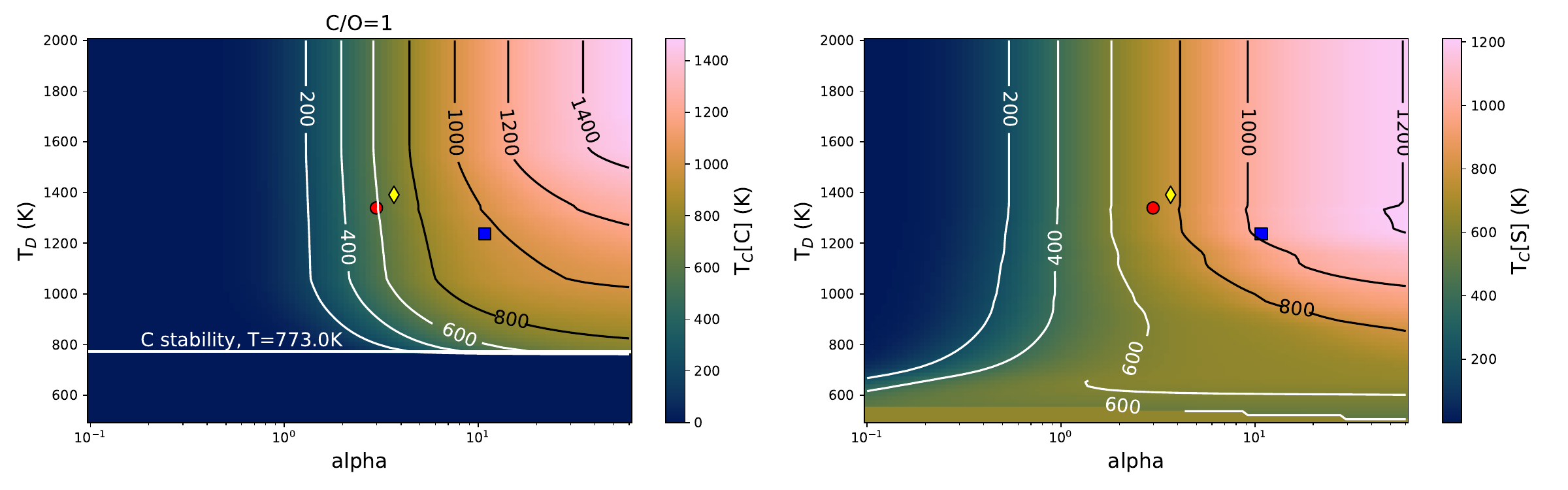}}
    \vspace{0.5cm}
    \resizebox{\hsize}{!}{\includegraphics{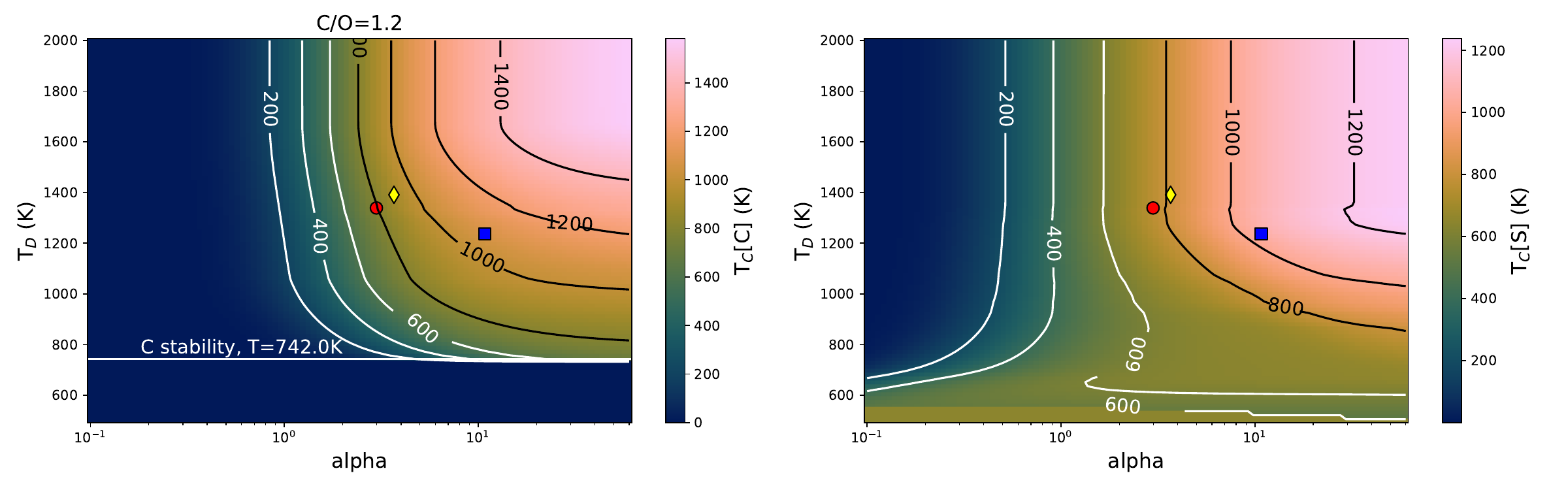}}
    \caption[Tc,eff of C and S as a function of devolatilisation trend, for C/O = 0.8 and 0.9]{Effective condensation temperatures (K) of C (left) and S (right) as a function of devolatilisation trend, in the shape of Eq.\ \ref{eq:devol_trend_Wang}. The bulk disk composition is solar \citep{Lodders2003}, adjusting only C and O abundances such that bulk disk C/O ratios are 1.0 (top) and 1.2 (bottom), while conserving $10^{A_C}+10^{A_O}$. We visualise $T_D=10^{-\beta/\alpha}$ rather than parameter $\beta$ on the y-axis. In addition, we show the lowest $T_D$ in which C may still condense. The devolatilisation trends of Earth (yellow diamond), Mars (red circle), and Vesta (blue square) are indicated.}
    \label{fig:App_Tc_CS_vs_devol_high_CO}
\end{figure*}
The effective condensation temperatures of elements are specific to a particular devolatilisation trend. This is specifically relevant to C and S, which have condensation curves that are not monotonically increasing. To generalise our effective condensation temperature approach to planets beyond Earth-like volatile depletion patterns, we show $\Tceff{C}$ and $\Tceff{S}$ values for a range of devolatilisation trends (Figs.\ \ref{fig:App_Tc_CS_vs_devol_low_CO}, \ref{fig:App_Tc_CS_vs_devol_high_CO}). The effective temperatures increase steadily with both increasing $\alpha$ and $\beta$, although we note that increased $\Tceff{}$ does not correspond to increased abundance of that element in the planet. Vesta uniformally has higher $\Tceff{C}$ and $\Tceff{S}$ than Earth, while Mars always has slightly lower temperatures (see also Fig.\ \ref{fig:Mars_Vesta}). Graphite also remains stable to lower temperatures as C/O increases.

\section{Devolatilisation controls on composition}\label{sec:App_Devol_heatmaps}
\begin{figure*}[h!]
    \resizebox{0.8\hsize}{!}{\includegraphics{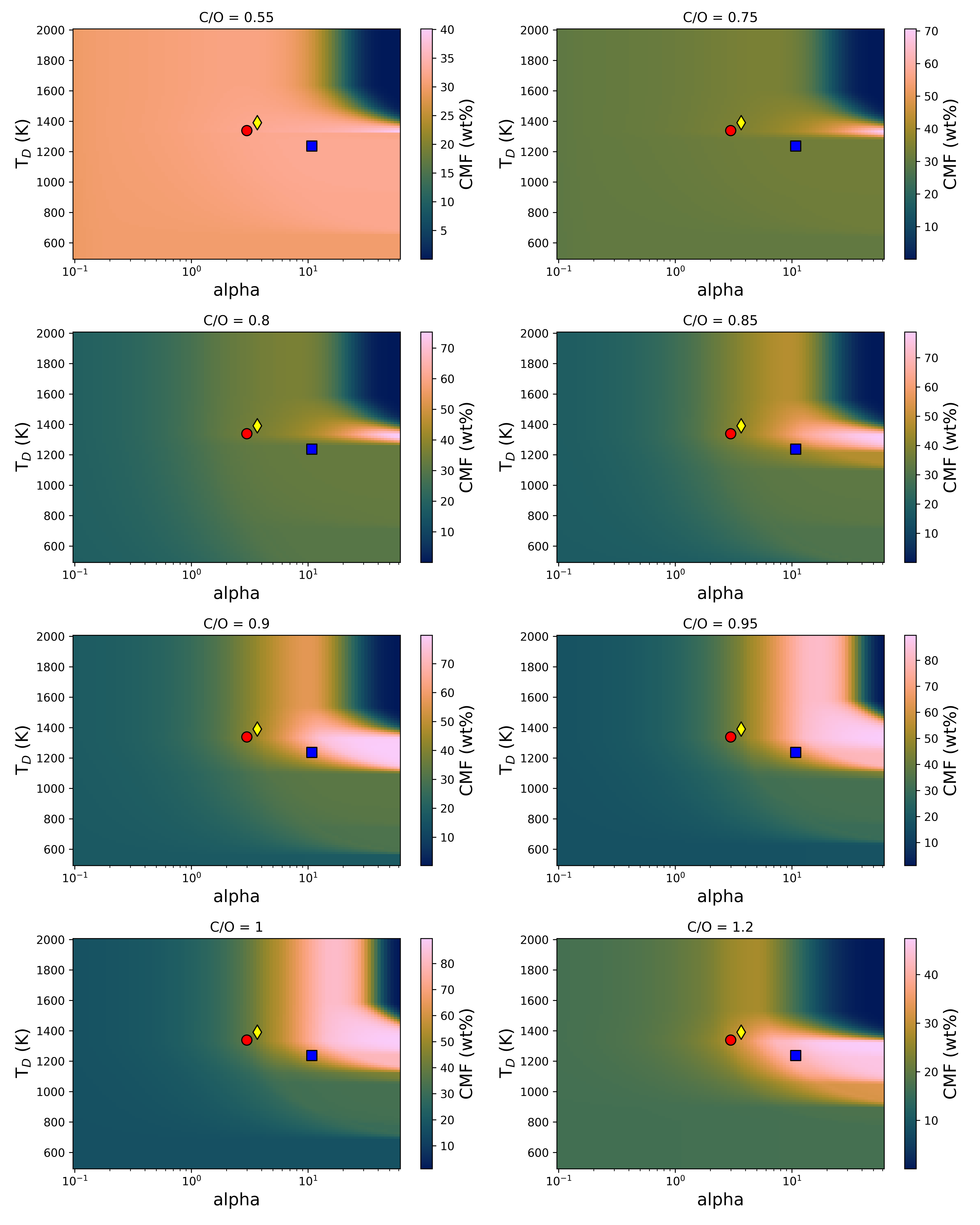}}
      \caption[Planet core mass fraction as a function of devolatilisation alpha and beta]{The core mass fractions (CMF, in wt.\%) of planets forming with various devolatilisation trend shapes, given in terms of $\alpha$ and $T_D = 10^{-\beta/\alpha}$, where $\alpha$ and $\beta$ are seen in Eq.\ \ref{eq:devol_trend_Wang}. disk compositions are largely solar, except in C and O abundances, where C/O is varied while preserving $\epsilon_{\mathrm{C}}$ and $\epsilon_{\mathrm{O}}$. The devolatilisation trends of Earth (yellow diamond), Mars (red circle), and Vesta (blue square) are indicated.}
      \label{fig:App_Devol_heatmap_CMF}
\end{figure*}

Throughout most of the main paper, we focus on applying the Earth-Sun devolatilisation trend to stellar abundances to simulate exoplanet compositions. While these calculations can be extended to exoplanets with any arbitrary devolatilisation trend, the mechanisms behind shaping these trends are poorly understood. Here, we illustrate how the shape of the devolatilisation pattern affects the planet composition. We maintain the general shape of the Earth-Sun trend (Eq.\ \ref{eq:devol_trend_Wang}), and vary parameters $\alpha$ and $\beta$. We perform this for a range of disk C/O ratios, keeping all other abundances equal to solar. At solar C/O ratio, core mass fractions (CMF, defined as bulk planet Fe+Ni mass fraction) are uniform in most of the parameter space (Fig.\ \ref{fig:App_Devol_heatmap_CMF}). There is a narrow region at $\alpha > 11$ and $\beta < -35$ where CMF goes to zero, as planets here only consist of Ca and Al oxides; however, it is very unlikely that such planets can actually form, given the small quantity of dust available for building planets. At higher C/O, a region appears at high $\alpha$ and Earth-like $T_D = 10^{-\beta/\alpha}$ where CMF can reach up to 70 wt\%. This is caused by the lower condensation temperatures of Mg and Si at higher C/O, creating a window where planets sampling a narrow region in temperature space accrete Fe-bearing material but very little Mg- and Si-bearing material. Especially at $\mathrm{C/O}\geq0.9$, a body with a similar devolatilisation trend to Vesta's can have CMF as high as 70 wt\% (Fig.\ \ref{fig:Mars_Vesta}). Thus, the shape of the devolatilisation trend plays a first-order role in shaping a planets structure.

\begin{figure*}[h!]
    \resizebox{0.8\hsize}{!}{\includegraphics{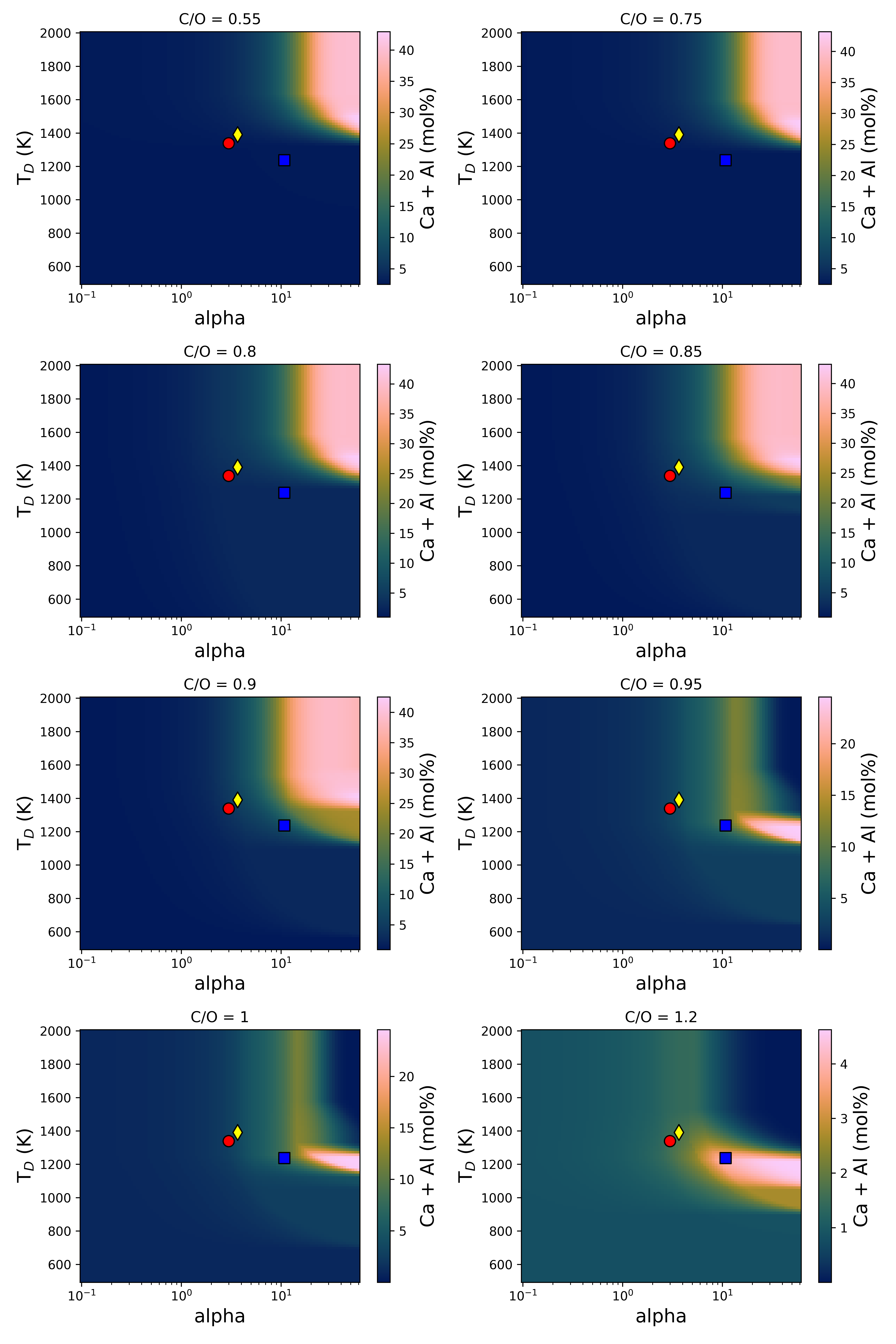}}
      \caption[Planet refractory abundance as a function of devolatilisation alpha and beta]{The molar refractory element (Ca and Al) abundance of planet mantles forming with various devolatilisation trend shapes, given in terms of $\alpha$ and $T_D = 10^{-\beta/\alpha}$, where $\alpha$ and $\beta$ are seen in Eq.\ \ref{eq:devol_trend_Wang}. disk compositions are largely solar, except in C and O abundances, where C/O is varied while preserving $\epsilon_{\mathrm{C}}$ and $\epsilon_{\mathrm{O}}$. The devolatilisation trends of Earth (yellow diamond), Mars (red circle), and Vesta (blue square) are indicated. Note that we plot mol\% of Ca and Al here, which is limited by the abundance of oxygen that appears together with the refractory elements.}
      \label{fig:App_Devol_heatmap_CaAl}
\end{figure*}
Planets accrete high reservoirs of the refractory rock-forming elements Ca and Al when they primarily sample high-temperature space in low-C/O disks ($\mathrm{C/O}\leq0.9$), with the Ca and Al concentrations decreasing quickly as the temperature-space feeding zone widens (i.e.\ $\alpha$ decreases), or when it samples only lower temperatures (i.e.\ $T_D = 10^{-\beta/\alpha}$) decreases (Fig.\ \ref{fig:App_Devol_heatmap_CaAl}). At higher C/O, both Ca and Al become less refractory, and planets forming with major Ca+Al reservoirs occupy a small region of ($\alpha$,$\beta$) parameter space, primarily at $\alpha>11$, similar to the high-CMF planets. This enrichment of Ca and Al disappears at high C/O, as graphite is stable to very high temperatures. At shallower devolatilisation trends (i.e.\ lower $\alpha$) or trends sampling only low temperature space, the CMF and abundances of Ca and Al remain fairly uniform; here, primarily the abundances of moderately volatile elements vary, where especially Na could have a major effect on planet properties. We predict the widest variation in planet compositions for devolatilisation trends steeper than that of Earth or sampling a higher temperature range. However, the shape of the devolatilisation trend is an important control on planet composition throughout the entire parameter space, especially when more volatile material (e.g.\ water ice) is considered than what we cover here. Formation models that accurately reproduce the devolatilisation patterns observed in the Solar System are required to predict how devolatilisation patterns can vary among exoplanets, and should be able to expand our understanding of compositional diversity of rocky exoplanets further.

\FloatBarrier
\clearpage

\end{appendix}

\end{document}